\documentclass[fleqn,usenatbib]{mnras}

\usepackage[flushleft]{threeparttable}
\usepackage{pifont}
\usepackage{longtable}

\usepackage[T1]{fontenc}
\usepackage{amsmath} 
\newcommand{\angstrom}{\textup{\AA}}

\DeclareRobustCommand{\VAN}[3]{#2}
\let\VANthebibliography\thebibliography
\def\thebibliography{\DeclareRobustCommand{\VAN}[3]{##3}\VANthebibliography}
\newcommand{\cmark}{\ding{51}}%
\newcommand{\xmark}{\ding{55}}%

\usepackage{graphicx}	
\usepackage{amsmath}	
\usepackage{amssymb}	
\usepackage{newtxtext,newtxmath}

\def\gax{\mathrel{\raise.3ex\hbox{$>$}\mkern-14mu\lower0.6ex\hbox{$\sim$}}}
\def\lax{\mathrel{\raise.3ex\hbox{$<$}\mkern-14mu\lower0.6ex\hbox{$\sim$}}}
\def\gtorder{\mathrel{\raise.3ex\hbox{$>$}\mkern-14mu
             \lower0.6ex\hbox{$\sim$}}}
\def\ltorder{\mathrel{\raise.3ex\hbox{$<$}\mkern-14mu
             \lower0.6ex\hbox{$\sim$}}}

\title[V723 Mon]{A Unicorn in Monoceros: the $3M_\odot$ dark companion to the bright, nearby red giant V723 Mon is a non-interacting, mass-gap black hole candidate}

\author[T. Jayasinghe et al.]{T. Jayasinghe$^{1,2}$\thanks{E-mail: jayasinghearachchilage.1@osu.edu},
K. Z. Stanek$^{1,2}$,
Todd A. Thompson$^{1,2}$,
C. S. Kochanek$^{1,2}$,
D. M. Rowan$^{1,2}$,
P. J. Vallely$^{1,2}$,
\newauthor 
K. G. Strassmeier $^{3}$,
M. Weber $^{3}$,
J. T. Hinkle$^{4}$,
F.-J. Hambsch$^{5,6,7}$,
D. Martin$^{1,8}$,
J. L. Prieto$^{9,10}$,
T. Pessi$^{9}$,
\newauthor 
D. Huber$^{4}$,
K. Auchettl$^{11,12,13}$,
L. A. Lopez$^{1,2}$,
I. Ilyin$^{3}$,
C. Badenes$^{14}$,
A. W. Howard$^{15}$,
\newauthor 
H. Isaacson$^{16,17}$,
S. J. Murphy$^{18,19}$
\\
$^{1}$Department of Astronomy, The Ohio State University, 140 West 18th Avenue, Columbus, OH 43210, USA\\
$^{2}$Center for Cosmology and Astroparticle Physics, The Ohio State University, 191 W. Woodruff Avenue, Columbus, OH 43210, USA\\
$^{3}$Leibniz Institute for Astrophysics Potsdam (AIP), An der Sternwarte 16, D-14482 Potsdam, Germany\\
$^{4}$Institute for Astronomy, University of Hawai`i, 2680 Woodlawn Drive, Honolulu, HI 96822, USA\\
$^{5}$Vereniging Voor Sterrenkunde (VVS), Oostmeers 122 C, 8000 Brugge, Belgium\\
$^{6}$Bundesdeutsche Arbeitsgemeinschaft für Veränderliche Sterne e.V. (BAV), Munsterdamm 90, D-12169 Berlin, Germany\\
$^{7}$American Association of Variable Star Observers (AAVSO), 49 Bay State Road, Cambridge, MA 02138, USA\\
$^{8}$Fellow of the Swiss National Science Foundation\\
$^{9}$N\'ucleo de Astronom\'ia de la Facultad de Ingenier\'ia y Ciencias, Universidad Diego Portales, Av. Ej\'ercito 441, Santiago, Chile\\
$^{10}$Millennium Institute of Astrophysics, Santiago, Chile\\
$^{11}$School of Physics, The University of Melbourne, Parkville, VIC 3010, Australia\\
$^{12}$ARC Centre of Excellence for All Sky Astrophysics in 3 Dimensions (ASTRO 3D)\\
$^{13}$Department of Astronomy and Astrophysics, University of California, Santa Cruz, CA 95064, USA\\
$^{14}$Department of Physics and Astronomy and Pittsburgh Particle Physics, Astrophysics and Cosmology Center (PITT PACC),\\
University of Pittsburgh, 3941 O‘Hara Street, Pittsburgh, PA 15260, USA\\
$^{15}$Department of Astronomy, California Institute of Technology, Pasadena, CA 91125, USA\\
$^{16}$Department of Astronomy,  University of California Berkeley, Berkeley CA 94720, USA\\
$^{17}$Centre for Astrophysics, University of Southern Queensland, Toowoomba, QLD, Australia\\
$^{18}$Sydney Institute for Astronomy (SIfA), School of Physics, University of Sydney, NSW 2006, Australia\\
$^{19}$Stellar Astrophysics Centre, Department of Physics and Astronomy, Aarhus University, 8000 Aarhus C, Denmark\\
}

\date{Accepted XXX. Received YYY; in original form ZZZ}

\pubyear{2021}

\begin{document}
\label{firstpage}
\pagerange{\pageref{firstpage}--\pageref{lastpage}}
\maketitle

\begin{abstract}
 We report the discovery of the closest known black hole candidate as a binary companion to V723 Mon. V723 Mon is a nearby ($d\sim460\,\rm pc$), bright ($V\simeq8.3$~mag), evolved ($T_{\rm eff, giant}\simeq4440$~K, and $L_{\rm giant}\simeq173~L_\odot$) red giant in a high mass function, $f(M)=1.72\pm 0.01~M_\odot$, nearly circular binary ($P=59.9$~d, $e\simeq 0$). V723 Mon is a known variable star, previously classified as an eclipsing binary, but its All-Sky Automated Survey (ASAS), Kilodegree Extremely Little Telescope (KELT), and Transiting Exoplanet Survey Satellite (\textit{TESS}) light curves are those of a nearly edge-on ellipsoidal variable. Detailed models of the light curves constrained by the period, radial velocities and stellar temperature give an inclination of $87.0^\circ{}^{+1.7^{\circ}}_{-1.4^{\circ}} $, a mass ratio of $q\simeq0.33\pm0.02$, a companion mass of $M_{\rm comp}=3.04\pm0.06~M_\odot$, a stellar radius of $R_{\rm giant}=24.9\pm0.7~R_\odot$, and a giant mass of $M_{\rm giant}=1.00\pm0.07~ M_\odot$. We identify a likely non-stellar, diffuse veiling component with contributions in the $B$ and $V$-band of ${\sim}63\%$ and ${\sim}24\%$, respectively. The SED and the absence of continuum eclipses imply that the companion mass must be dominated by a compact object. We do observe eclipses of the Balmer lines when the dark companion passes behind the giant, but their velocity spreads are low compared to observed accretion disks. The X-ray luminosity of the system is $L_{\rm X}\simeq7.6\times10^{29}~\rm ergs~s^{-1}$, corresponding to $L/L_{\rm edd}{\sim}10^{-9}$. The simplest explanation for the massive companion is a single compact object, most likely a black hole in the ``mass gap''.   
\end{abstract}

\clearpage



\section{Introduction}

The discovery and characterization of neutron stars and black holes in the Milky Way is crucial for understanding core-collapse supernovae and massive stars. This is inherently challenging, partly because isolated black holes are electromagnetically dark and partly because compact object progenitors (OB stars) are rare. To date, most mass measurements for neutron stars and black holes come from pulsar and accreting binary systems selected from radio, X-ray, and gamma-ray surveys (see, for e.g., \citealt{Champion2008,Liu2006,Ozel2010,Farr2011}), and from the LIGO/Virgo detections of merging systems (see, for e.g., \citealt{Abbott2016,Abbott2017NS}). Interacting and merging systems are however a biased sampling of compact objects. A more complete census is needed to constrain their formation pathways.

One important component of such a census is to identify non-interacting compact objects in binaries around luminous companions. By their very nature, interacting black holes only sample a narrow range of binary configurations, and almost the entire parameter space of binaries with black holes that are non-interacting remains unexplored. Interacting compact object binaries are only active for relatively short periods of time, so most systems are quiescent or non-interacting. The discovery and characterization of these non-interacting black holes are important for understanding the birth mass distribution of black holes and their formation mechanisms. 

With advances in time-domain astronomy and precision \textit{Gaia} astrometry \citep{GaiaMAIN}, a significant number of these systems should be discoverable. For example, \citet{Breivik2017} estimated that ${\sim}10^3-10^4$ non-interacting black holes are detectable using astrometry from \textit{Gaia}. Similarly, \citet{Shao2019} used binary population synthesis models to estimate that there are ${\sim}10^3$ detached non-interacting black holes in the Milky Way, with $10^2$ of these systems having luminous companions that are brighter than $G{\sim}20$ mag. 

\citet{Thompson2019} recently discovered the first low-mass ($M_{\rm BH}\simeq3.3_{-0.7}^{+2.8}~M_\odot$) non-interacting black hole candidate in the field. It is in a circular orbit with $\rm P_{\rm orb}\sim83\,d$ around a spotted giant star. Other non-interacting BH candidates have been discovered in globular clusters: one by \citet{Giesers2018} in NGC 3201 (minimum black hole mass  $M_{\rm BH} = 4.36 \pm0.41$\,M$_\odot$), and two by \citet{Giesers2019} in NGC 3201 ($M_{\rm BH}\sin(i) = 7.68 \pm0.50$\,M$_\odot$ and $M_{\rm BH}\sin(i) = 4.4 \pm2.8$\,M$_\odot$). While obviously interesting in their own right, these globular cluster systems likely have formation mechanisms that are very different from those of field black hole binaries.

Other claims for non-interacting BH systems have been ruled out. For example, LB-1, which was initially thought to host an extremely massive stellar black hole ($M_{\rm BH}\simeq68_{-3}^{+11}~M_\odot$, \citealt{Liu2019}), was later found to have a much less massive companion that was not necessarily a compact object (see, for e.g., \citealt{Shenar2020,Irrgang2020,Abdul-Masih2020,El-Badry2020a}). Similarly, the naked-eye star HR 6819 was claimed to be a triple system with a detached black hole with $M_{\rm BH} = 6.3 \pm0.7~M_\odot$ \citep{Rivinius2020}, but was later argued to contain a binary system with a rapidly rotating Be star and a slowly rotating B star \citep{El-Badry2020b,Bodensteiner2020}.

Here we discuss our discovery that the bright red giant V723 Mon has a dark, massive companion that is a good candidate for the closest known black hole. We discuss the current classification of this system in Section \ref{section:v723}, and describe the archival data and new observations used in our analysis in Section \ref{section:v723obs}. In Section \ref{section:v723res}, we analyze photometric and spectroscopic observations to derive the parameters of the binary system and the red giant secondary. In Section \ref{section:v723disc}, we discuss the nature of the dark companion. We present a summary of our results in Section \ref{section:v723conc}. 

\section{V723 Mon}
\label{section:v723}
V723 Mon (HD 45762, SAO 133321, TIC 43077836) is a luminous ($m_V\simeq8.3)$ red-giant\footnote{V723 Mon has previously been assigned a spectral type of G0 II \citep{Houk2000}, however, in this work we find that it is more consistent with a K0/K1 III spectral type.} in the Monoceros constellation with J2000 coordinates $(\alpha,\delta)=(97.269410^\circ,-5.572286^\circ)$. It was classified as a likely long period variable in the General Catalogue of Variable Stars (GCVS; \citealt{Kazarovets1999}) after it was identified as a variable source in the Hipparcos catalogue with period $\rm P=29.97\,d$ \citep{ESA1997}. Subsequently, the All-Sky Automated Survey (ASAS) \citep{Pojmanski1997,Pojmanski2002} classified it as a contact/semi-detached binary with $\rm P=59.87\,d$. The Variable Star Index (VSX; \citealt{Watson2006}) presently lists it as an eclipsing binary of the $\beta$-Lyrae type (EB) with $\rm P=59.93\,d$. 

V723 Mon has a well determined spectroscopic orbit and is included in the the Ninth Catalogue of Spectroscopic Binary Orbits \citep{Pourbaix2004}.  In particular, \citet{Griffin2010} identified V723 Mon as a single-lined spectroscopic binary (SB1) with a nearly circular $P{\sim}60$~d orbit. \citet{Strassmeier2012} (hereafter S12) refined the orbit to $P_{\rm orb}=59.9363\pm0.0016$~d and $e_{\rm orb}{\simeq}0.0150\pm0.0009$ using a large number of high precision RV measurements from STELLA \citep{Weber2008}. S12 also argue that this is the outer orbit of a triple system, and that the more massive
(inner) component is an SB1 consisting of another giant star and an unseen companion with a period $P_{\rm inner}\simeq P_{\rm outer}/3$. \citet{Griffin2014} (hereafter G14) was unable to find a spectral feature indicative of a second companion in their cross-correlation functions. G14 discusses several peculiarities in the S12 RV solution. In particular, the radial velocity curve associated with the second component is unusual in structure compared to any other system characterized by S12 and a triple system with this period ratio would almost certainly be dynamically unstable.

The most striking feature of the well-measured $60$~day RV curve is its large mass function of 
\begin{equation}
    f(M) = \frac{P_{\rm orb}K^3(1-e^2)^{3/2}}{(2 \pi G)}= \frac{M_{\rm comp}^3 \sin^3i}{(M_{\rm giant}+M_{\rm comp})^2} \simeq1.73\, M_\odot,
\end{equation} given $P_{\rm orb}=59.9363$~d, $e=0.015$ and $K=65.45~\rm km~s^{-1}$ from S12. If the observed giant has a mass of $M_{\rm giant}\sim1\, M_\odot$, the mass function implies a massive companion with a minimum mass of $M_{\rm comp}\sim 3\, M_\odot$. Since the observed light is clearly
dominated by the giant and the companion has to be both much less luminous and significantly more massive than the giant, V723~Mon is a prime candidate for a non-interacting, compact object binary.  This realization led us to investigate V723~Mon in detail as part of a larger project to identify non-interacting, compact object binaries.

\clearpage
\section{Observations}
\label{section:v723obs}
Here we present observations, both archival and newly acquired, that are used in our analysis of V723 Mon.

\subsection{Distance, Kinematics and Extinction}

In \textit{Gaia} EDR3 \citep{GaiaEDR3MAIN}, V723 Mon is {\verb"source_id="3104145904761393408}. Its EDR3 parallax of $\varpi_{\rm EDR3}=2.175 \pm0.033 \,\rm mas$ implies a distance of $d=1/\varpi=460\pm 7$~pc.  This is little changed from its \textit{Gaia} DR2 parallax of $\varpi_{\rm DR2}=1.9130 \pm0.0563$~mas.  At these distances, there is little difference between $d=1/\varpi=519 \pm 15$~pc (for DR2) and the more careful estimate of  $d=515.6 ^{+15.6} _{-14.6}\,\rm pc$ by \citet{BailerJones2018}.  The astrometric solution has significant excess noise of $\Delta=0.22$~mas, which is not surprising given that the motion of the giant should be $\sim 0.8$~mas.  However, its renormalized unit weight error (RUWE) of $1.39$, while larger than unity, is not indicative of problems in the parallax. We adopt a distance of $d=460$~pc for the remainder of the paper. The distance uncertainties are unimportant for our analysis.

V723 Mon has Galactic coordinates $(l,b)\simeq(215.372^\circ,-7.502^\circ)$, close to the Galactic disk, but away from the Galactic center. At the EDR3 distance, V723 Mon is ${\sim}32$~pc below the midplane. Its proper motion in EDR3 is $\mu_{\alpha}=-1.347\pm0.032 \, \rm mas\;yr^{-1}$, and $\mu_{\delta}=16.140\pm0.031 \, \rm mas\;yr^{-1}$.  Combining this with the 
 systemic radial velocity from $\S$\ref{section:orbit} and the definition of the local standard of rest (LSR) from \citet{Schonrich2010}, the 3D space motion of V723 Mon relative to the LSR as $(U,V,W)_{\rm LSR}=(-10.8\pm4.3,36.9\pm5.3,20.1\pm3.4)$ $\rm km~s^{-1}$ using \verb"BANYAN" \citep{Gagne2018} for the calculations. We calculated the thin disk, thick disk and halo membership probabilities based on the $UVW$ velocities following \citet{Ramirez2013} to obtain $P(\rm thin)\simeq97\%$, $P(\rm thick)\simeq3\%$ and $P(\rm halo)\simeq0\%$, respectively. This suggests that this system is a kinematically normal object in the thin disk.

\textit{Gaia} DR2 \citep{GaiaMAIN} also reports a luminosity $L_{\rm Gaia}=165.5\pm6.5\,\rm L_\odot,$ temperature $T_{\rm eff,Gaia}=4690^{+40}_{-30}$~K, and radius $R_{\rm Gaia}=19.5^{+0.2}_{-0.4}\,\rm R_\odot,$ for the star that are consistent with an evolved red giant. While \textit{Gaia} DR2 does not report a value for the reddening towards V723 Mon, \citet{Gontcharov2017} reports $E(B-V)\simeq0.10$. The three-dimensional dust maps of \citet{Green2019} give $E(B-V) \simeq 0.10 \pm0.04$ at the \textit{Gaia} distance, consistent with this estimate.

\subsection{Light Curves}

We analyzed well-sampled light curves from the All-Sky Automated Survey (ASAS) and the Kilodegree Extremely Little Telescope (KELT), a densely sampled but phase-incomplete light curve from the Transiting Exoplanet Survey Telescope (TESS), $BVR_cI_c$ light curves from the Remote Observatory Atacama Desert (ROAD) and a sparse ultraviolet (UV) light curve from the Neil Gehrels Swift Observatory.

ASAS \citep{Pojmanski1997,Pojmanski2002} obtained a $V$-band light curve of V723 Mon spanning from November 2000 to December 2009 (${\sim}54$ complete orbits). We selected $580$ epochs with \verb"GRADE=A" or \verb"GRADE=B" for our analysis. V723 Mon clearly varies in the ASAS light curve, with two equal maxima but two unequal minima, when phased with the orbital period from S12.
We determined the photometric period using the \verb"Period04" software \citep{Lenz2005}. The dominant ASAS period of $P_{\rm ASAS}{\simeq}29.9674\pm0.0138$~d corresponds to $P_{\rm orb}/2$. Once the ASAS light curve was whitened using a sinusoid, we find an orbital period of
\begin{equation}
     P_{\rm orb,ASAS}=59.9863\pm0.0551\,{\rm d},
	\label{eq:period}
\end{equation} which agrees well with the spectroscopic periods from S12 and G14.  Unfortunately, V723~Mon is saturated in the Automated Survey for SuperNovae (ASAS-SN;
\citealt{Shappee2014,Kochanek2017}) images, and we could not use it to extend the time span of the V-band data.

The KELT (\citealt{Pepper2007}) light curve contains 1297 epochs
which we retrieved from the Exoplanet Archive\footnote{\url{https://exoplanetarchive.ipac.caltech.edu/}}. The KELT $R_K$ filter can be considered as a very broad Johnson R-band filter \citep{Siverd2012}. However,  there can be significant color corrections compared to a standard Johnson R-band filter for very blue and very red stars \citep{Pepper2007,Siverd2012}. KELT observations were made between September 2010 and February 2015 (${\sim}26$ complete orbits).
The dominant period in the KELT data ($P_{\rm KELT}{\simeq}29.9682\pm0.0279$~d) again corresponds to $P_{\rm orb}/2$. We find an orbital period of
\begin{equation}
     P_{\rm orb,KELT}=60.0428\pm0.1121\,{\rm d}.
	\label{eq:period}
\end{equation} 
The difference between the ASAS and KELT photometric period estimates is not statistically significant.

V723~Mon (\verb"TIC" 43077836) was observed by \textit{TESS} \citep{Ricker2015} in Sector 6, and the 27 days of observations correspond to [0.46,0.82] in orbital phase where the phase of the RV maximum is $0.75$. V723 Mon was also observed in Sector 33, with the observations spanning [0.94,0.36] in orbital phase. We analyzed the \textit{TESS} data using the adaptation of the ASAS-SN image subtraction pipeline for analyzing \textit{TESS} full-frame images (FFIs) described in \cite{Vallely2020}.
While this pipeline produces precise differential flux light curves, the large pixel scale of \textit{TESS} makes it difficult to obtain reliable measurements of the reference flux of a given source. We normalized the light curve to have the reference $TESS$-band magnitude of 7.26 in the \textit{TESS} Input Catalog \citep{2019TIC}.  Conveniently, the mean of the Sector 6 observations is approximately the mean for a full orbital cycle (see Figure~\ref{lcs}).
We use a zero point of 20.44 electrons (\textit{TESS} Instrument Handbook\footnote{\url{https://archive.stsci.edu/files/live/sites/mast/files/home/missions-and-data/active-missions/tess/_documents/TESS_Instrument_Handbook_v0.1.pdf}}).
The light curve does not include epochs where the observations were compromised by the spacecraft's momentum dump maneuvers.

We obtained $BVR_cI_c$ light curves at the Remote Observatory Atacama Desert (ROAD; \citealt{Hambsch2012}). All observations were acquired through Astrodon Photometric filters with an Orion Optics, UK Optimized Dall Kirkham 406/6.8 telescope and a FLI 16803 CCD camera. Twilight sky-flat images were used for flatfield corrections. Reductions were performed with the \verb"MAXIM DL" program\footnote{\url{http://www.cyanogen.com}} and the photometry was carried out using the \verb"LesvePhotometry" program.\footnote{\url{http://www.dppobservatory.net/}}

We obtained Swift UVOT \citep{roming05} images in the $UVM2$ (2246 \AA) band \citep{poole08} through the Target of Opportunity (ToO) program (Target ID number 13777). We only obtained images in the $UVM2$ 
band because the \textit{Swift} $UVW1$ and $UVW2$ filters have significant red leaks that make them unusable in the presence of the emission from the cool giant, and the star is too bright to obtain images in the optical UVOT filters. Each epoch of UVOT data includes multiple observations per filter, which we combined using the \texttt{uvotimsum} package. We then used \texttt{uvotsource} to extract source counts using a 5\farcs{0} radius aperture centered on the star. We measured the background counts using a source-free region with radius of 40\farcs{0} and used the most recent calibrations \citep{poole08, breeveld10} and taking into account the recently released update to the sensitivity correction for the Swift UV filters\footnote{\url{https://www.swift.ac.uk/analysis/uvot/index.php}}. The Swift $UVM2$ observations are summarized in Table \ref{tab:swift}. We report the median and standard deviation of these $UVM2$ observations in Table \ref{tab:phot}.

\subsection{Radial Velocities}

We used two sets of radial velocity (RV) measurements.  The first set, from \citet{Griffin2014},  was obtained between December 2008 and November 2013 as part of the Cambridge Observatory Radial Velocity Program and span 1805 days. The median RV error for this dataset is ${\sim}0.80\, {\rm km\,s^{-1}}$.  These 41 RV epochs were retrieved from the Ninth Catalogue of Spectroscopic Binary Orbits (SB9; \citealt{Pourbaix2004}),\footnote{\url{https://sb9.astro.ulb.ac.be/DisplayFull.cgi?3936+1}} converting the reported
epochs to Barycentric Julian Dates (BJD) on the TDB system \citep{Eastman2010} using \verb"barycorrpy" \citep{Kanodia2018}.

The second set of RV data consists of 100 epochs obtained by S12 with the high resolution ($R \approx 55000$) STELLA spectrograph. STELLA spectra have a wavelength coverage of 390-880 nm and a spectral resolution of $0.12$~\AA~ at 650 nm. Spectra were obtained between November 2006 and April 2010, spanning a baseline of 1213 days. The spectra were reduced following the standard procedures described in \citet{Strassmeier2012} and \citet{Weber2008}. Of the 100 spectra, 75 had $\rm S/N>30$ near 650 nm. There were 87 epochs with good RV measurements for the giant and the median RV error is ${\sim}0.19\, {\rm km\,s^{-1}}$. The STELLA RV measurements are listed in Table \ref{tab:stellarv}.

\subsection{Additional Spectra}

To better understand the V723 Mon system, and to test for possible systematic errors, we obtained a number of additional high and medium resolution spectra   These observations are summarized in Table \ref{tab:spec}. 
Using the HIRES instrument \citep{Vogt1994} on Keck I, we obtained 7 spectra with $R \approx 60000$ between Oct 20 2020 and Dec 26 2020 using the standard California Planet Search (CPS) setup \citep{Howard2010}. The exposure times ranged from 35 to 60 seconds.
We also obtained a very high resolution ($R \approx 220000$) spectrum on 
29 Nov. 2020 using the Potsdam Echelle Polarimetric and Spectroscopic Instrument (PEPSI; \citealt{Strassmeier2015}) on the Large Binocular Telescope.  We used the $100~\mu$m fiber and 6 cross-dispersers (CD). The data were processed as described in \citet{Strassmeier2018}. The total integration time was 90 minutes, and the combined spectrum covers the entire wavelength range accessible to PEPSI ($3840-9070~{\angstrom}$). The spectrum has a signal-to-noise ratio (S/N) of 70 in the wavelength range $4265-4800~{\angstrom}$ (CD2) and S/N$=260$ in the range $7419-9070~{\angstrom}$ (CD6).

Using the medium resolution ($R\approx2000$) Multi-Object Double Spectrographs mounted on the twin 8.4m Large Binocular Telescope \citep[MODS1 and MODS2;][]{PoggeMODS}, we obtained a series of spectra from Nov 18 to Nov 22 2020 as the dark companion moved into eclipse behind the observed giant.
Exposure times were typically 15 seconds, but this was extended to 24 seconds for the Nov 22 observation to compensate for clouds. We used a $2\farcs{4}$ wide slit to ensure that all the light was captured.
We reduced these observations using a standard combination of the \verb"modsccdred"\footnote{\url{http://www.astronomy.ohio-state.edu/MODS/Software/modsCCDRed/}} \textsc{python} package, and the \verb"modsidl" pipeline\footnote{\url{http://www.astronomy.ohio-state.edu/MODS/Software/modsIDL/}}.
The blue and red channels of both spectrographs were reduced independently, and the final MODS spectrum for each night was obtained by averaging the MODS1 and MODS2 spectra. The best weather conditions during this observing run occurred on Nov 20.
The HIRES, PEPSI and MODS observations are summarized in Table \ref{tab:spec}.

\subsection{X-ray data}

We analyzed X-ray observations from the \textit{Swift} X-Ray Telescope (XRT; \citealt{Burrows2005}) and \textit{XMM-Newton} \citep{Jansen2001}. The XRT data were taken simultaneously with the $UVM2$ observations, with individual exposure times of 160 to 1015 seconds. Additionally, two longer ${\sim}5~\rm ks$ XRT exposures were taken on 2021-01-21 ($\phi\simeq0.5$) and 2021-02-20 ($\phi\simeq0$). In total, V723 Mon was observed with \textit{Swift} XRT for 19865 seconds. All XRT observations were reprocessed using the \textit{Swift} \textsc{xrtpipeline} version 0.13.2 and standard filter and screening criteria\footnote{\url{http://swift.gsfc.nasa.gov/analysis/xrt_swguide_v1_2.pdf}} and the most up-to-date calibration files. To increase the signal to noise of our observations, we combined all cleaned individual XRT observations using \textsc{XSELECT} version 2.4g. To place constraints on the presence of X-ray emission, we used a source region with a radius of 30 arcsec centered on the position of V723 Mon and a source-free background region with a radius of 150 arcsec located at RA = 06:28:53.1, Dec =$-$05:38:49.6 (J2000).

We also retrieved archival \textit{XMM-Newton} data obtained during a ${\sim}10$~ks observation of the nearby ultraluminous infrared galaxy IRAS 06269-0543 (Observation ID 0153100601; PI: N. Anabuki). However, V723 Mon is ${\sim}12$\arcmin\ off-axis in these observations, resulting in a non-optimal PSF with a $90\%$ enclosed energy radius of ${\sim}1$\arcmin. We reduced the data using the {\it XMM-Newton} Science System (SAS) Version 15.0.0 \citep{gabriel04}. We removed time intervals with proton flares or high background after identifying them by producing count-rate histograms using events with an energy between 10--12~keV. For the data reduction, we used the standard screening procedures and the FLAGS recommended in the current SAS analysis threads\footnote{\url{https://www.cosmos.esa.int/web/xmm-newton/sas-threads}} and \textit{XMM-Newton} Users Handbook\footnote{\url{https://xmm-tools.cosmos.esa.int/external/xmm_user_support/documentation/uhb/}}.

\section{Results}
\label{section:v723res}

Here we present our analyses of the observations described in $\S$\ref{section:v723obs}. In $\S$\ref{section:giant}, we characterize the red giant using its spectral energy distribution (SED) and spectra. In $\S$\ref{section:orbit}, we fit Keplerian models to the radial velocities and derive a spectroscopic orbit for V723 Mon. In $\S$\ref{section:phoebe}, we model the ellipsoidal variations of the red giant using multi-band light curves and the binary modeling tool \verb"PHOEBE" to derive the masses of the red giant and the dark companion. We also derive limits on companion eclipse depths. In $\S$\ref{section:veiling}, we characterize the veiling component in this system. In $\S$\ref{section:limits}, we place constraints on luminous stellar companions. In $\S$\ref{section:balmer}, we characterize the Balmer emission lines and their variability. In $\S$\ref{section:xrays}, we discuss the X-ray observations.

\subsection{Properties of the Red Giant Secondary} \label{section:giant}

We characterized the red giant using both fits to its overall SED
and analyses of the available spectra.  For the SED we used photometry from APASS DR10 \citep{Henden2018}, SkyMapper DR2 \citep{Onken2019}, 2MASS \citep{Skrutskie2006} and AllWISE \citep{Wright2010AJ}. We used the \textit{Swift} UVM2 photometry only as an upper limit ($\S3.5$). The compilation of the multi-band photometry used in these fits are given in Table \ref{tab:phot}. 

\begin{figure}
	\vspace{-2cm}
	\includegraphics[width=0.5\textwidth]{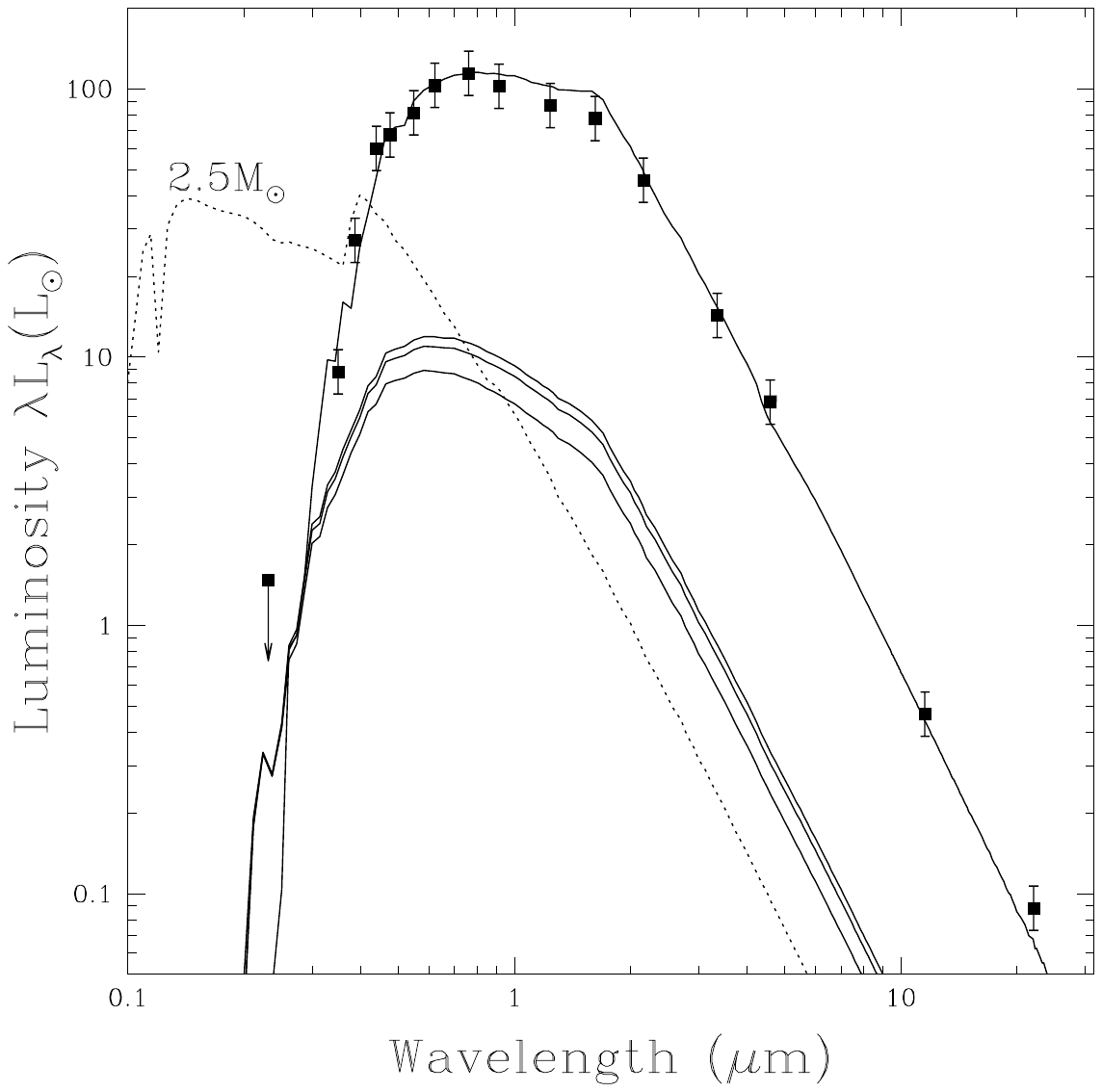}
	\vspace{-2cm}
    \caption{The best-fitting, extinction-corrected SED model for V723 Mon without considering veiling (see $\S$\ref{section:veiling}). The \textit{Swift} \textit{UVM2} detection (arrow; see Table \ref{tab:phot}) is treated as an upper limit and the error bars are expanded to give $\chi^2/N_{dof} \simeq 1$. The SEDs for a main sequence companion of mass $1.29~M_\odot$, an equal mass binary consisting of two main sequence stars each with $0.99~M_\odot$, and a binary with companion masses of $1.25~M_\odot$ and $0.77~M_\odot$ are also shown (see $\S$\ref{section:limits}). These SEDs closely overlap and are hard to distinguish. As shown by the SED of a single $2.5~M_\odot$ main sequence star (dashed line), hotter stars are easily ruled out.}
    \label{sedfit}
\end{figure}

We fit the SED of V723 Mon using DUSTY \citep{Ivezic1997,Elitzur2001} inside a MCMC wrapper \citep{Adams2015}. We assumed foreground extinction due to $R_V=3.1$ dust \citep{Cardelli1989} and used the ATLAS9  \citet{Castelli2003} model atmospheres for the star. We assume that the source is at the Gaia EDR3 distance and used minimum luminosity uncertainties of 19\% for each band to obtain $\chi^2/N_{dof} \simeq 1$. The expanded uncertainties needed to reach $\chi^2/N_{dof} \simeq 1$ with respect to the reported photometric errors in each measurement are likely driven by using single epoch photometry for a variable source plus any systematic problems in the models and photometry. We used a weak $T_{\rm eff, giant}=4500 \pm500$~K prior on the temperature and a prior of $E(B-V)=0.10\pm0.04$ on the extinction from
\citet{Green2019}. The SED fit yields $T_{\rm eff, giant} \simeq 4440\pm90$~K, $L_{\rm giant}= 173 \pm 8~ L_\odot$, $R_{\rm giant}\simeq22.2\pm0.8~R_\odot$ and $E(B-V) \simeq 0.085 \pm 0.034$ mag. Figure \ref{sedfit} shows the SED and the best fitting model. The SED well-constrains the temperature and is consistent with the extinction estimates.
If we force the model to have smaller stellar radii, the goodness of fit worsens rapidly, from
$\chi^2=13.6$ for the best fit to $20.3$ for $R_{\rm giant}=20 R_\odot$ and $40.9$ for $R_{\rm giant}=18 R_\odot$.  Decreasing the radius forces the star to become hotter ($4800$~K for $18R_\odot$) to fit the SED at long wavelengths and more extincted to fit it at short wavelengths. Constraints on a stellar companion to the giant from the SED are discussed in Section $\S$\ref{section:limits}.

The equivalent width ($\rm EW$) of the $\ion{Na}{i}~\rm D$ doublet provides an independent estimate of the Galactic reddening towards V723 Mon (see for e.g., \citealt{Poznanski2012}). Using the very high resolution PEPSI spectrum, we measure $\rm EW(D_1)=125.1\pm12.0 ~\rm m$\AA~ and $\rm EW(D_2)=85.7\pm7.5 ~\rm m$\AA. From the equivalent width---$E(B-V)$ calibration in \citet{Poznanski2012}, we find that $E(B-V)=0.025\pm0.005$~mag. This is consistent with the low foreground extinction that was derived from the SED models and the \citet{Green2019} extinction maps.

\begin{table*}
	\centering
	\caption{Multi-band photometry measurements used in the construction of the SED for V723 Mon. Luminosities in each band are calculated assuming a nominal distance of $d\simeq460$~pc.}
	\label{tab:phot}
\begin{tabular}{rrrrrr}
		\hline
		 Filter & Magnitude & $\sigma$ & $F_\lambda\,[\rm ergs~ s^{-1}~ cm^{-2} ~ \angstrom^{-1}]$ & $\lambda L_\lambda\,[L_\odot]$ &  Reference\\
		\hline

\textit{Swift} UVM2 & 14.11 & 0.07 & $1.0\times10^{-14}$ & 0.15  & This work \\
SkyMapper u & 10.30 & 0.01 & $2.5\times10^{-13}$  & 5.7   & \citet{Onken2019} \\
SkyMapper v & 9.74 & 0.01 & $7.1\times10^{-13}$ & 18.1 &  \citet{Onken2019} \\
Johnson B & 9.24 & 0.05 & $1.3\times10^{-12}$ & 36.5  &  \citet{Henden2018} \\
Johnson V & 8.30 & 0.04 & $1.7\times10^{-12}$  & 61.8  &  \citet{Henden2018} \\
Sloan g' & 8.73 & 0.01 & $1.6\times10^{-12}$ & 49.1   & \citet{Henden2018} \\
Sloan r' & 7.87 & 0.12 & $2.0\times10^{-12}$  & 81.2 & \citet{Henden2018} \\
Sloan i' & 7.49 & 0.14 & $1.9\times10^{-12}$  & 94.9  & \citet{Henden2018} \\
Sloan z' & 7.36 & 0.05 & $1.1\times10^{-12}$  & 77.8 &  \citet{Henden2018} \\
2MASS J & 6.26 & 0.03 & $9.8\times10^{-13}$ & 80.2   & \citet{Skrutskie2006} \\
2MASS H & 5.58 & 0.03 & $6.7\times10^{-13}$ & 73.3  &  \citet{Skrutskie2006} \\
2MASS $K_s$ & 5.36 & 0.02 & $3.1\times10^{-13}$  & 43.7  & \citet{Skrutskie2006} \\
WISE W1 & 5.28 & 0.16 & $6.3\times10^{-14}$  & 14.0  &  \citet{Wright2010AJ} \\
WISE W2 & 5.10 & 0.07 & $2.2\times10^{-14}$  & 6.7  &  \citet{Wright2010AJ} \\
WISE W3 & 5.01 & 0.02 & $6.5\times10^{-16}$   & 0.5  &  \citet{Wright2010AJ} \\
WISE W4 & 4.82 & 0.04 & $6.0\times10^{-17}$   & 0.1 & \citet{Wright2010AJ} \\

\hline
\end{tabular}
\end{table*}

All high resolution spectra indicate that the giant is rapidly rotating. \citet{Strassmeier2012} derived a projected rotational velocity of $v_{\rm rot} \, \textrm{sin} \, i=16\pm2 \, \rm km \,s^{-1}$ using the SES spectra. \citet{Griffin2014} found a similar average $v_{\rm rot} \, \textrm{sin} \, i\sim 15 \, \rm km \,s^{-1}$ but noted that the values seemed to depend on the orbital phase and ranged from $v_{\rm rot} \, \textrm{sin} \, i\sim 10 \, \rm km \,s^{-1}$ to $v_{\rm rot} \, \textrm{sin} \, i\sim 20 \, \rm km \,s^{-1}$. We also find that $v_{\rm rot} \, \textrm{sin} \, i$ varies from ${\sim}17.6 \, \rm km \,s^{-1}$ to ${\sim}21.7 \, \rm km \,s^{-1}$ in the SES spectra. The HIRES CPS pipeline \citep{Petigura2015} reports $v_{\rm rot} \, \textrm{sin} \, i=21.6\pm1.0 \, \rm km \,s^{-1}$ and we found $v_{\rm rot} \, \textrm{sin} \, i=17.9 \pm 0.4 \, \rm km \,s^{-1}$ from the PEPSI spectrum using \verb"iSpec" \citep{Blanco-Cuaresma2014,Blanco-Cuaresma2019}.  Assuming that the rotation of the giant is tidally synchronized with the orbit, the SES, PEPSI and HIRES measurements yield stellar radii of $R\sin\, i=19.0\pm2.4 \, R_\odot$, $R\sin\, i=21.2\pm0.5 \, R_\odot$ and $R\sin\, i=25.6\pm1.2 \, R_\odot$, respectively, consistent both with estimates from the SED and the radius derived for the giant from the \verb"PHOEBE" models ($\S$\ref{section:phoebe}). The match to the estimate of the radius from the SED indicates that $\sin i \simeq 1$ independent of the \verb"PHOEBE" models.

To derive the surface temperature ($T_{\rm{eff}}$), surface gravity ($\log(g)$), and metallicity ($\rm [Fe/H]$) of the giant, we use the spectral synthesis codes \verb"FASMA" \citep{Tsantaki2020,Tsantaki2018} and \verb"iSpec". \verb"FASMA" generates synthetic spectra based on the ATLAS-APOGEE atmospheres \citep{Kurucz1993,Meszaros2012} with \verb"MOOG" \citep{Sneden1973} and outputs the best-fit stellar parameters following a $\chi^2$ minimization process. \verb"iSpec" carries out a similar minimization process with synthetic spectra generated by \verb"SPECTRUM" \citep{Gray1994} and MARCS model atmospheres \citep{Gustafsson2008}. The line lists used in this process span the wavelength range from 480~nm to 680~nm. Since we are confident the companion should undergo eclipses (see $\S$\ref{section:limits}), we fit the SES spectrum near conjunction ($0.48\ltorder  \phi\ltorder 0.52$) when any companion would be eclipsed by the giant. For the detailed fits, we fix the rotational velocity to the value of $v_{\rm rot} \, \textrm{sin} \, i=19.1\pm0.4 \, \rm km \,s^{-1}$ found by \verb"iSpec" for this spectrum.  

\begin{figure*}
	\includegraphics[width=\textwidth]{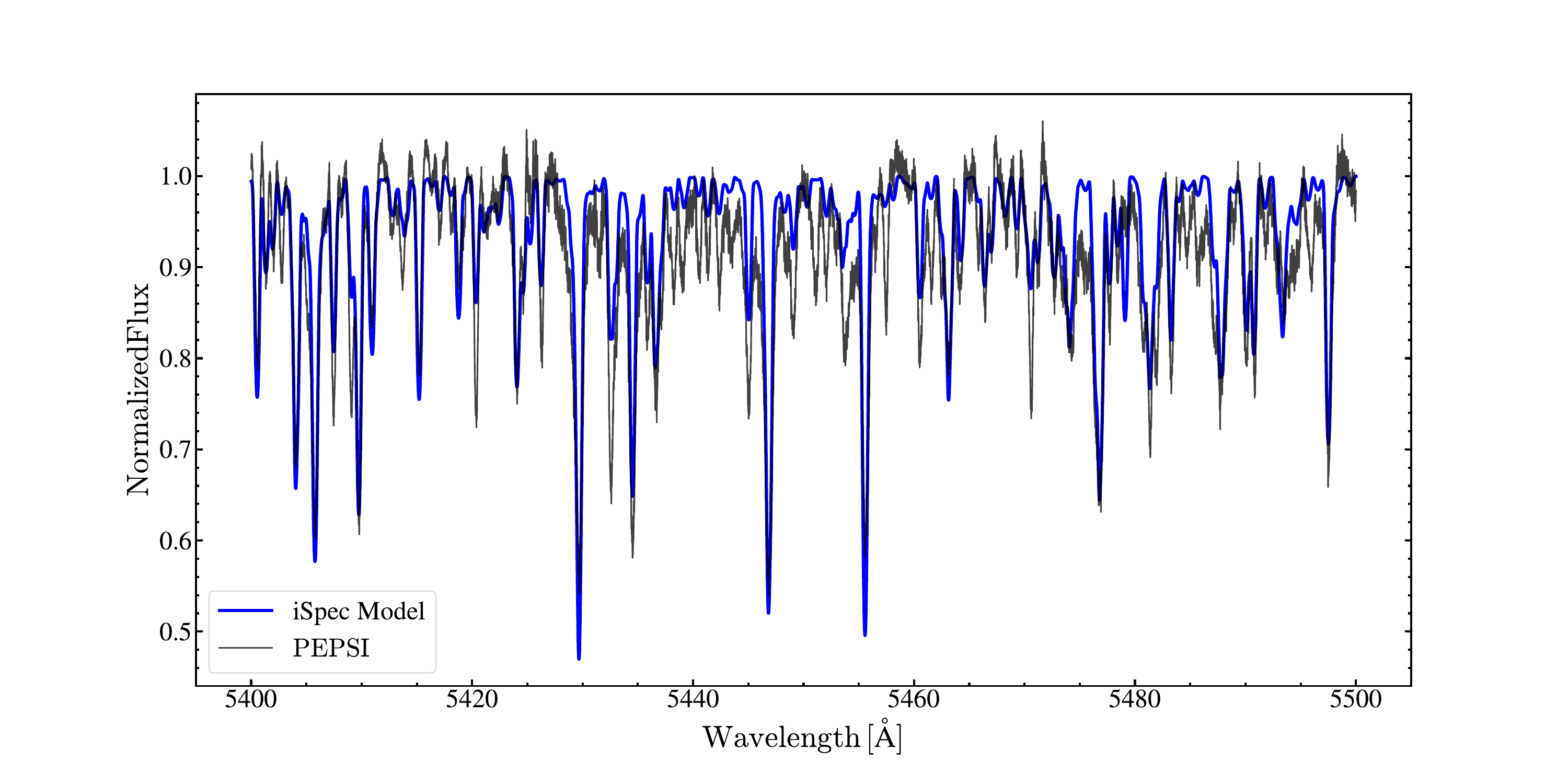}
	
    \caption{The LBT/PEPSI spectrum from 5400~\AA~ to 5600~\AA ~(black). A model spectrum using the atmospheric parameters derived from iSpec (Table \ref{tab:specpar}) is shown in blue.}
    \label{pepsisynth}
\end{figure*}

For the \verb"FASMA" fits we initially keep the macroturbulent broadening ($v_{\rm mac}$) and the microturbulence ($v_{\rm micro}$) fixed at $v_{\rm mac}=5\, \rm km \,s^{-1}$ and $v_{\rm mic}=2\, \rm km \,s^{-1}$, but then allow them to be optimized once we have a reasonable fit. This fits yield $T_{\rm{eff, giant}}=4480\pm50$~K, $\log(g)=1.7\pm0.2$, and $\rm [Fe/H]=-1.1\pm0.1$. In the \verb"iSpec" fits, $v_{\rm mic}$ was kept as a free parameter and we obtain similar results with $T_{\rm{eff, giant}}=4570\pm60$~K, $\log(g)=1.7\pm0.1$, $\rm [Fe/H]=-0.9\pm0.1$, $\rm [\alpha/Fe]=0.7\pm0.1$, and $v_{\rm micro}=0.6\pm0.1\, \rm km \,s^{-1}$. We adopt the parameters from \verb"iSpec" as our standard. Figure \ref{pepsisynth} compares a model spectrum generated using the \verb"iSpec" parameters to the LBT/PEPSI spectrum. The model spectrum is a reasonable fit to the PEPSI data. The spectroscopic parameters derived for the giant are summarized in Table \ref{tab:specpar}. These estimates of the spectroscopic parameters do not consider the effects of veiling on the observed spectrum (see $\S$\ref{section:veiling}). Veiling introduces systematic uncertainties on these parameters and will lower the temperature estimate for the giant.

The spectroscopic surface gravity is consistent with that inferred from the \verb"PHOEBE" model in $\S$\ref{section:phoebe} ($\log(g)_{\rm PHOEBE}=1.63\pm0.06$). Given the spectroscopic $\log(g)$ and the radius of the giant from $\S$\ref{section:phoebe}, the spectroscopic mass is $M_{\rm giant,spec}=1.1\pm0.5~M_\odot$, consistent with the mass of the giant derived from the \verb"PHOEBE" model. The spectroscopic temperature is also consistent with that obtained from the SED fits. Based on the \citet{vanBelle1999} temperature scale for giants, our temperature estimate is more consistent with a K0/K1 giant than the archival classification of G0 (${\sim}6000$~K) from \citet{Houk2000}. The absolute $V$-band magnitude ($M_V\simeq-0.3\pm0.1$) is consistent with a luminosity class of III \citep{Straizys1981}. From single-star evolution, the spectroscopic measurement of $\log(g)$ suggests that the giant is currently evolving along the upper red giant branch. The giant has a luminosity larger than red clump stars, suggesting that it has not yet undergone a helium flash. 

We used the spectroscopic parameters in Table \ref{tab:specpar} and the luminosity constraint from the SED fit as priors to infer the physical properties of the giant using MESA Isochrones and Stellar Tracks (MIST; \citealt{Dotter2016,Choi2016}). We used the \verb"isochrones" package for the fitting \citep{Morton2015}. We find that $M_{\rm giant,MIST}=1.07\pm0.24~M_\odot$ and $R_{\rm giant,MIST}=24.8\pm4.7~R_\odot$. The age of the giant from the MIST models is ${\sim}5.4^{+5.1}_{-2.6}~{\rm Gyr}$. These results are consistent with the properties of the giant derived from the spectra and the SED. 


\begin{table}
\caption{Properties of the red giant in V723 Mon (not accounting for veiling, see $\S$\ref{section:veiling})}
\begin{threeparttable}
	\begin{tabular}{r c c }
	\hline	
	Parameter & \verb"FASMA"   &   \verb"iSpec" \\
	\hline
	\vspace{2mm}
	$T_{\rm{eff}}~(\rm K)$ & $4480\pm50$ & $4570\pm60$ \\
	\vspace{2mm}
	$\log(g)$ & $1.7\pm0.2$ & $1.7\pm0.1$\\	
	\vspace{2mm}
	$\rm [Fe/H]$ & $-1.1\pm0.1$ & $-0.9\pm0.1$\\	
	\vspace{2mm}
	$\rm [\alpha/Fe]$ & --- & $0.7\pm0.1$\\	
	\vspace{2mm}
	$v_{\rm mic}~({\rm km~s^{-1}})$ & 1.64 (fixed) & $0.64\pm0.07$\\	
	\vspace{2mm}
	$v_{\rm mac}~({\rm km~s^{-1}})$ & 5.1 (fixed) & 5.7 (fixed)\\	
	\vspace{2mm}	
	$v_{\rm rot} \, \textrm{sin} \, i~({\rm km~s^{-1}})^*$ & \multicolumn{2}{c}{$19.1\pm0.4$} \\ 
	\vspace{2mm}	
	$R~(R_\odot)$ & \multicolumn{2}{c}{$24.9\pm0.7 $} \\ 
	\vspace{2mm}	
	$L~(L_\odot)$ & \multicolumn{2}{c}{$173\pm8 $} \\ 

	\vspace{2mm}	
	$M~(M_\odot)$ & \multicolumn{2}{c}{$1.00\pm0.07 $} \\ 
	\hline	
\end{tabular}
    \begin{tablenotes}
      \small
      \item $^*v_{\rm rot} \, \textrm{sin} \, i$ varies with orbital phase
    \end{tablenotes}

\end{threeparttable}
\label{tab:specpar}
\end{table}

\begin{figure*}
	\includegraphics[width=0.98\textwidth]{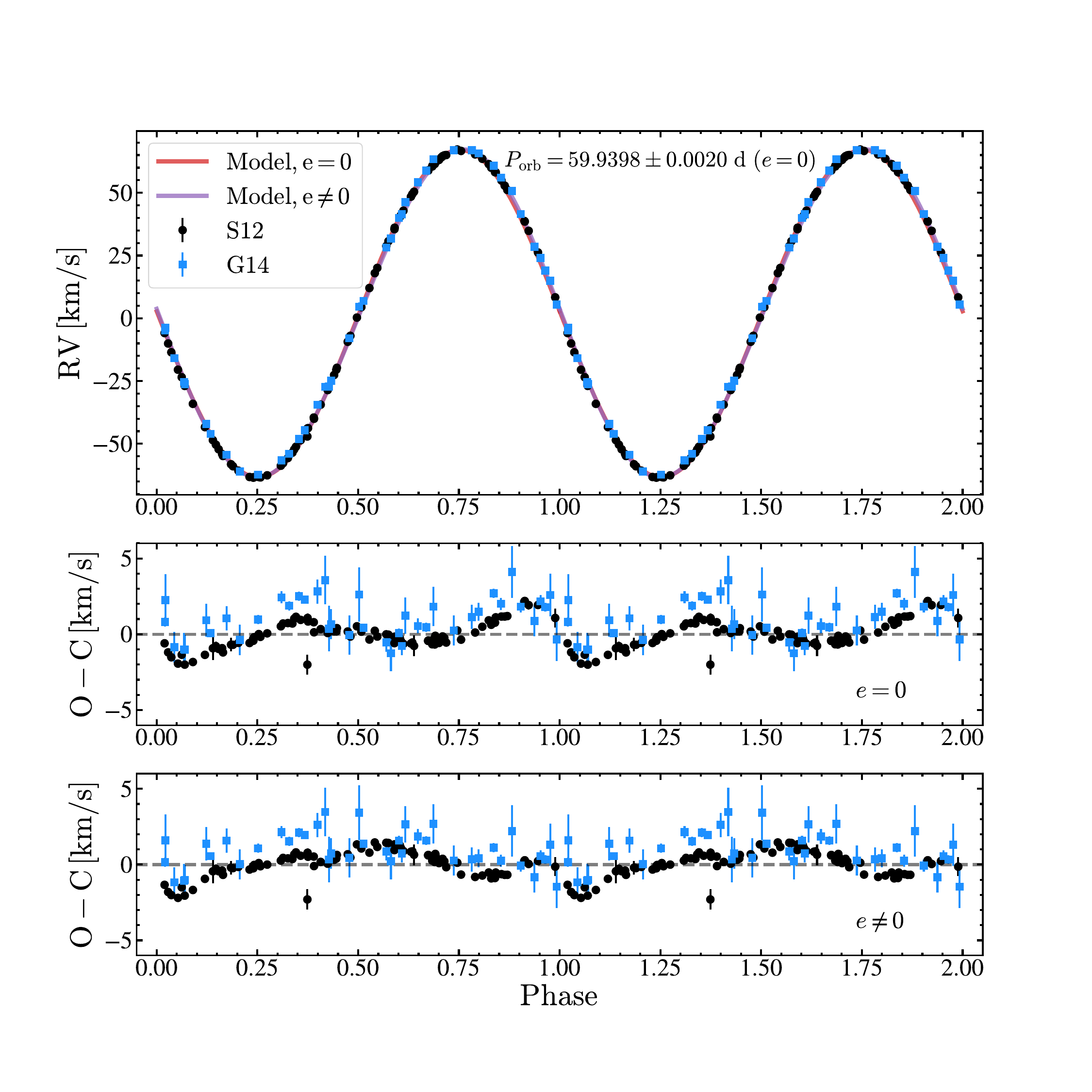}
    \caption{The observed radial velocities for V723 Mon as a function of orbital phase, defined with the epoch of maximal RV at $\phi=0.75$ (top). The RVs from \citet{Strassmeier2012} [S12] are shown as black circles and the RVs from \citet{Griffin2014} [G14] are shown as blue squares. The joint, circular (elliptical) RV fit from Table \ref{tab:orbit} is shown in red (purple). The two models closely overlap and are hard to distinguish. The residuals from both the circular and elliptical fit to the combined data are shown in the lower panels. The RV residuals are most likely a result of the ellipsoidal variability rather than the result of a triple system (see appendix $\S$\ref{app:tertiary}.)}
    \label{rvs}
\end{figure*}

\subsection{Keplerian orbit models}\label{section:orbit}

We fit Keplerian models both independently and jointly to the S12 and G14 radial velocities using the Monte Carlo sampler \verb"TheJoker" \citep{Price-Whelan2017}. The results for the four fits are summarized in Table~\ref{tab:orbit}.  We first fit each data set independently as an elliptical orbit to verify that we obtain results consistent with the published results. We then fit the joint data set using either a circular or an elliptical orbit. In the joint fits we include an additional parameter to allow for any velocity zero point offsets between the S12 and G14 data. For the circular orbit we also set the argument of periastron $\omega=0$. 
Since the orbit is nearly circular even for the elliptical models, we derived the epoch of maximal radial velocity $T_{\rm RV, max}$ instead of the epoch of periastron. We define phases so that $T_{\rm RV, max}$ (BJD/TDB) corresponds to $\phi=0.75$, the companion eclipses at $\phi=0.5$ and the giant would be eclipsed at $\phi=0$. After doing a first fit for the elliptical models, we did a further fit with  $P$, $K$, $e$, $\gamma$ and $T_{\rm RV, max}$ fixed to their posterior values and further optimized $\omega$ using least-squares minimization.

The results of the fits are summarized in Table \ref{tab:orbit} and shown in Fig.~\ref{rvs}. The fits to the individual data sets agree well with the published results, and the mass functions are well-constrained and mutually consistent.  The elliptical models all yield a small, non-zero
ellipticity, consistent with G14's arguments.  We do find a small velocity offset 
of $\Delta V=0.58\pm0.14~\rm km ~s^{-1}$ for the elliptical model and $\Delta V=0.84\pm0.21~\rm km ~s^{-1}$ for the circular model between the S12 and G14 data.  While the velocity residuals of the fits are small compared to $K$ ($\lesssim 1.7~\rm km ~s^{-1}$ versus $65~\rm km ~s^{-1}$), they are large compared to the measurement uncertainties. Thus, while the RV curve is clearly completely dominated by the orbital model, Fig.~\ref{rvs} also shows that there are significant velocity residuals for both joint fits. The circular fit is dominated by a residual of period $P_{\rm orb}/2$ and the elliptical fit is dominated by a residual of period $P_{\rm orb}/3$.  Fitting an elliptical orbit with a circular orbit will show a dominant $P_{\rm orb}/2$ residual, but an elliptical orbit should not show a large $P_{\rm orb}/3$ residual. This $P_{\rm orb}/3$ residual to the fit of an elliptical orbit is likely the origin of the S12 hypothesis that the companion is an SB1 with this period. The residuals do not however, resemble those of a Keplerian orbit. We discuss this hypothesis and the velocity residuals in Appendix \ref{app:tertiary}.

Binaries with evolved components ($\log(g)<2.5$) and orbital periods shorter than ${\sim}100$~days are expected to have gone through tidal circularization and have circular orbits (e.g., \citealt{Verbunt1995,Price-Whelan2018}). However, in the joint fits, the models with ellipticity are a better fit and have smaller RMS residuals than the circular models. \citet{Griffin2014} carried out an $F$-test and noted that the ellipticity in their best-fit orbit for V723 Mon was significant, and concluded that it was very likely real. While we use the circular orbit for the \verb"PHOEBE" models in $\S$\ref{section:phoebe}, the differences in the mass function and semi-major axis compared to the elliptical orbit are small and have no effect on our conclusions.

Independent of this issue with the origin of the RV residuals, the companion to the red giant in V723 Mon must be very massive given a mass function of $1.72-1.74~M_\odot$. The mass function itself is greater than the Chandrasekhar mass of ${\sim}1.4M_\odot$, immediately ruling out a white dwarf companion. For an edge on orbit (see $\S$\ref{section:phoebe}) and $M_{\rm giant} \simeq 1 M_\odot$, the companion mass is $M_{\rm comp} \simeq 3 M_\odot $ and the semi-major axis is $a \simeq 100 R_\odot$.  The Roche limits are approximately $R_{\rm L,giant}=0.29 a \simeq 30 R_\odot$ for the giant and $R_{\rm L,comp}= 0.48 a \simeq 49 R_\odot$ for the companion.  Based on the radius estimate from the SED fits ($\S$\ref{section:giant}), the giant is comfortably inside its Roche lobe ($R_{\rm giant}/R_{\rm L,giant} \simeq 0.66$) but in the regime where we should be seeing strong ellipsoidal variability due to the tidal deformation of the giant by the gravity of the companion (e.g., \citealt{Morris1985}). Any stellar companion also has to be well within its Roche lobe or it would dominate the SED because its Roche lobe is significantly larger. Hence, we can be confident that we have a detached binary whose light is dominated by a giant that should show ellipsoidal variability and might show eclipses.

\begin{table*}
	\centering
	\caption{Orbital Elements for V723 Mon}
	\label{tab:orbit}
\begin{tabular}{rrrrr}
		\hline
		 Parameter & S12 & G14 & S12 + G14 & S12 + G14 \\
		\hline
	$P_{\rm orb} (\rm d)$ & $59.9358\pm0.0017$ & $59.9391\pm0.0020$ & $59.9394\pm0.0014$ & $59.9398\pm0.0020$ \\
	$K~(\rm km ~s^{-1})$ & $65.483\pm0.068$ &  $65.209\pm0.117$ & $65.360\pm0.081$ & $65.150\pm0.125$ \\	
	$e$ & $0.0152\pm0.0010$ & $0.0186\pm0.0014$ & $0.0158\pm0.0012$ & 0 (fixed) \\
	$\omega~(^\circ)$ & $89.1\pm4.1$ & $88.2\pm5.8$ & $88.4\pm4.6$ & 0 (fixed) \\	
	${\gamma}~(\rm km~ s^{-1})$  &  $1.81\pm0.053$ &  $3.02\pm0.07$ & $1.95\pm0.07$ & $1.88\pm0.10$ \\
	$a_{\rm giant}\sin i$ $(R_{\odot})$  & $77.567\pm0.081$ & $77.242\pm0.139$ & $77.425\pm0.096$ & $77.187\pm0.148$ \\	
	$T_{\rm RV, max}$ (BJD-2450000)  &  $4096.196\pm0.696$ &  $4098.946\pm0.954$ & $4096.529\pm0.773$ & $4096.168\pm0.038$ \\
	
    \hline
	RMS Residual $(\rm km~ s^{-1})$& 0.519 & 0.969 & 0.876 & 1.17 \\	    
    \hline    
	$f(M)$ $(M_{\odot})$  & $1.743\pm0.005$ & $1.721\pm0.009$ & $1.734\pm0.006$ & $1.717\pm0.010$ \\

\hline
\end{tabular}
\end{table*}

\subsection{PHOEBE and ELC binary models}\label{section:phoebe}

While it has been previously claimed that V723 Mon is a contact/semi-detached binary of the $\beta$-Lyrae type, we find that this is very unlikely. As we just argued, both the giant and any companion must lie well within their Roche lobes given the properties of the giant and the orbit. Additionally, the morphology of the light curve is inconsistent with those of detached and most semi-detached eclipsing binaries. Here we interpret the light curves as ellipsoidal variability and deduce limits on any eclipses of the companion for use in $\S$\ref{section:limits}.

We fit the ASAS $V$-band, KELT $R_K$-band and \textit{TESS} $T$-band light curves (Figure \ref{lcs}) using \verb"PHOEBE 2.3" \citep{Phoebe2016,Horvat2018,Conroy2020}. Since the companion appears to be dark and producing no eclipses, we fix it to be a small ($R=3\times10^{-6}~R_\odot$), cold ($T_{\rm eff} =300$~K) black body, use the simplest and fastest eclipse model (\verb"eclipse_method=only_horizon") and do not include the effects of irradiation and reflection. We adopt the joint-circular RV solution from Table \ref{tab:orbit} (period, $e=0$, systemic velocity, semi-major axis, and argument of periastron $\omega=0$). We set the bolometric gravity darkening coefficient as $\beta=0.54$, comparable to those obtained by \citet{Claret2011} in the $R$-band for stars with $T_{\rm eff} {\sim}4600$~K, $-1\lesssim\rm [Fe/H]\lesssim-0.5$ and $0\lesssim\log(g)\lesssim2$. We did not find significant differences in the final parameters when we varied $\beta$ by $\pm20\%$ in the \verb"PHOEBE" models. A $20\%$ change in $\beta$ roughly corresponds to $\Delta T_{\rm eff}\simeq500$~K or $\Delta \log(g)\simeq1.5$.

We initially performed trial fits to the KELT and \textit{TESS} light curves simultaneously using the Nelder-Mead simplex optimization routine \citep{Lagarias1998}. The parameters from the trial fit were then used to initialize a MCMC sampler with 20 walkers, which was then run for 3000 iterations using the \verb"emcee" \citep{Foreman-Mackey2013} solver in \verb"PHOEBE 2.3". We fit for the binary mass ratio $q=M_{\rm giant}/M_{\rm comp}$, orbital inclination ($i$), the radius ($R_{\rm giant}$) and the effective temperature of the giant ($T_{\rm eff, giant}$). We marginalize over the semi-major axis of the red giant ($a_{\rm giant}\sin i$), the passband luminosities and nuisance parameters that account for underestimated errors in the KELT $R_K$-band and \textit{TESS} $T$-band light curves. We adopt Gaussian priors on the effective temperature ($T_{\rm eff, giant}=4440\pm90$~K) and the radius ($R_{\rm giant}=22.2\pm0.8~R_\odot$) based on the SED fits in $\S$\ref{section:giant}. We adopt uniform priors of [$70^\circ$, $90^\circ$] and [0, 1] for the orbital inclination and mass ratio respectively.  

\begin{figure}
	\includegraphics[width=0.5\textwidth]{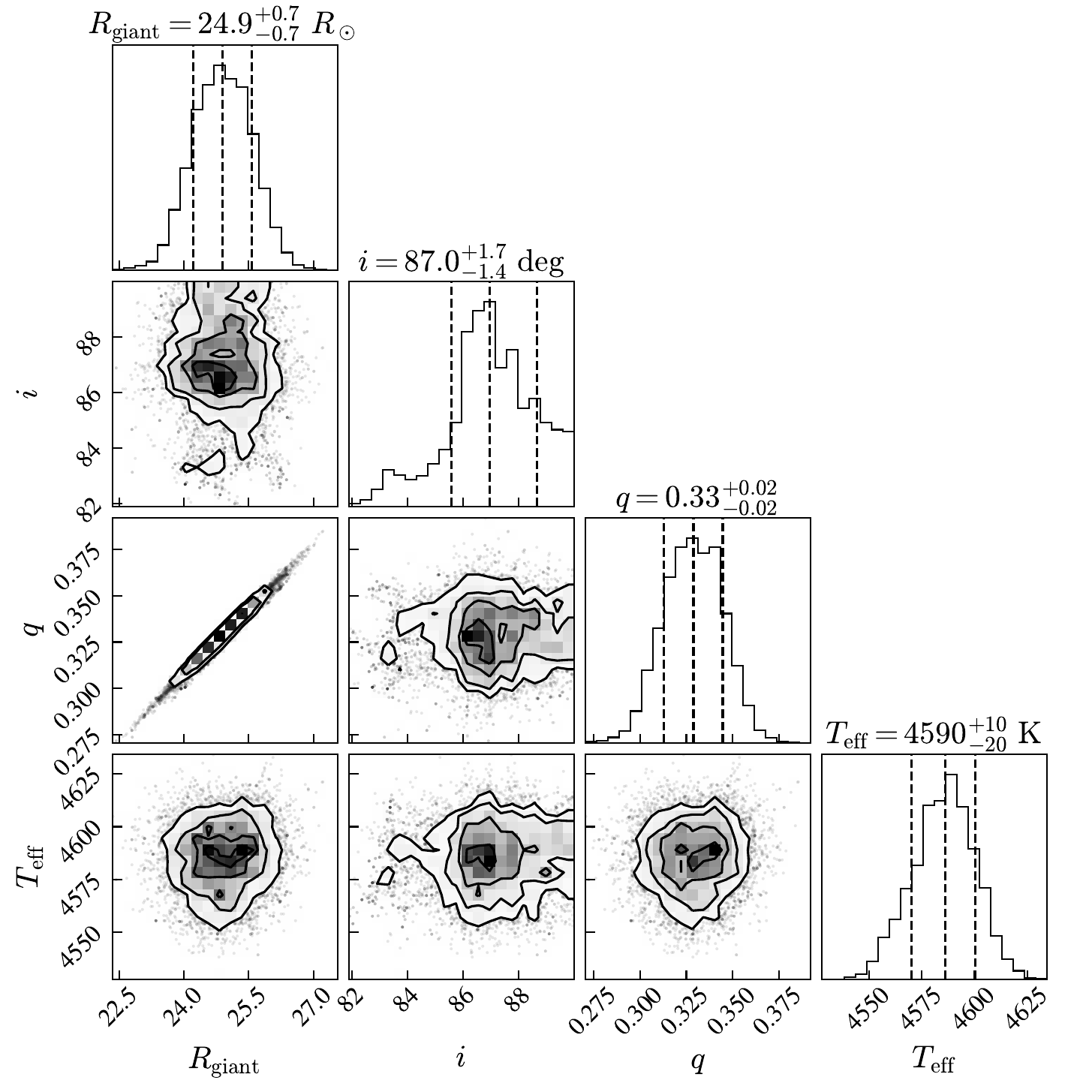}
    \caption{The corner plot for the posterior samples for $q$, $i$, $R_{\rm giant}$ and $T_{\rm eff, giant}$ in the best-fitting PHOEBE model to the KELT and \textit{TESS} light curves.}
    \label{corner}
\end{figure}

\begin{figure*}
	
	\includegraphics[width=0.7\textwidth]{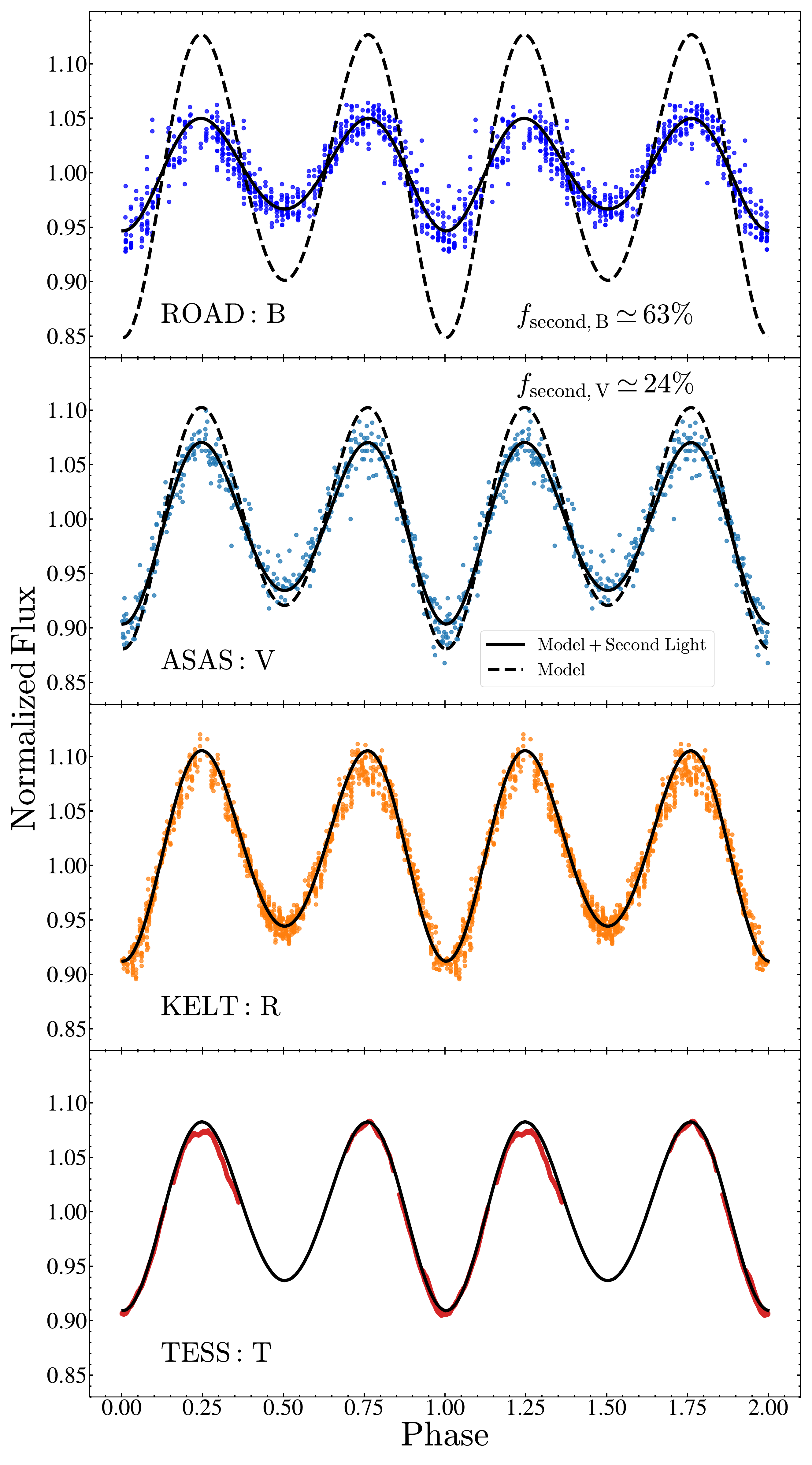}
		\vspace{-0.2cm}	
    \caption{The normalized ROAD $B$-band, ASAS, KELT and \textit{TESS} light curves for V723 Mon as a function of orbital phase (defined with the epoch of maximal RV at $\phi=0.75$). The best-fitting binary models from PHOEBE are also shown. Second light from the veiling component (see $\S$\ref{section:veiling}) is considered in the models for the $B$ ($f_{\rm second,B}{\sim}63\%$) and $V$-band ($f_{\rm second,V}{\sim}24\%$) data.}
    \label{lcs}
\end{figure*}

The corner plot of the posterior samples for $q$, $i$, $R_{\rm giant}$ and $T_{\rm eff, giant}$ are shown in Figure \ref{corner} and the results of the best-fitting \verb"PHOEBE" model are listed in Table \ref{tab:params}. The errors in the parameters were derived from the MCMC chains. Our model is a good fit to the KELT $R_K$-band and \textit{TESS} $T$-band light curves, shown in Figure \ref{lcs}. The \verb"PHOEBE" model indicates that the orbital inclination of V723 Mon is nearly edge on ($87.0^\circ{}^{+1.7^{\circ}}_{-1.4^{\circ}} $). The semi-major axis of the binary is $a_{\rm orb}\sin i =a_{\rm giant}\sin i+ a_{\rm c}\sin i =102.7\pm1.3~R_{\odot}$. The radius derived from the \verb"PHOEBE" model ($24.9\pm0.7~R_\odot$) agrees well with those obtained from the SED fits and the MIST evolutionary models in $\S$\ref{section:giant}. 

To verify the results of the \verb"PHOEBE" models, we independently fit the KELT $R_K$-band light curve and the S12 radial velocities using the differential evolution Markov Chain optimizer \citep{TerBraak2006} and the \verb"ELC" binary modelling code \citep{Orosz2000}. The optimizer was initialized using the parameters from the Nelder-Mead simplex optimization from \verb"PHOEBE" and the optimizer was run for 10,000 iterations using 18 chains. We fit for the mass, radius and temperature of the red giant ($M_{\rm giant}$, $R_{\rm giant}$, $T_{\rm eff, giant}$), the orbital inclination ($i$), the binary mass ratio ($q$), and the observed radial velocity semi-amplitude ($K$) assuming uniform priors. We included priors for the temperature ($T_{\rm eff, giant}=4440\pm90$~K), radius ($R_{\rm giant}=22.2\pm0.8~R_\odot$), surface gravity($\log(g)=1.7\pm0.2$) and $v_{\rm rot} \, \textrm{sin} \, i=19.1\pm1~\rm km ~s^{-1}$. From the \verb"ELC" posteriors, we obtain $M_{\rm giant}=0.99^{+0.04}_{-0.05}~M_\odot$, $T_{\rm eff, giant}=4570^{+60}_{-40}$~K, $R_{\rm giant}=24.5^{+0.4}_{-0.5}~R_\odot$, $i=88.9^\circ{}^{+0.8^\circ}_{-1.6^\circ}$, $q=0.33443^{+0.09654}_{-0.06113}$ and $K=65.213\pm0.025~\rm km ~s^{-1}$. This implies that the mass of the companion is $M_{\rm comp}=3.07^{+0.44}_{-0.28}~M_\odot$. The parameters obtained from \verb"ELC" agree well with those obtained from \verb"PHOEBE".

\begin{table}
\begin{center}
\caption{PHOEBE parameter estimates for the dark companion primary (DC) and red giant secondary (RG)  }\label{tab:params}
	\begin{tabular}{r c c c }
	\hline	
	& Parameter & DC   &   RG \\
	\hline
	\vspace{2mm}
	& $P_{\rm orb}$  (d) & \multicolumn{2}{c}{59.9398 (fixed) } \\
	\vspace{2mm}
	& $\omega~(^\circ)$ & \multicolumn{2}{c}{$0$ (fixed)}\\
	\vspace{2mm}
	& $e$ & \multicolumn{2}{c}{$0$ (fixed)}\\
	\vspace{2mm}
	& ${\gamma}~(\rm km ~s^{-1})$ &  \multicolumn{2}{c}{1.88 (fixed)}\\
	\vspace{2mm}
	& $a\sin i ~(R_{\odot})$ & $25.376 ^{+1.221}_{-1.251}$ & $77.178^{+0.160}_{-0.171}$ \\
	\vspace{2mm}	
	& $i~(^\circ)$  & \multicolumn{2}{c}{$87.0^{+1.7}_{-1.4} $ }\\
	\vspace{2mm}	
	& $T_{\rm{eff}}~(K)$ & $300$ (fixed) & $4590\pm10$\\
	\vspace{2mm}
	& $R~(R_{\odot})$ & $3\times 10^{-6}$ (fixed) & $24.9\pm0.7$\\	
	\vspace{2mm}
	& $q$ & \multicolumn{2}{c}{$0.32889_{-0.01627}^{+0.01576} $} \\
	\vspace{2mm}	
	& $M~(M_{\odot})$ & $3.04\pm0.06$ & $1.00\pm0.07$\\
	\hline
	\vspace{2mm}	
\end{tabular}
\end{center}
\end{table}

With the well-constrained Keplerian model for the RVs and the \verb"PHOEBE" model for the ellipsoidal variations, we are able to directly determine the masses of the two components. The mass of the companion is

\begin{equation}
    M_{\rm comp}=\frac{f(M)(1+q)^2}{\sin^3 i}\simeq \frac{2.9~M_\odot}{\sin^3 i} \bigg(\frac{f(M)}{1.7~M_\odot}\bigg)\bigg(\frac{1+q}{1.3}\bigg)^2
\end{equation} and from the \verb"PHOEBE" models, we find that the red giant has a mass $M_{\rm giant}\simeq1.00\pm0.07~M_\odot$ and the companion has a mass $M_{\rm comp}\simeq3.04\pm0.06~M_\odot$. The reported errors are purely statistical and do not consider systematic effects (veiling, etc.) in the derivation of the binary solution. However, our results in $\S$\ref{section:veiling} indicate that the veiling in the KELT and TESS light curves is minimal and should not significantly bias the mass estimates.
In the absence of evidence for a stellar companion in the SED, spectra, or in eclipse, perhaps the simplest explanation for the mass of the companion ($M_{\rm comp}\simeq3.04\pm0.06~M_\odot$) is that it is a non-interacting black hole in the ``mass-gap'' between $3-5~M_\odot$ \citep{Ozel2010,Farr2011}. We discuss other scenarios in $\S$\ref{section:v723disc}. The estimated mass of the red giant ($M_{\rm giant}\simeq1.00\pm0.07~M_\odot$) places it towards the lower end of measured red giant masses in the APOKASC catalog \citep{Pinsonneault2018}. Of the APOKASC sources with measured asteroseismic masses, only ${\sim}13\%$ had masses lower than $1.0~M_\odot$. The mass derived from the binary modelling is also very similar to the MIST estimate in $\S$\ref{section:giant}.

\begin{figure*}
	\includegraphics[width=0.90\textwidth]{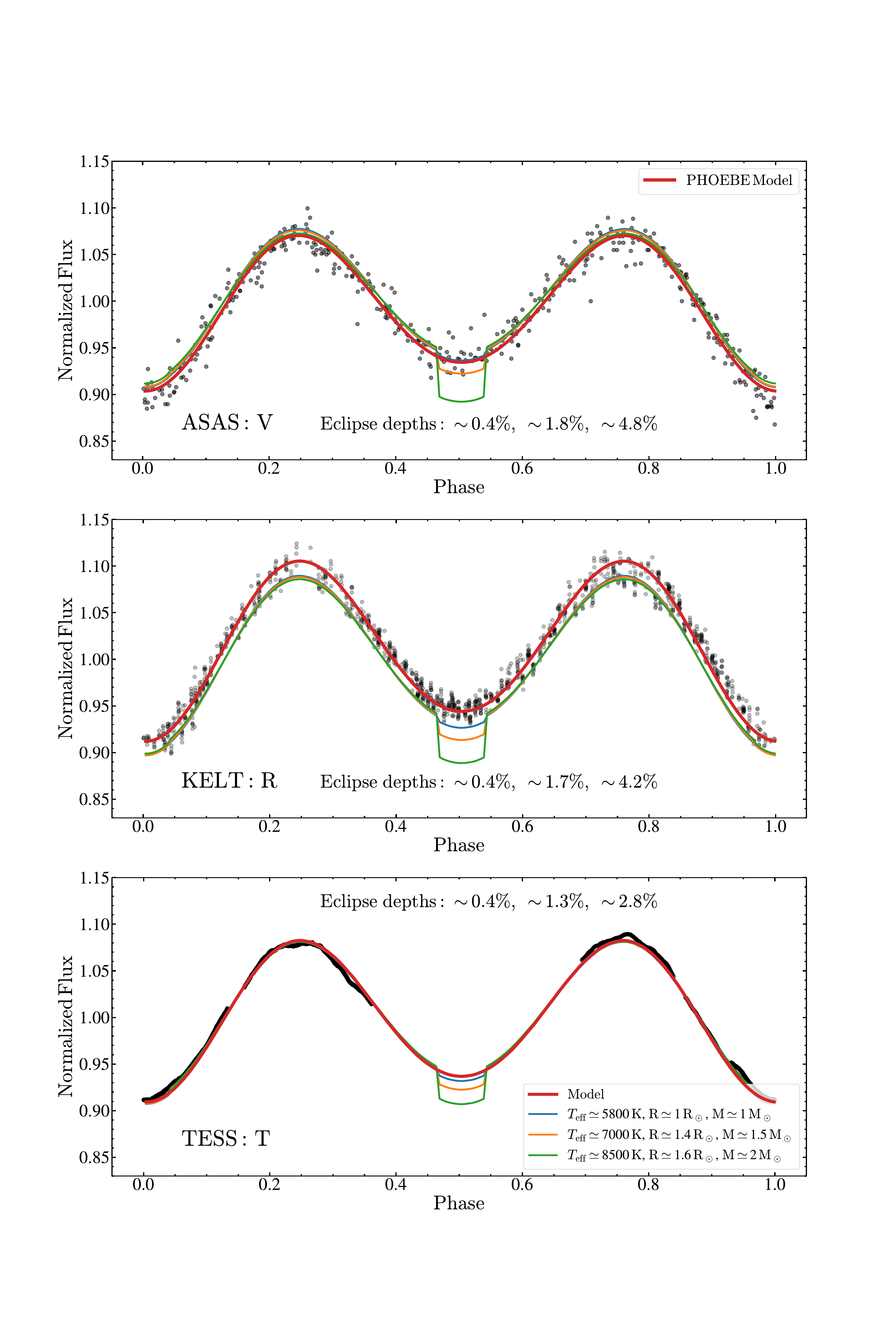}
	\vspace{-1.5cm}	
    \caption{The observed ASAS, KELT and \textit{TESS} light curves compared to various eclipsing models with main sequence companions to the red giant. The ellipsoidal model is shown in red. The depths of the eclipses at $\phi=0.5$ compared to the non-eclipsing ellipsoidal model is also shown. Eclipse constraints rule out single and binary stellar companions with $M_{\rm single}>0.80~M_\odot$ and $M_{\rm binary}>1.51~M_\odot$ (see $\S$\ref{section:limits}).}

    \label{ecllc}
\end{figure*}

\begin{figure*}
	\includegraphics[width=0.90\textwidth]{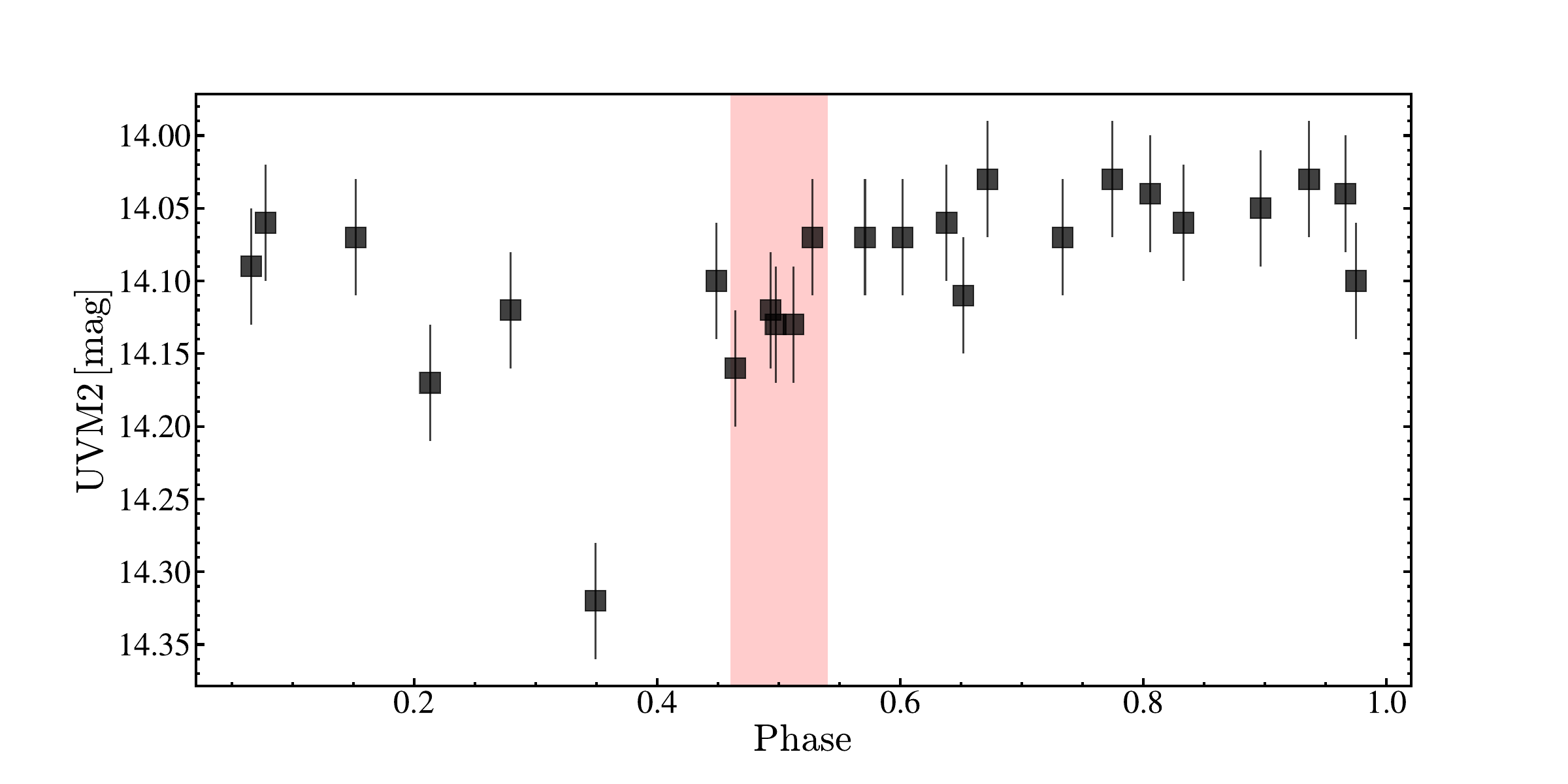}
    \caption{The \textit{Swift} UVM2 light curve. The expected eclipse duration ($\sim5$~days) of the companion is highlighted in red.}
    \label{swiftlc}
\end{figure*}

Given the configuration of the binary and the large size of the giant (${\sim}25~R_\odot$), a companion with a radius of $1~R_\odot$ should be eclipsed for inclination angles $i\gtrsim76^\circ$. Figure \ref{ecllc} shows the eclipses predicted for the ASAS, KELT and \textit{TESS} light curves for main sequence stars with masses of $1~M_\odot$, $1.5~M_\odot$ and $2~M_\odot$.  Any such stars would have produced relatively easy to detect eclipses. We will focus on the KELT limits because these data have higher S/N compared to ASAS photometry, and \textit{TESS} only observed the eclipse of the red giant and not of the companion. At the expected phase of the eclipse, the RMS of the KELT data relative to the eclipse-free ellipsoidal (ELL) model is only $0.84\%$ for phases from 0.46 to 0.54.  If we bin the data 0.01 in phase, the RMS of the binned data is only $0.38\%$. We will adopt a very conservative KELT band eclipse limit of 1\%. Similarly, based on the \textit{Swift}, ASAS and ROAD light curves, we will adopt eclipse limits of $10\%$, $3\%$ and $2\%$ for the $UVM2$, $B$ and $V$ bands, respectively. The \textit{Swift} $UVM2$ light curve is shown in Figure \ref{swiftlc}. The $UVM2$ light curve appears somewhat variable, with one larger outlier on UT 2020-11-13 ($\phi\simeq0.35$). However, we do not see any evidence for an eclipse at the (conservative) level of ${\sim}10\%$ at the expected phases $0.46\lesssim\phi\lesssim0.54$. While the \textit{TESS} light curve does not cover the eclipse of the companion at $\phi=0.5$, it does cover the eclipse of the giant at $\phi=0$. For this eclipse, we estimate a conservative limit on the \textit{TESS} band eclipse depth of ${\sim}0.3\%$.

\subsection{Veiling} \label{section:veiling}

After the KELT $R_K$-band and \textit{TESS} $T$-band light curves were jointly fit, we compared the resulting model $V$-band light curve to the ASAS observations, and found that the observed $V$-band variability amplitude was smaller than predicted by our model. Lower amplitudes at (usually) bluer wavelengths are generally attributed to an additional source of diluting (``second'') light. For X-ray binaries, it is also known as ``disk veiling'' due to additional light from accretion (e.g., \citealt{Hynes2005,Jianfeng2015}). The amount of additional flux is usually characterized either by the ratio $r$ of the veiling flux to the stellar flux or the veiling flux as a fraction of the total flux $f=r/(1+r)$. Interacting X-ray binaries are frequently observed to have veiling factors upwards of $f_{\rm disk}{\sim}50\%$, corresponding to $r{\sim}1$ (see, for e.g., \citealt{Jianfeng2015}). 

We fixed the system parameters to those from the joint fit to the KELT+\textit{TESS} light curves and fit for additional flux (second light) in the $V$-band. The fit to the ASAS $V$-band light curve is consistent with an additional $V$-band flux ($F_2$) amounting to $f_{\rm second,V}=F_2/(F_2+F_{\rm giant})=24\pm2\%$. Figure \ref{lcs} shows the \verb"PHOEBE" model fit for the $V$-band after accounting for this additional flux. The redder ROAD $R_c$ and $I_c$ light curves agree well with the \verb"PHOEBE" model without additional flux. The ROAD observations also agree with the \verb"PHOEBE" model for the $V$-band with the additional flux ($f_{\rm second,V}=24\pm2\%$) added. The discrepancy between the $B$-band light curve and a model without extra flux is even larger than for the $V$-band. Fitting the $B$-band data requires a larger second light contribution of $f_{\rm second,B}=63\pm2\%$. The color of the extra light $B-V=-0.2\pm0.1$ is very blue even before applying any extinction corrections. This corresponds to stars with $T_{\rm eff}$ ranging from ${\sim}13,000$~K to ${\sim}30,000$~K \citep{Papaj1993}. However, no hot star could have a rapidly rising SED from $V$ to $B$ and then drop rapidly from $B$ to the $UVM2$-band. Stellar SEDs are intrinsically broader in wavelength, like the model in Figure \ref{sedfit}. Additionally, stars with these temperatures are easily ruled out by constraints from the SED and eclipses (see $\S$\ref{section:limits}). 

Veiling also affects the spectrum of the star (``line veiling'') by reducing the observed depths of stellar absorption features. This is a well-known problem in interacting X-ray binaries (see, for e.g., \citealt{Casares1993}) and T-Tauri systems \citep{Gahm2008}. Veiling in spectroscopic observations is characterized through the fractional veiling 
\begin{equation}
    \label{eq:veil}
    r(\lambda)= \frac{V(\lambda)}{S(\lambda)},
\end{equation} where $V(\lambda)$ is the veiling spectrum and $S(\lambda)$ is the uncontaminated spectrum of the star \citep{Casares1993}. The ratio $r(\lambda)$ is related to the second light ($F_2$) as \begin{equation}
    \label{eq:second}
    F_2= r(\lambda)F_{\rm giant},
\end{equation} where $F_{\rm giant}$ is the flux from the giant. The observed spectrum of the star is a function of the veiling factor $F_{\rm veil}(\lambda)=1+r(\lambda)$
\begin{equation}
    r(\lambda)= A(\lambda)S(\lambda)F_{\rm veil}(\lambda),
\end{equation} where $A(\lambda)$ is a normalization factor. The monochromatic veiling factor 
\begin{equation}
    F_{\rm veil}=\frac{\rm EW_{s}}{\rm EW_{obs}}
\end{equation} is the ratio of the observed equivalent widths to those predicted from the standard/synthetic spectrum.

We calculated the monochromatic veiling factors for various $\ion{Ca}{i}$ and $\ion{Fe}{i}$ absorption lines in the SES spectra and the red side of the PEPSI spectrum ($\lambda>500$~nm). For the Ca lines, we assumed an abundance of $\rm [Ca/H]=0.02\pm0.02$ derived from the PEPSI spectrum. We generated a synthetic \verb"iSpec"/\verb"SPECTRUM" \citep{Gray1994} spectrum for the red giant using the atmospheric parameters in Table \ref{tab:specpar}. Since the $R\approx220,000$ PEPSI spectrum has lower S/N at blue wavelengths, we use the mean veiling factors in the SES spectra for the absorption lines blue-ward of 500~nm. We use the standard deviations of the estimates from the individual SES spectra to estimate the uncertainties. For the redder lines at $>500$~nm, we use the veiling factors derived from the PEPSI spectrum at $\phi\simeq0.63$ and assign errors of $\pm0.10$ in $r(\lambda)$. This method can have large systematic uncertainties (see \citealt{Casares1993}), but it provides a way to independently test the estimates from the dependence of the ellipsoidal variability amplitudes on wavelength.

\begin{figure*}
	\includegraphics[width=0.95\textwidth]{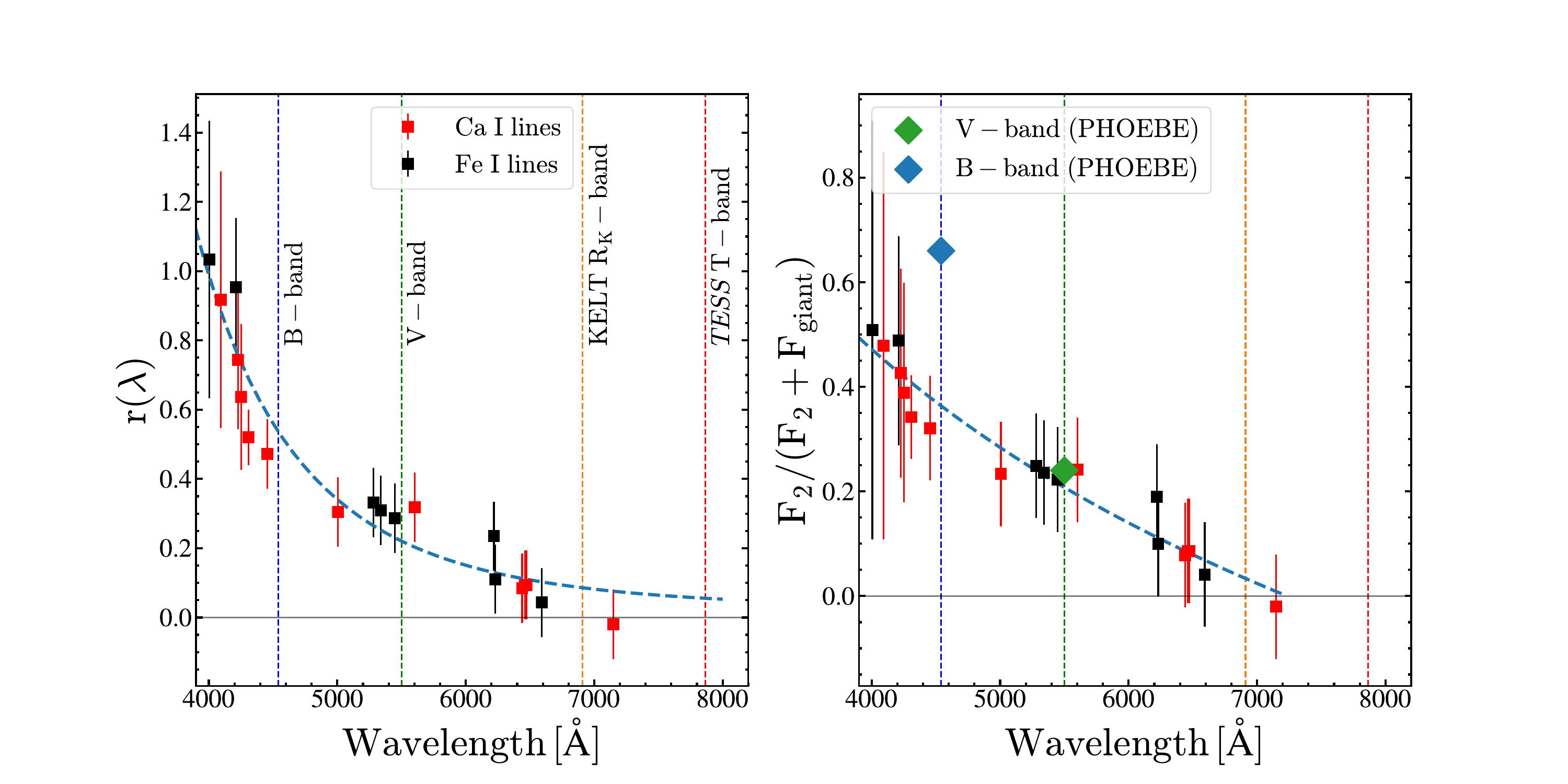}
    \caption{The fractional veiling $r(\lambda)$ (Equation \ref{eq:veil}) [left] and second light (Equation \ref{eq:second}) [right] calculated using $\ion{Ca}{i}$ and $\ion{Fe}{i}$ absorption lines as a function of wavelength. The fractional second light in the $BV$-bands as calculated from PHOEBE are shown as filled diamonds. Power-law fits are shown as blue dashed lines. The effective wavelengths of the $BVR_KT$-bands are shown as dashed lines.}
    \label{veil}
\end{figure*}

Figure \ref{veil} shows that the fractional veiling and second light as a function of wavelength $r(\lambda)$ rises steeply towards bluer wavelengths, with a power-law ($\lambda^\alpha$) index of $\alpha\simeq-5.0\pm1.5$, steeper than that expected from the Rayleigh-Jeans law ($\alpha=-4$). These results also indicate minimal contamination in the KELT $R_K$ and \textit{TESS} $T$ filters that are used for our primary fits to the ellipsoidal variations. As a precaution, we ran \verb"PHOEBE" models adding $10-20\%$ extra flux in the KELT $R_K$ band, and found that the masses of the red giant and dark companion are comparable to those obtained without considering veiling given the error estimates. The fractional second light in the $V$-band from the \verb"PHOEBE" fits is comparable to that derived from the line veiling estimates. The \verb"PHOEBE" estimate of the second light in the $B$-band is somewhat larger than that seen in the line veiling, however both estimates indicate a large contribution to the $B$-band flux. Based on the lack of eclipses in the light curves (see $\S$\ref{section:limits}), we can also infer that this emission component has to be diffuse. 

\begin{figure}
	\vspace{-2cm}
	\includegraphics[width=0.5\textwidth]{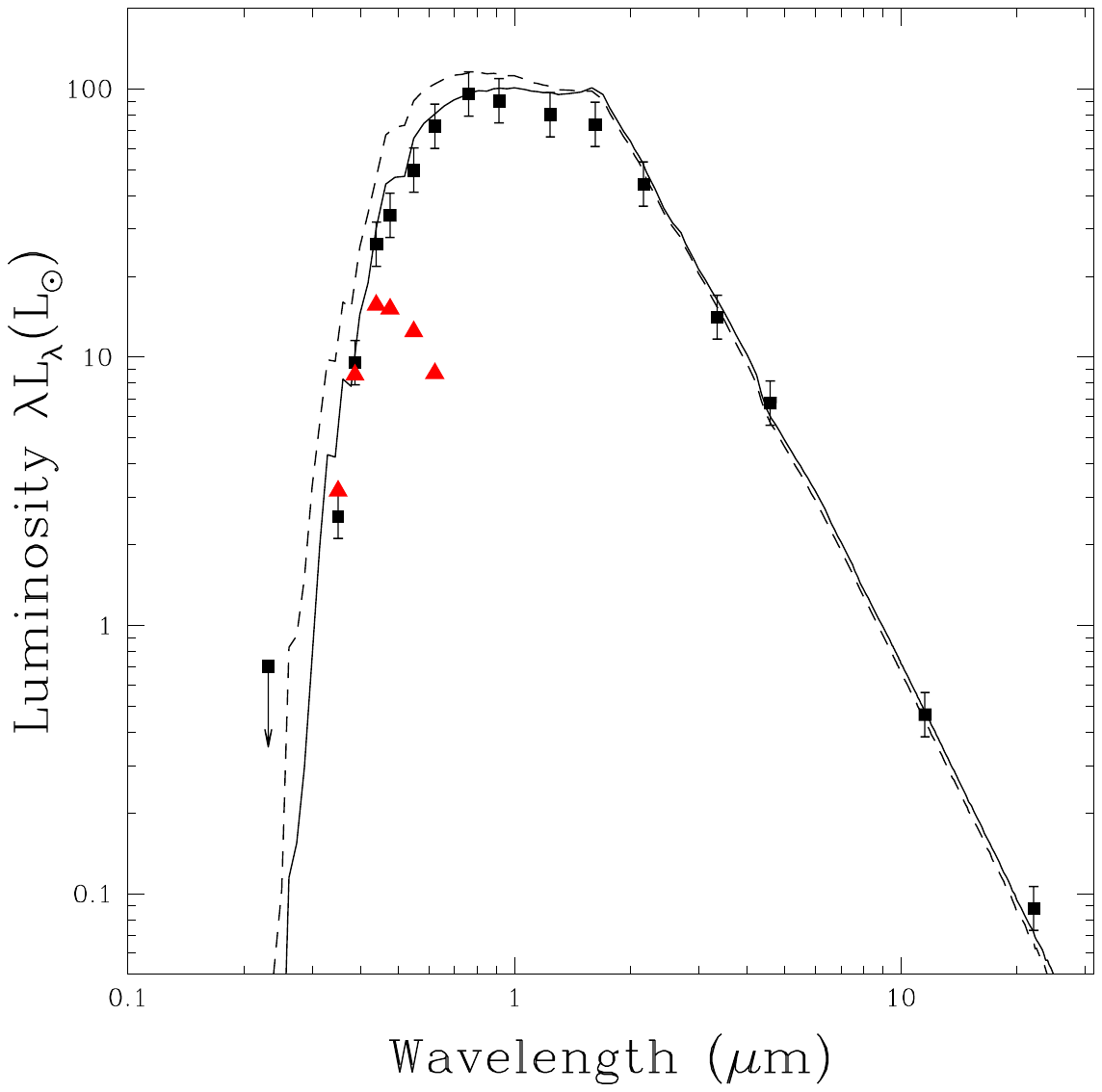}
	\vspace{-2cm}
    \caption{The best-fitting SED model for V723 Mon after correcting for the veiling component (red triangles). The uncorrected SED from Figure \ref{sedfit} is also shown (dashed line). Note that the SED of the veiling component is very different from the SEDs of stars, as can be seen by comparing to Figure \ref{sedfit}.}
    \label{veiledsedfit}
\end{figure}

Since the second light has a non-negligible contribution to the total SED of V723 Mon for $\lambda\lesssim600$~nm, the inferred properties of the star from $\S$\ref{section:giant} can also be affected. We refit the SED of the giant after removing the estimated veiling component to obtain $T_{\rm eff, giant} \simeq 4150\pm100$~K, $L_{\rm giant}= 153 \pm 9~ L_\odot$, $R_{\rm giant}\simeq24.0\pm0.9~R_\odot$ and $E(B-V)\simeq0.069\pm0.044$ (Figure \ref{veiledsedfit}). The veiling component contributes ${\sim}12\%$ of the total SED flux, with most of the flux in the bluer wavelengths. We perform \verb"iSpec" fits to the PEPSI spectrum after truncating it to the redder wavelengths at $\lambda>600$~nm that are minimally affected by veiling. We obtain $T_{\rm{eff}}=4350\pm50$~K, $\log(g)=2.0\pm0.1$, $\rm [Fe/H]=-0.7\pm0.1$, $\rm [\alpha/Fe]=0.3\pm0.1$, and $v_{\rm micro}=0.8\pm0.1\, \rm km \,s^{-1}$. From the parameters listed in Table \ref{tab:specpar}, these differ by $\Delta T_{\rm eff, giant}\simeq220$~K, $\Delta \log(g)\simeq0.3$, and $\Delta \rm [Fe/H]\simeq0.3$. In contrast, the parameters derived from the \verb"PHOEBE" fits (Table \ref{tab:params}) do not change significantly if we change $T_{\rm eff, giant}$. Because the models of the veiled SED are driven to larger, cooler stars, models forcing the star to be more compact are even more disfavored than in the unveiled models. The
best model of the veiled SED has $\chi^2=14.2$, rising to $33.9$ for $R_{\rm giant}=20~ R_\odot$ and to $64.0$ for $R_{\rm giant}=18~R_\odot$.

The origin of this veiling component remains unclear although it is clearly non-stellar in nature. However, the morphology of the veiling component is broadly compatible with the spectra of advection dominated accretion flows (ADAF). ADAF spectra can be described by contributions due to synchrotron emission, Compton scattering and Bremsstrahlung radiation \citep{Quataert1999}. The most luminous feature in the ADAF models come from the synchrotron peak and for stellar mass black holes, this peak falls in the optical wavelengths \citep{Quataert1999}. In their ADAF models for quiescent black hole binaries ($M_{\rm BH}=6~M_\odot$) with low accretion rates ($\log(\dot{M}/\dot{M}_{\rm edd})\sim-4$), \citet{Esin1997} show that the SED peaks at optical wavelengths and rapidly decays at both longer and shorter wavelengths. For V723 Mon, detailed ADAF models are necessary to determine whether the veiling component is related to accretion flows around the dark companion, but such analysis is beyond the scope of this paper.

\subsection{Limits on Luminous Stellar Companions} \label{section:limits}

We can constrain the presence of luminous companions using either the SED or the absence of eclipses. The limits using only the SED will be weaker than those using the eclipses, but are also independent of any knowledge of the inclination.  For the SED constraints, we require that the companion contributes less than 100\%, 60\% and 20\% of the light in the $UVM2$, $B$ and $V$ bands respectively.  The $B$ and $V$ band limits correspond to the estimated veiling source from \S\ref{section:veiling} and are conservative since the SED of the veiling light appears to be inconsistent with a star, and the $UVM2$ band limit is simply the total observed flux because we lack any constraint on the amount of veiling for this band.  For the eclipse constraints, we require that the
companion contributes less than 10\%, 3\%, 2\% and 1\% in the $UVM2$, $B$, $V$ and $R$ bands based on the eclipse models in \S\ref{section:phoebe}. While we did not use the $UVM2$ band in the SED fits because the SED models cannot account for any possible chromospheric emission from the rapidly rotating giant, there is no issue with using them to obtain these limits.

To provide models for the relationships between luminosity, temperature and mass, we sampled stars from $\rm [Fe/H]=-1$ PARSEC \citep{Bressan2012} isochrones with ages from 1 to 10~Gyr, and we logarithmically interpolated along each isochrone to sample more densely in mass. We considered models in which the companion is either a single star or two stars. In the latter case, we considered all combinations of two stars on the same isochrone. The age of the stars is the principal variable leading to changes in the mass limits, but the differences are not large. Two star models will generically allow higher limiting masses for two reasons.  First, since luminosity is a steeply rising function of mass, dividing a single star into two lower mass stars of the same total mass leads to a lower total luminosity. Second, the two lower mass stars are also cooler on the main sequence, so they produce less blue light per unit luminosity than the single star, which also allows the total mass be larger for the same constraints on
the bluer fluxes.

\begin{figure}
	\vspace{-2cm}	
	\includegraphics[width=0.50\textwidth]{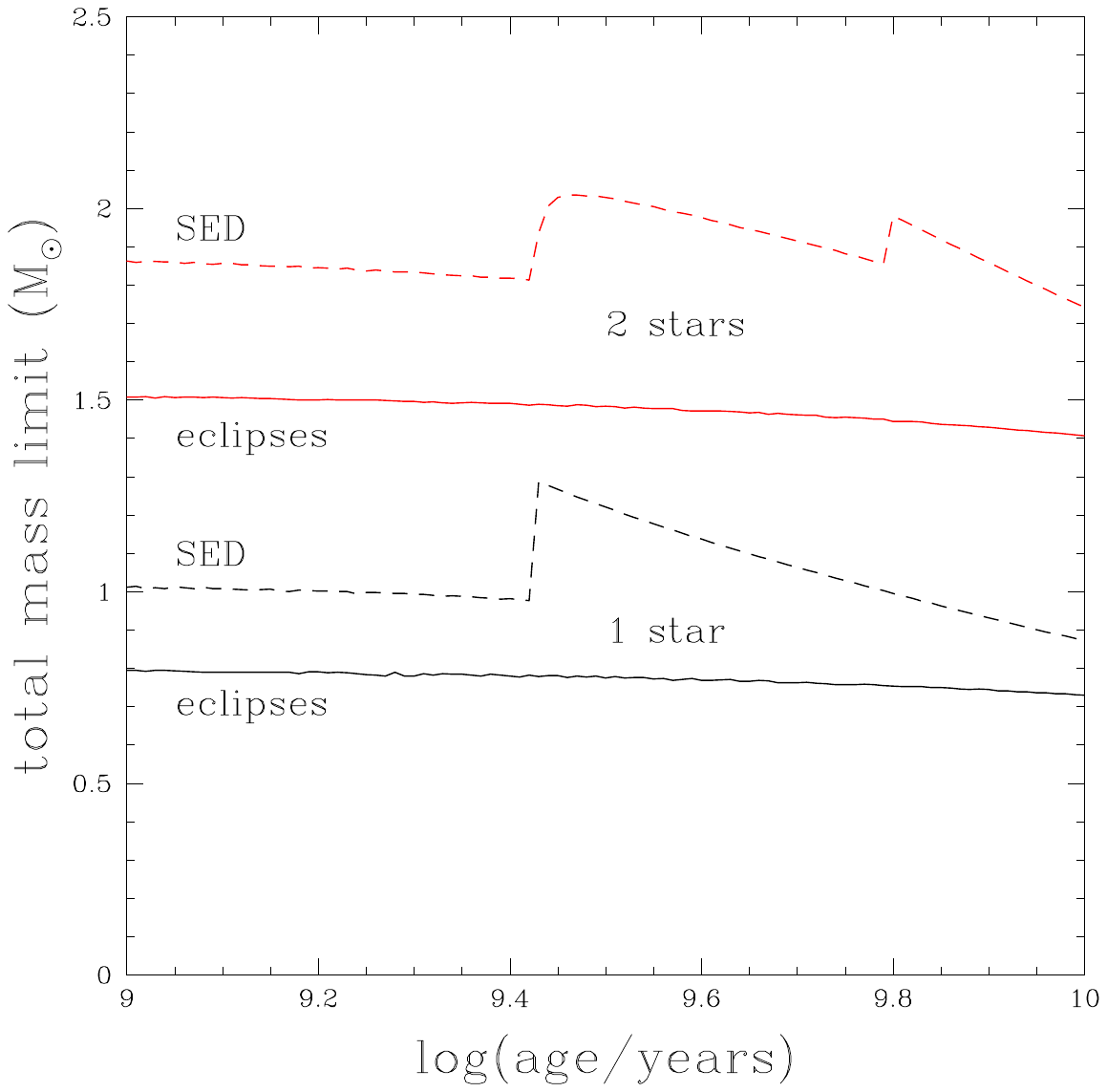}
	\vspace{-2cm}	
    \caption{Limits on the mass of a companion that is comprised of either one (black) or two (red)
      luminous stars based on either the lack of KELT eclipses (solid) or the SED (dashed) as a function of the age of the stars. From these limits, the maximum allowed mass for a single star is $1.29~M_\odot$ (SED) and $0.80~M_\odot$ (eclipses). For two stars, the maximum allowed mass is $1.98~M_\odot$ (SED) and $1.51~M_\odot$ (eclipses).}

    \label{fig:llimit}
\end{figure}

We start with the weaker but inclination independent limits from the SED, where Figure~\ref{fig:llimit} shows the mass limits as a function of stellar age. For a single star, the maximum allowed mass is 
$1.29~M_\odot$ ($\log(\rm age)=9.49$, $L=15.5~L_\odot$, $T=5430$~K), and it occurs where the star is just starting to evolve off the main sequence because the initial drop in temperature weakens the constraints more than the rise in luminosity (see Figure \ref{sedfit}).  There are two maximum masses for models with two stars.  The lower
age peak corresponds to pairing a slightly evolved star with a lower mass main sequence star.  The
total mass of $1.98~M_\odot$ comes from combining a $1.25~M_\odot$ ($L=12.0~L_\odot$, $T=5510$~K) star with a $0.77M_\odot$ ($L=0.5~L_\odot$, $T=5900$~K) star at $\log(\rm age)=9.45$. The higher age peak comes from
combining two stars of the same mass.  The total mass is again $1.98~M_\odot$ and the two components
each have $M=0.99~M_\odot$ ($\log(\rm age)=9.80$, $L=8.5~L_\odot$, $T=5430$~K).  Fig.~\ref{sedfit} shows the SEDs of these three models as compared to that of the giant, and they are nearly identical. 

For the BH candidates LB1 and HR 6819, the stellar companions seemed ``dark'' because they were hot. This cannot be the case here because of the tight $UVM2$ luminosity limit of $\lambda L_\lambda \approx2~L_\odot$ (after correcting for extinction). Models normalized by this luminosity only have total luminosities of $4~L_\odot$, $5~L_\odot$ and $75~L_\odot$, for temperatures of $10^4$~K, $3\times10^4$~K and $10^5$~K respectively, much too low for a massive He star companion. For example, a Helium star with $T\sim10^5$~K will have $M\sim8~M_\odot$ and $L\sim90,000~L_\odot$ \citep{Grafner2012}. For any hot companion with a luminosity approaching that of the giant, we would also expect to see the effects of irradiation of the giant on its light curve and phase dependent spectra, but no such perturbations are seen.

Given the size of the orbit and the size of the giant, companions with radii similar to those allowed by the SED will be eclipsed provided the inclination angle is $i\gtrsim 76^\circ$.  While the systematic uncertainties on the estimated inclination of $87.0^\circ{}^{+1.7^{\circ}}_{-1.4^{\circ}} $ may be moderately larger, the light curve shapes at inclinations anywhere approaching this upper limit for seeing eclipses are grossly incompatible with the observations.  Fig.~\ref{fig:llimit} also shows the limits on masses using the flux limits on the companion required to avoid visible eclipses, and they are stronger than those from the SED as expected. The biggest change is that they eliminate the ``bumps'' associated with stars starting to move off the main sequence because they more directly constrain the size of any companion.  The improvements are otherwise modest because the (blue) luminosities are such strong functions of mass that moderate changes in the limits on the luminosity produce very modest changes in the limits on the mass.  The single star limit is now $0.80 M_\odot$ and the two star limit is $1.51M_\odot$ where the two star limit is always weakest for the stars being twins. These limits are smaller than the binary mass function for this system (see $\S$\ref{section:orbit}).

\subsection{Balmer $\rm H\alpha$ and $\rm H\beta$ emission} \label{section:balmer}

\begin{figure*}
	\includegraphics[width=0.98\textwidth]{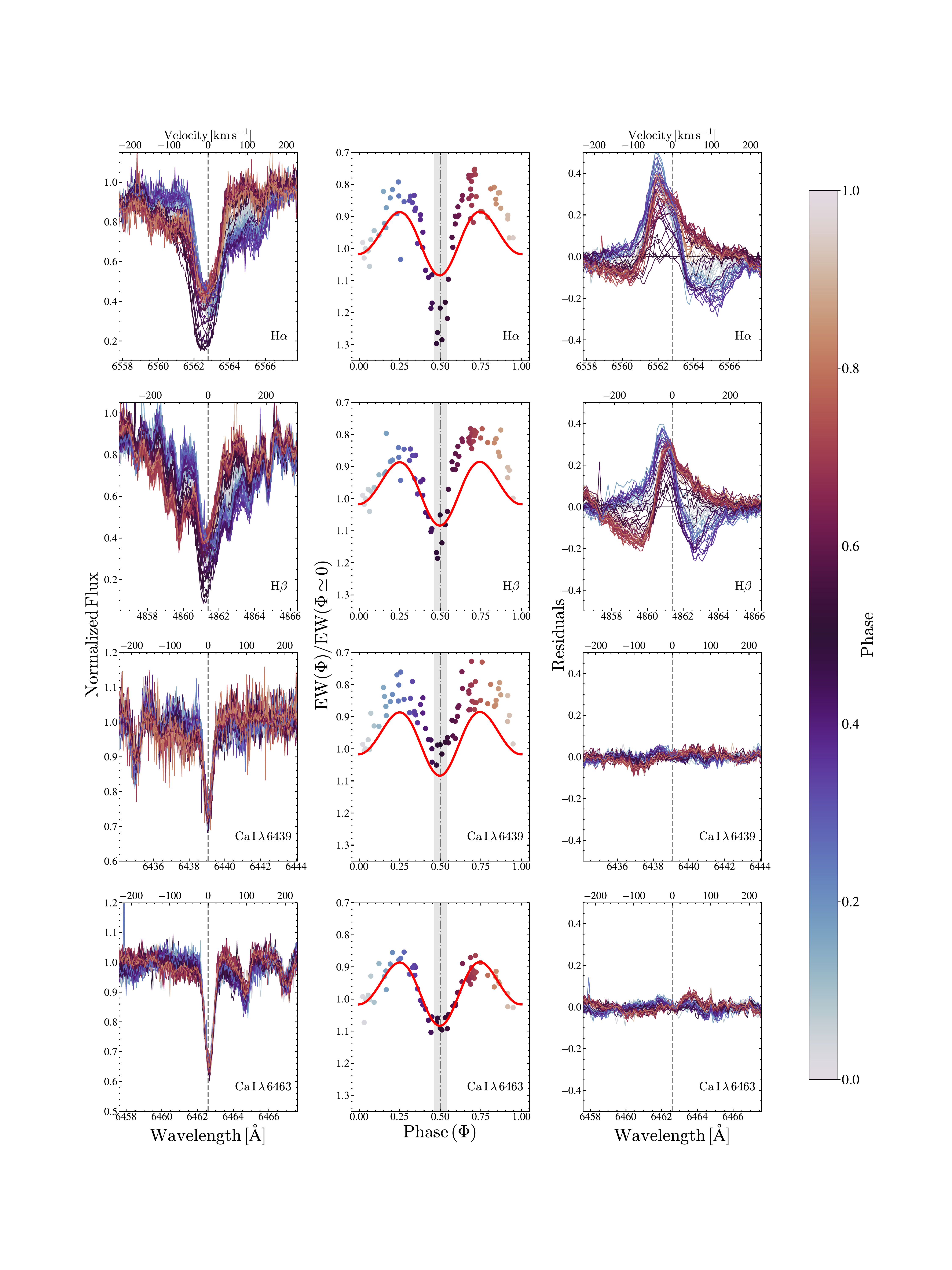}
		\vspace{-1.5cm}		
    \caption{SES line profiles for the Balmer $\rm H\alpha$, $\rm H\beta$, $\ion {Ca}{i}\, \lambda 6439$ and $\ion {Ca}{i}\, \lambda 6463$ lines as a function of orbital phase (left). The middle column shows the equivalent width normalized to $\phi\simeq0$ as a function of phase. A simple sinusoidal model (red) is fit to the variations in the $\ion {Ca}{i}\, \lambda 6463$ line to illustrate the variations in the equivalent width due to ellipsoidal variations. The shaded gray region illustrates the duration of the eclipse (${\sim}5$~days). The right column shows the residuals of the line profiles after a template for the giant is subtracted.}
    \label{spec}
\end{figure*}

We find that the Balmer $\rm H\alpha$ and $\rm H\beta$ lines appear to significantly vary with phase (Figure \ref{spec}) and this is unusual for a red giant. Given our phase convention, the dark companion will be eclipsed by the giant at $\phi=0.50$ with an eclipse duration of $t_{\rm ecl}\simeq5$~days. If a companion is responsible for Balmer emission, the subtraction of a template spectrum near phase $\phi=0.50$ should isolate its contribution. To explore this, we subtracted the SES spectrum of the red giant near $\phi \simeq 0.5$ from the remaining spectra with $S/N>30$.

\begin{figure*}
	\includegraphics[width=\textwidth]{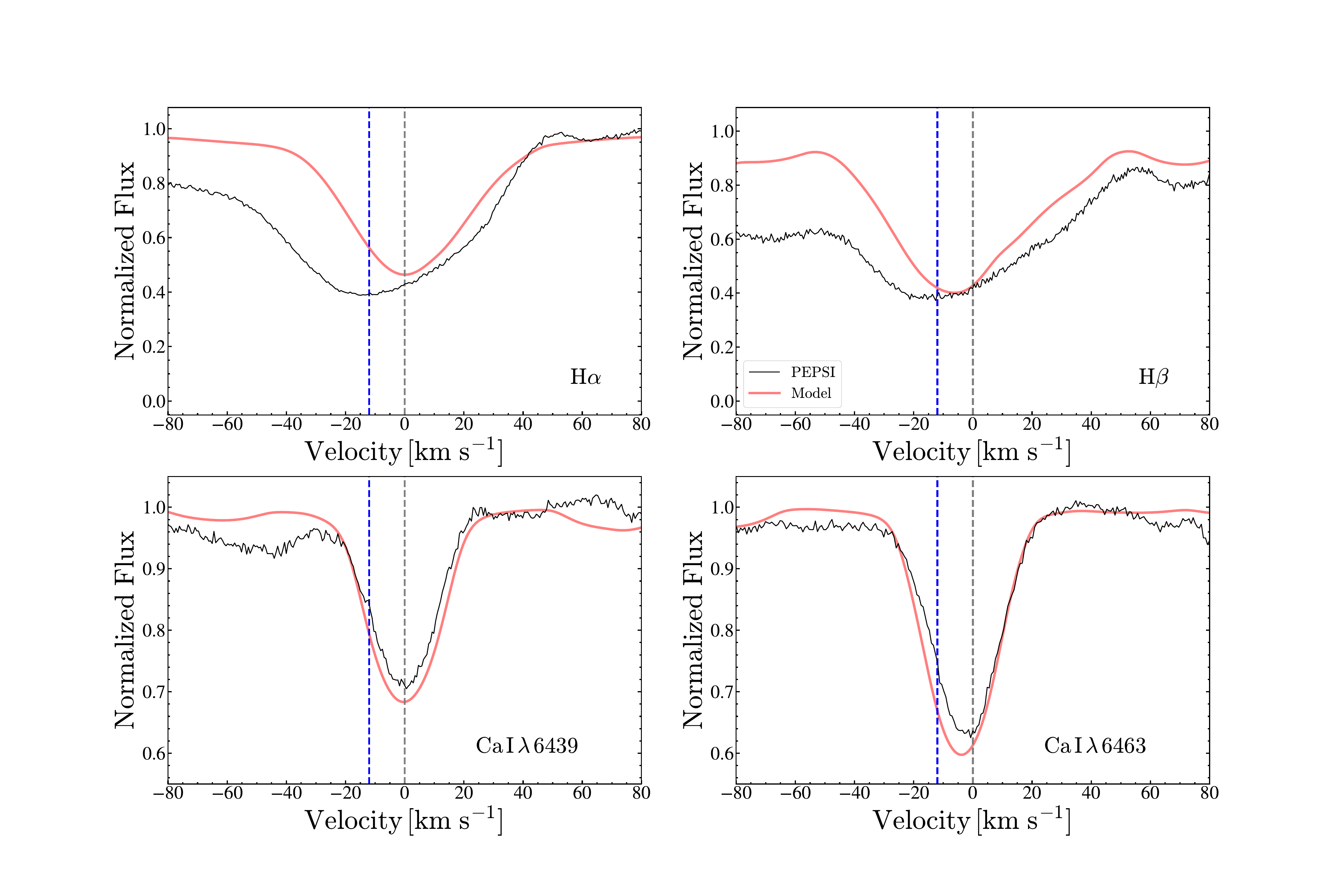}
	
    \caption{LBT/PEPSI line profiles for the Balmer $\rm H\alpha$, $\rm H\beta$, $\ion {Ca}{i}\, \lambda 6439$ and $\ion {Ca}{i}\, \lambda 6463$ lines (black). A model spectrum using the atmospheric parameters derived in $\S$\ref{section:giant} is shown in red. The blue lines show the velocity offset of the Balmer absorption lines (${\sim}12\, \rm km\,s^{-1}$) from the rest frame of the giant.}
    \label{pepsil}
\end{figure*}

Figure \ref{spec} shows the changes in the $\rm H\alpha$, $\rm H\beta$, $\ion {Ca}{i}\, \lambda 6439$ and $\ion {Ca}{i}\, \lambda 6463$ lines with orbital phase. The left panel shows the continuum normalized spectra, the middle panel shows the equivalent widths and the right panel shows the residuals relative to the spectrum observed when the companion should be in eclipse. A first point to note is that all four lines show a phase-dependent change in equivalent width which mirrors the ellipsoidal variability - this effect is well-known (e.g., \citealt{Neilsen2008}) and further confirms the origin of the variability ($\S$\ref{section:phoebe}). The Balmer absorption lines appear to be blue shifted by ${\sim}12 \, \rm km\,s^{-1}$ (Figure \ref{pepsil}). This is also seen in the spectra close to conjunction at $\phi=0.5$ (Figure \ref{spec}). However, we do not see a similar shift in the other photospheric lines. The second point to note is that while photospheric absorption lines like $\ion {Ca}{i}\, \lambda 6439$ and $\ion {Ca}{i}\, \lambda 6463$ lines are cleanly subtracted, the $\rm H\alpha$ and $\rm H\beta$ emission clearly varies with orbital phase. We do not see the $\ion {Ca}{ii}$ H and K lines in emission, so the changes in the Balmer lines are unlikely to be caused by chromospheric activity. The Balmer emission could be caused by mass loss from the red giant through a stellar wind.
 
The typical Gaussian FWHM of the $\rm H\alpha$ and $\rm H\beta$ emission profiles is ${\sim}290\, \rm km\,s^{-1}$. The median equivalent width of the residual $\rm H\alpha$ and $\rm H\beta$ lines is $\rm EW(H\alpha)=1.79\pm0.02$ and $\rm EW(H\beta)=2.72\pm0.02$ respectively. We can convert the $\rm H\alpha$ equivalent width to the flux at the stellar surface using $F_{\rm H\alpha}=F_{c}~\rm EW(H\alpha)$ (see for e.g., \citealt{Soderblom1993,GH2010}). $F_c$ is the continuum flux which we derive using \citet{Hall1996} as $\log(F_c)=7.538-1.081(B-V)_0$. For V723 Mon, we find $(B-V)_0=0.85\pm0.06$~mag, using the APASS DR10 photometry and the $E(B-V)$ from the SED fits. We obtain $\log(F_c)=6.61\pm0.06$ and $F_{\rm H\alpha}=7.3\pm1.0~\rm ergs~cm^{-2}~s^{-1}$. Normalizing the $\rm H\alpha$ flux to the bolometric flux (i.e. $R_{\rm H\alpha}=F_{\rm H\alpha}/\sigma T^4$), we obtain $\log(R_{\rm H\alpha})=-3.53\pm0.06$. $\log(R_{\rm H\alpha})$ is usually compared to the Rossby number $R_0=P_{\rm rot}/\tau_c\sim2.84$, where $\tau_c\approx21.1$~d based on \citet{Noyes1984} and assuming that the giant's rotation is fully synchronized with the orbital period. At this value of $R_0$, V723 Mon has a value of $\log(R_{\rm H\alpha})$ higher than any of the chromospherically active single stars from \citet{LS2010} (see figure 5 in \citealt{GH2010}). This also indicates that the observed $\rm H\alpha$ emission is not just from chromospheric activity. 

\begin{figure*}
	\includegraphics[width=0.85\textwidth]{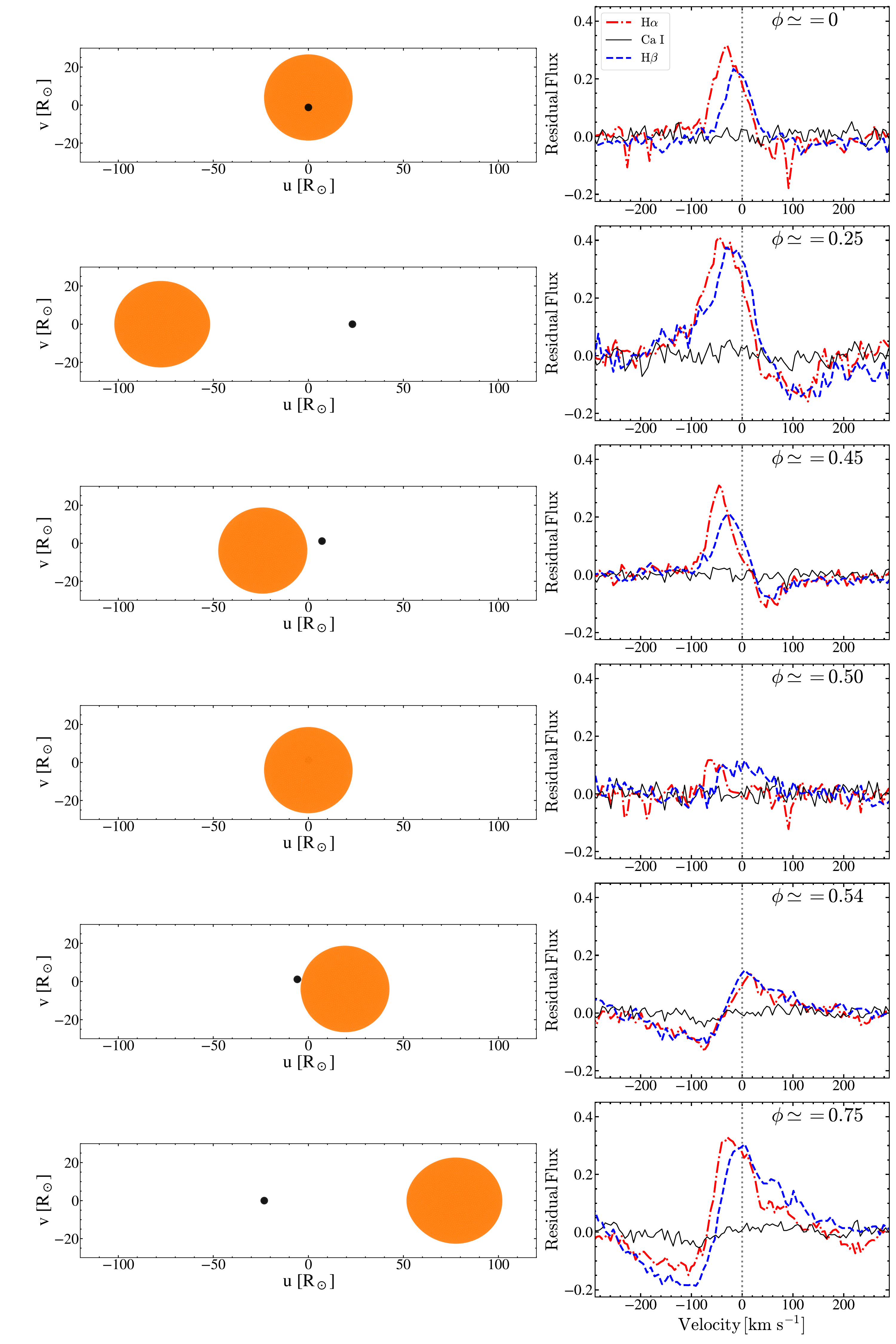}
	
    \caption{The right panels show the template subtracted $\rm H\alpha$ (red dot-dashed), $\rm H\beta$ (blue dashed) and $\ion {Ca}{i}\, \lambda 6439$ (black) line profiles at various orbital configurations illustrated in the left panels. The size of the compact object is simply chosen to make it easily visible.}
    \label{orbhalpha}
\end{figure*}

\begin{figure}
	\includegraphics[width=0.5\textwidth]{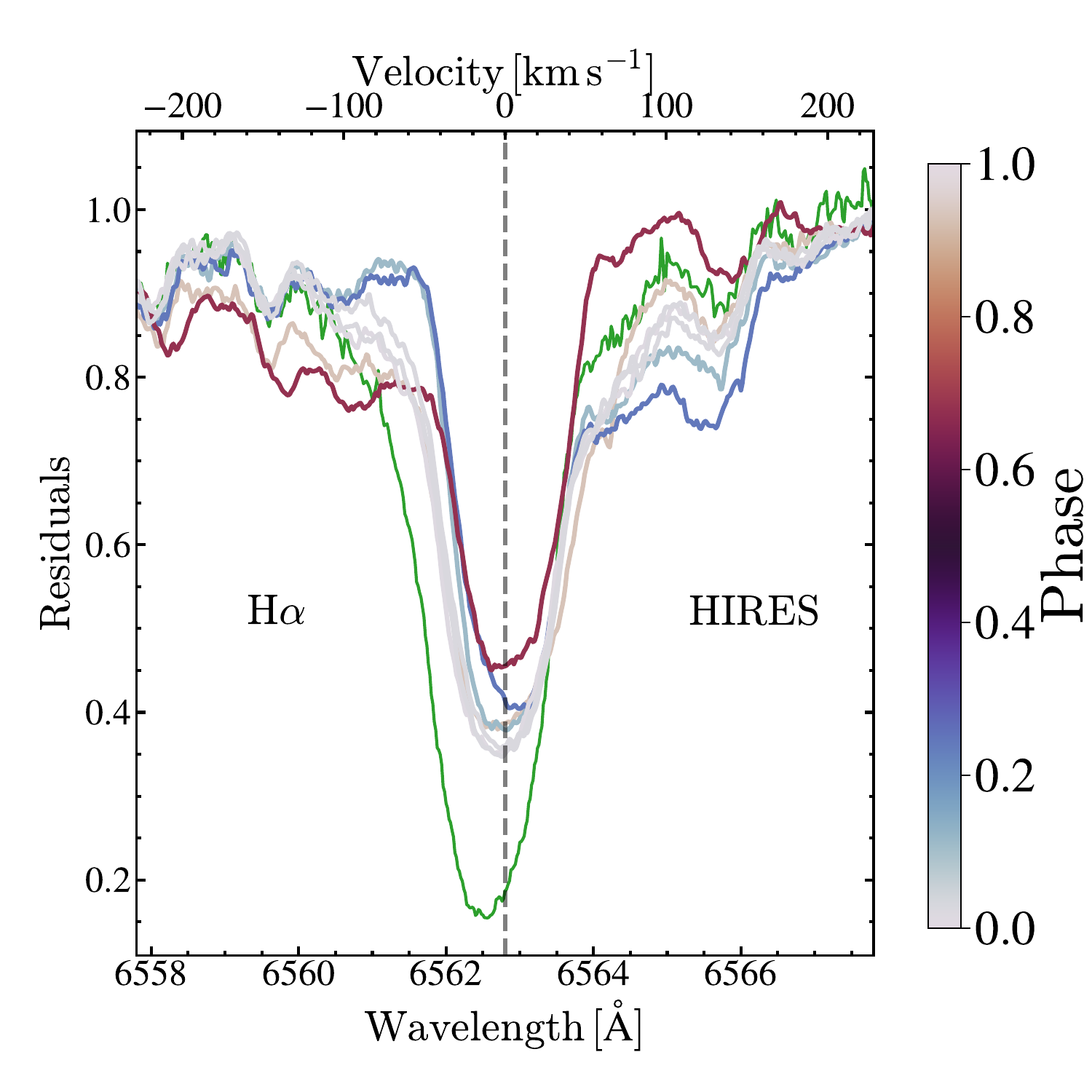}
    \caption{The HIRES $\rm H\alpha$ line profiles as a function of orbital phase. The template SES spectrum at $\phi\simeq0.5$ is shown in green. The variability in the line profiles is the same as that seen in the SES spectra (see Figure \ref{spec}, $\S$\ref{section:balmer}).}
    \label{hiresbalmer}
\end{figure}

Another argument against chromospheric activity is that the changes in the Balmer lines with phase are coherent over the ${\sim}3.5$~years spanned by the SES data. If the Balmer emission was dominated by chromospheric activity, we would expect the structure to change with time as the spot patterns evolve. At RV quadrature ($\phi\ \simeq0.25$ and $\phi\ \simeq0.75$), the emission profiles of the Balmer lines resemble P-Cygni profiles with both absorption and emission components. This is illustrated in Figure \ref{orbhalpha}. In comparison, the $\ion {Ca}{i}\, \lambda 6439$ line is cleanly subtracted and does not show a similar correlation with the phase of the binary orbit. P Cygni profiles tend to be associated with mass outflow and stellar winds. At RV minimum ($\phi\ \simeq0.25$), the absorption component in the Balmer lines is red shifted whereas at RV maximum ($\phi\ \simeq0.75$), the absorption is blue shifted. In the rest-frame of the giant, the peak of the $\rm H\alpha$ emission component is relatively stationary at ${\sim}-35\, \rm km\,s^{-1}$. The median separation between the absorption and emission components in $\rm H\alpha$ is $\Delta V{\sim}115\, \rm km\,s^{-1}$, however this appears to vary with orbital phase. The separation is largest for $0\leq\phi\leq0.25$, with $\Delta V{\sim}148\, \rm km\,s^{-1}$, and drops thereafter. The smallest separation between the two components is seen at phases $0.5\leq\phi\leq0.75$, with $\Delta V{\sim}104\, \rm km\,s^{-1}$. Much like with the SES spectra, we also see clear variability in the $\rm H\alpha$ line profiles in the HIRES spectra (Figure \ref{hiresbalmer}), and the asymmetry in the HIRES line profiles also reverses after $\phi=0.5$.

Remarkably, near conjunction ($\phi\simeq0.5$) there is very little Balmer emission (Figure \ref{orbhalpha}). Both the Balmer and photospheric lines are modulated with the ellipsoidal variations (Figure \ref{spec}), however, near $\phi \simeq0.5$, the equivalent width of the Balmer lines increases abruptly, signalling a dramatic drop in Balmer emission. A similar EW increase is not seen in the photospheric lines. This feature is coincident with the eclipse of the unseen companion by the red giant and has the expected duration. The MODS spectra were taken around the eclipse of the dark companion by the red giant at $\phi\simeq0.5$ and the depths of the MODS Balmer $\rm H\alpha$ and $\rm H\beta$ absorption features also deepen during the eclipse.

While we have clear evidence to show the presence of Balmer emission that is correlated with the orbital motion of the putative black hole ($\S$\ref{section:balmer}), the exact origin of the Balmer emission is unclear. One explanation for this additional luminosity and the variability in the Balmer lines is from an accretion flow. Alternative explanations for the Balmer emission can come from a stellar wind and/or mass outflow at the inner Lagrangian point ($L_1$). At $\phi=0$, the inner Lagrangian point is directed toward the observer. This also coincides with the deeper minimum in the light curve because the surface gravity and brightness is smallest at $L_1$ \citep{Beech1985}. Conversely, at $\phi=0.5$, $L_1$ points away from the observer. It is possible that the Balmer emission is associated with the photoionization of matter streaming through the inner Lagrangian point. The nearly-constant velocity offset of the emission peak from the secondary is also consistent with this interpretation. However, this scenario requires a source of photoionization. The neutron star binary 1FGL J1417.7$-$4402 also has a persistent $\rm H\alpha$ line with a complex morphology \citep{Strader2015,Swihart2018}. Instead of an accretion disk, the authors attributed this behavior to the interaction between the magnetically driven wind from the secondary and the pulsar wind. Balmer photons originating from an interbinary shock or the wind from the secondary will have a velocity offset from the secondary. This is also seen in V723 Mon. However, the $H\alpha$ line in 1FGL J1417.7$-$4402 is significantly different from what is observed in V723 Mon as it has a double-peaked emission profile that is observed directly in the spectra without the need for template subtraction. \citet{Swihart2018} also did not find independent evidence of a disk through their light curves, whereas we see clear evidence of second light in the $B$ and $V$-band light curves for V723 Mon.

In summary, we see evidence of broad Balmer emission in the spectra. The Balmer line profiles prior to when the unseen companion is eclipsed by the giant ($0\leq\phi\lesssim0.46$) resemble P-Cygni profiles, with a blue shifted emission component and a red shifted absorption component in the rest-frame of the giant. When the unseen companion is behind the giant at $\phi\simeq0.5$, both the Balmer components disappear and the Balmer absorption features from the giant become deeper. When the unseen companion re-emerges after the eclipse at phases $0.54\lesssim\phi\leq1$,  we see a clear change in the P-Cygni-like line profile, with both the absorption and emission components blue shifted. However, the absorption component is blue shifted more than the emission and the P-Cygni profile is reversed. We also see significant changes in the Balmer line depths at $\phi=0.5$ when the companion is eclipsed. The exact origin of the Balmer emission, be it from an accretion disk around the black hole, a stellar wind, mass flow into an inner Lagrange point in the binary, or a combination of these, remains unclear. 

\subsection{X-ray detection} \label{section:xrays}

In the individual \textit{Swift} XRT exposures, we only detected a source above the background in the exposures taken on 2020-11-05 ($\phi\simeq0.2$), 2021-01-21 ($\phi\simeq0.5$) and 2021-02-20 ($\phi\simeq0.0$) with $4.2\pm2.2$, $6.8\pm2.9$ and $3.7\pm2.2$ background-subtracted counts in the 0.3-2.0 keV energy range, respectively. In the merged \textit{Swift} XRT data, we detect a source with $18.5\pm4.9$ background-subtracted counts in the 0.3-2.0 keV energy range, corresponding to an aperture corrected count rate of $(1.4\pm0.4)\times10^{-3}$ counts/second. Assuming that $E(B-V)\simeq0.086$ from $\S$\ref{section:giant}, and using the relationship between reddening and atomic hydrogen column density from \citet{Liszt2014}, the Galactic column density towards V723 Mon is $N_{\rm H, V723}\simeq7.1\times10^{20}~\rm cm^{-2}$. This is considerably smaller than the total column density in the line of sight towards V723 Mon ($N_{\rm H, LOS}=3.42\times10^{21}~\rm cm^{-2}$). Assuming an absorbed power-law with a photon index of 2 and the Galactic column density estimate $N_{\rm H, V723}$, the \textit{Swift} XRT count rate corresponds to an absorbed flux of $(3.0\pm0.8)\times10^{-14}~ \rm ergs~cm^{-2}~s^{-1}$ or an unabsorbed flux of $(4.1\pm1.0)\times10^{-14}~ \rm ergs~cm^{-2}~s^{-1}$ in the 0.3-2.0 keV energy range. At the \textit{Gaia} EDR3 distance, we obtain absorbed and unabsorbed X-ray luminosities of $(7.6\pm2.0)\times10^{29}~ \rm ergs~s^{-1}$ and $(1.0\pm0.3)\times10^{30}~ \rm ergs~s^{-1}$, respectively. If we use a 0.3-10.0 keV energy range, we obtain a similar number of counts as there is very little emission $>$2.0 keV\footnote{We derive a 3$\sigma$ upperlimit of $1.2\times10^{-4}$ counts/second to the 2.0-10.0 keV count rate, which corresponds to an absorbed flux of $1.4\times10^{-14} \rm ergs~s^{-1}$ and a luminosity of $3.6\times10^{29} \rm ergs~s^{-1}$.}. We also derive a limit on the hardness ratio\footnote{Here the hardness ratio is defined as $HR=(H-S)/(H+S)$ where $H$ is the number of counts in the 2.0-10.0 keV energy range and $S$ is the number of counts in the 0.3-2.0 keV energy range. } of $-0.21$ which indicates a relatively soft emission spectrum that is consistent with the lack of hard X-ray emission above 2.0 keV. After grouping the XRT observations by orbital phase, we calculate the out-of-eclipse and in-eclipse absorbed X-ray luminosities as $(6.1\pm2.2)\times10^{29}~ \rm ergs~s^{-1}$ and $(9.3\pm3.1)\times10^{29}~ \rm ergs~s^{-1}$ respectively. The out-of-eclipse and in-eclipse X-ray luminosities are consistent given the measurement uncertainties.

In the {\it XMM-Newton} observation, V723~Mon is not detected above the background, so we derive an upper-limit on $L_{\rm X}$. We find a 3$\sigma$ upper-limit count rate of 4.17$\times10^{-2}$ counts/second in the 0.5-8.0 keV band within a 60\arcsec\ aperture (consistent with the off-axis PSF at the source position). Assuming an absorbed power-law with a photon index of 2 and the Galactic column density of $N_{\rm H, V723}$, we find an upper-limit absorbed flux of 1.0$\times10^{-13}$~erg~cm$^{-2}$~s$^{-1}$ and an unabsorbed flux of 1.1$\times10^{-13}$~erg~cm$^{-2}$~s$^{-1}$. Adopting the {\it Gaia} EDR3 distance, the latter quantity yields an X-ray luminosity upper-limit of 2.7$\times10^{30}$~erg~s$^{-1}$ in the 0.5-8.0 keV band.

The \textit{Swift} XRT estimate of the X-ray luminosity is comparable to the $L_{\rm X}\sim10^{30}-10^{31}~ \rm ergs~s^{-1}$ observed for quiescent X-ray binaries \citep{Dincer2018} and the $L_{\rm X}\sim10^{29}-10^{30}~ \rm ergs~s^{-1}$ observed for chromospherically active RS CVn systems \citep{Demircan1987}. If the X-ray luminosity originates from an accretion disk, the putative black hole accretes at a very low luminosity of ${\sim}10^{-9}~L_{\rm edd}$. There is observational evidence which indicates that the X-ray luminosity from quiescent black holes is significantly fainter than that from neutron stars (see for e.g., \citealt{Asai1998,Menou1999}). While debated, this has been attributed to advection-dominated accretion flows (ADAF; see, for e.g., \citealt{Narayan1995}) and the existence of event horizons for black holes. However, given the rapid rotation of the giant, some of the observed X-ray luminosity may originate from the giant's chromosphere \citep{Gondoin2007}. The X-ray spectra of X-ray binaries are generally harder (hotter), with average temperatures $kT>5$ keV, compared to stellar coronae that have lower temperatures, with $kT<2$~keV \citep{Kong2002}. Given the results of the \textit{Swift} XRT analysis, it appears that the X-ray emission is relatively soft, which appears to be consistent with a chromospheric origin. However, to fully characterize this X-ray emission, followup X-ray spectra of significantly better S/N are necessary.

We can estimate the mass loss rate from the giant as
\begin{equation}
    \dot{M}=4\times10^{-13}~\eta_R ~\frac{LR}{M}\,M_\odot {\rm yr^{-1}} \approx 8\times10^{-10} \,M_\odot {\rm yr^{-1}},
\end{equation} \citep{Reimers1975} where $L$, $R$ and $M$ are in solar units and $\eta_R\simeq0.477$ \citep{McDonald2015}. The velocity of the wind from the giant is assumed to be the escape velocity, which is $V_{\rm wind}\approx V_{\rm esc}\sim120\rm \,km~s^{-1}$. For a scenario where the black hole accretes mass through the stellar wind, \citet{Thompson2019} approximated the amount of material gathered at the sphere of influence of the black hole as \begin{equation}
\begin{aligned}
& \dot{M}_{\rm acc}\sim \frac{\dot{M}}{(4\pi a_{\rm giant}^2)}\,\pi \left(\frac{G M_{\rm BH}}{V_{\rm wind}^2}\right)^2\\
& \sim7\times10^{-11}\,{\rm M_\odot\,\,yr^{-1}}\,\,\, \dot{M}_{-9}\bigg(\frac{M_{\rm BH}}{3 M_\odot}\bigg)^2 \bigg(\frac{V_{\rm wind}}{120 \rm \,km~s^{-1}}\bigg)^{-4},
\end{aligned}
\end{equation} where $\dot{M}_{-9}=\dot{M}/10^{-9}$\,M$_\odot$ yr$^{-1}$. For radiatively efficient accretion onto the black hole, the accretion luminosity can be approximated by
\begin{equation}
    L_{\rm acc}\sim0.1~\dot{M}_{\rm acc}~c^2\simeq83~L_\odot.
\end{equation}
For radiatively inefficient accretion, the luminosity can be much lower \citep{Narayan1995}. This estimate of the accretion luminosity is much larger than the observed X-ray luminosity of this system ($\S$\ref{section:xrays}), but closer to the luminosity of the veiling component ($L_{\rm acc}\approx 4L_{\rm veil}$). 

\section{The Nature of the Dark Companion}
\label{section:v723disc}

Given the observed properties of the system and the modeling results from $\S$\ref{section:v723obs} and $\S$\ref{section:v723res}, we next systematically discuss the nature of the companion. We first discuss the uncertainties in the mass
of the companion and then systematically consider the possible single and binary possibilities
for its composition. 

Ultimately, our knowledge of the mass of the companion is determined by how well we can constrain
the properties of the giant. Estimates of the mass through either the radius and gravity or
stellar evolution models give relatively crude limits of $1.1\pm 0.5~ M_\odot$ and $1.1 \pm 0.2~ M_\odot$,
respectively ($\S$\ref{section:giant}). If the giant has a degenerate core, which is likely to 
be true because it is more luminous than the red clump, then there is a very firm lower limit to 
its mass.
The luminosity of a giant with a degenerate core of mass $M_{\rm core}$ is
\begin{equation}
    \bigg(\frac{L}{L_\odot}\bigg)\approx 675 \bigg(\frac{M_{\rm core}}{0.4~M_\odot}\bigg)^{7}.
\end{equation} \citep{Boothroyd1988}, so we must have $M_{\rm giant}>M_{\rm core}\simeq0.33~M_\odot$ for $L\sim170~L_\odot$. This implies a lower bound for the companion mass of $M_{\rm comp}\gtrsim2.26~M_\odot$, above the mass of the most massive neutron star observed and larger than the limits for single and binary stellar companions in $\S$\ref{section:limits}.

The stronger limits on the mass come from the \verb"PHOEBE" models of the ellipsoidal variability in $\S$\ref{section:phoebe}. Ignoring inclination, ellipsoidal variability constrains the mass ratio $q$ because (to leading order)
the amplitude of the $P_{\rm orb}/2$ term depends on $\epsilon_2 \sim (R/a)^3/q$ while the amplitudes of the $P_{\rm orb}$ and $P_{\rm orb}/3$ terms depend on $(R_{\rm giant}/a)^4/q$ (e.g., \citealt{Morris1985,Gomel2020}).  Since the semi-major axis is determined by the period, mass function and mass ratio, $a^3 \propto P^2 f (1+q)^2$, the ellipsoidal variability can constrain both $R_{\rm giant}$ and $q$.  This also means that the radius of the giant is an important independent constraint.  Given the period, mass function and amplitude $\epsilon_2$, the mass of the giant $M_{\rm giant} = f q (1+q)^2 \propto R_{\rm giant}^3$ simply scales with the radius of the
giant. The mass of the companion, $M_{\rm comp} = f(1+q)^2$, is less dependent on the radius because the mass
ratio $q$ is small. We can verify this correlation using \verb"PHOEBE" models with the radius of the giant fixed to
$R_{\rm giant}=18$, 20, 22, 24, and $26~R_\odot$.  The mass of the giant increases monotonically
and fairly rapidly with radius, $M_{\rm giant}\simeq0.42$, 0.55, 0.72, 0.91, and $1.12~ M_\odot$, while
the mass of the companion increases more slowly, $M_{\rm comp} \simeq 2.38$, 2.55, 2.75, 2.95, and $3.16~M_\odot$,
as expected.  In fact, the numerical results almost exactly track the expected scalings from holding
$\epsilon_2$ fixed while varying the radius of the star. 

The \verb"PHOEBE" models on their own found a radius of $R_{\rm giant}=24.9 \pm 0.7~ R_\odot$, which
agrees well with the independent results from the SED fits.  Without the correction for veiling, these fits gave $22.2 \pm 0.8~ R_\odot$ and with the correction for veiling they gave $24.0\pm 0.9~ R_\odot$.
These fits use a much more complete SED model than used for the Gaia DR2 estimate of ${\sim}20~ R_\odot$,
but even for this smaller estimate the companion mass is $M_{\rm comp}=2.6~ M_\odot$. Since the radius estimates
from the SED and the ellipsoidal variability agree, we will proceed under the assumption that the
resulting mass estimates of $M_{\rm giant}=1.00 \pm 0.07~ M_\odot$ and $M_{\rm comp} = 3.04 \pm 0.06~ M_\odot$  are essentially correct.

We next consider all the possible scenarios for the composition of the companion, with Table~\ref{tab:scenarios} providing a summary. If the companion is a single object, the
options are a star, a white dwarf (WD), a neutron star (NS) or a black hole (BH). A
star is ruled out by the eclipse and SED limits from \S\ref{section:limits}, with
$M_{\rm comp} < 0.8 M_\odot$ based on the lack of eclipses.  A WD is ruled out because its
mass would exceed the Chandrasekhar mass limit of $1.4 M_\odot$.  A NS is unlikely, since a companion mass of $M_{\rm comp}\simeq3.04~M_\odot$ is still slightly
larger than the maximum passable mass of a neutron star ($M\sim3~M_\odot$; \citealt{Lattimer2001}).
Additionally, the mass is well above the maximum observed masses of $M\simeq2.01\pm0.04~M_\odot$ and $M\simeq2.14_{-0.09}^{+0.10}~M_\odot$ found for the neutron stars PSR J0348+0432 and MSP J0740+6620, respectively \citep{2013Sci...340..448A, Cromartie2020}.  The mass of the dark companion is slightly larger than the $M\simeq2.59_{-0.08}^{+0.09}~M_\odot$ mass-gap compact object in the LIGO/VIRGO gravitational wave merger event GW190814 \citep{Abbott2020}, and it
is comparable in mass to the $M_{\rm BH}\simeq3.3_{-0.7}^{+2.8}~M_\odot$ non-interacting black hole identified around the red giant 2MASS~J05215658+4359220 by \citet{Thompson2019}. The mass estimates for V723~Mon, due to its ellipsoidal variability, are much tighter than for 2MASS~J05215658+4359220. Thus,
the simplest explanation of observed V723~Mon system properties is that the binary companion is a black hole near the lower end of the mass gap.

There are many more possible scenarios if the companion is a binary.  The orbit of such a binary
has to be fairly compact (see Appendix \ref{app:tertiary}), but not too compact, or the system
would merge too rapidly (see below).  While a single phase of common envelope evolution can likely 
lead to the simple binary models, any of these triple solutions likely requires a more complex evolutionary pathway which we will not attempt to explore here.   Based on our analyses in $\S$\ref{section:limits}, the dark companion cannot be a stellar binary given that the two star 
mass limit is $M_{\rm binary}<1.5~M_\odot$ from the eclipse constraints (Figure \ref{fig:llimit}).  The companion must contain at least one compact
object.

Combining a star with a compact object seems unlikely.  With the stellar mass $<0.8~ M_\odot$
based on the lack of eclipses, the compact object mass must be $>2.2~ M_\odot$.  A WD is ruled
out because this exceeds the Chandrasekhar limit.  A NS is possible, but the mass is at or above 
the maximum observed NS masses.  A BH is also possible, but a single more massive BH seems
far more plausible than a BH with a NS-like mass combined with a star to form the inner binary
of a triple system.

A double WD binary requires that both WDs are very close to the Chandrasekhar limit or above
it. Given that such massive WDs are quite rare \citep{Tremblay2016}, putting two of them into such a system is unlikely even if the masses can be kept just below the limit.  Combining a WD and a NS
is allowed, and the NS mass is in the observed range if the WD mass is $>0.9M_\odot$.  Combining
a WD with a BH is possible, but has the same plausibility problems as combining a star with a BH.
A double NS binary is feasible.  They would need to be of similar mass, since a mass ratio
$>0.67$ is needed to keep the lower mass NS above the theoretical minimum of $1.2~ M_\odot$ \citep{Suwa2018}. Combining a NS with
a BH and a double BH binary seem implausible because they both require BH masses in the range
observed for neutron stars.

\begin{table*}
	\centering
	\caption{Comparison of the possible scenarios involving stellar companions (*), white dwarfs (WD), neutron stars (NS) and black holes (BH) that can explain the nature of the dark companion. We assumed $M_{\rm comp}=3.04\pm 0.06~M_\odot$ from $\S$\ref{section:phoebe} (Table \ref{tab:params}). A `\cmark' indicates that the scenario is possible, a `?' indicates that while the scenario is technically possible, it is very unlikely and a `\xmark' indicates that the scenario is ruled out. The simplest explanation is that of a single low mass black hole, indicated with `\cmark \cmark'.}
	\label{tab:scenarios}
\begin{tabular}{rrr}
		\hline
		 Dark Companion & Possibility & Comment \\
		 \hline
		 \vspace{1mm}		 
		 Single Star & \xmark & Ruled out by SED/eclipse limits from $\S$\ref{section:limits} ($M_*\lesssim0.8~M_\odot$). \\
		 \vspace{1mm}
		 Single WD & \xmark & WD will exceed Chandrasekhar limit ($M_{\rm WD}>1.4~M_\odot$).\\
		 \vspace{1mm}		 
		 Single NS & ? & Requires an extreme NS equation of state. \\		 
		 \vspace{1mm}		 
		 Single BH & \cmark \cmark & Simplest explanation. \\
		\hline
				 \vspace{1mm}
		 Star + Star & \xmark &  Ruled out by SED/eclipse limits from $\S$\ref{section:limits} ($M_{\rm binary}\lesssim1.5~M_\odot$).\\
		 		 \vspace{1mm}
		 Star + WD & \xmark & For $M_*\lesssim0.8~M_\odot$ ($\S$\ref{section:limits}), the WD mass exceeds Chandrasekhar limit. \\		
		 		 \vspace{1mm}
		 Star + NS & ? & For $M_*<0.8~M_\odot$, the NS mass exceeds $2.2 M_\odot$.\\
		 		 \vspace{1mm}
		 Star + BH & ? & BH mass is even lower than with no star. \\	
		\hline
		 		 \vspace{1mm}		
		 WD + WD & \xmark & Both WD components near or above Chandrasekhar limit. \\
		 		 \vspace{1mm}
		 NS + WD & \cmark & NS mass is in the observed range if $M_{\rm WD}>0.8~M_\odot$ \\	
		 		 \vspace{1mm}
		 BH + WD  & ? & BH mass is even lower than with no WD. \\		 		 
		 		 \vspace{1mm}		
		 NS + BH & \xmark & The BH must have a NS-like mass. \\
		 		 \vspace{1mm}		 
		 NS + NS & \cmark & Both NS components should have $M_{\rm NS}\gtrsim1.2~M_\odot$, so $q_{\rm inner}\gtrsim0.67$. \\
		 		 \vspace{1mm}
		 BH + BH & \xmark & The BHs have NS masses. \\

\hline
\end{tabular}
\end{table*}

An additional consideration for any compact object binary model for the companion is its lifetime
due to the emission of gravitational waves.  For an equal mass, $3~M_\odot$ binary in a circular
orbit with semi-major axis $a_{\rm in}$, the merger time is

\begin{equation}
      t_{\rm merge} = 2.2 \times 10^7~\hbox{years} \left( { a_{\rm in} \over R_\odot }\right)^4
          \left( { 3~ M_\odot \over M_{\rm comp} }\right)^3.
\end{equation}
If the age of the system must be $>1~(10)$~Gyr to allow time for the red giant to 
evolve, then $a_{\rm in} > 2.6~R_\odot~(4.6~R_\odot)$. Unfortunately, a binary with such 
a long life time is too weak a gravitational wave source to be detected by the Laser Interferometer 
Space Antenna (LISA, \citealt{Robson2019}). As discussed in 
Appendix~\ref{app:tertiary}, dynamical stability requires $a_{\rm in} < 31 R_\odot$
(significantly inside the Roche lobe radius of the companion, ${\sim}49~R_\odot$), so
a long-lived, dynamically stable binary is possible for $4~ R_\odot \ltorder a_{\rm  in} \ltorder 31~R_\odot$.  This does not guarantee secular stability, and for much of this range of 
semi-major axes we would also expect to see dynamical perturbations of the outer orbit and additional tidal interactions (see Appendix \ref{app:tertiary}).

In summary, while it is difficult to rule out more complex scenarios where the companion to the giant is a binary consisting of at least one compact object, the simplest explanation is that the dark companion is a low-mass black hole. This would make V723 Mon a unique system, as it would contain both the lowest mass BH in a binary and be the closest known black hole yet discovered. The low X-ray luminosity ($L/L_{\rm edd}{\sim}10^{-9}$) of this system likely also suggests a black hole companion, as quiescent black holes are known to be X-ray faint (see $\S$\ref{section:xrays}).
  
\section{Conclusions}
\label{section:v723conc}

The nearby ($d\simeq460\,\rm pc$), bright ($V\simeq8.3$~mag), evolved red giant ($T_{\rm eff, giant}\simeq4440$~K, $L_{\rm giant}\simeq173~L_\odot$) V723 Mon is in a high mass function, $f(M)=1.72\pm 0.01~M_\odot$, nearly circular binary ($P=59.9$~d, $e\simeq 0$) with a dark companion of mass $M_{\rm comp}=3.04\pm 0.06~M_\odot$. V723 Mon is a known variable that had been typically classified as an eclipsing binary, but the ASAS, KELT and \textit{TESS} light curves indicate that is in fact a nearly edge-on ellipsoidal variable ($\S$\ref{section:phoebe}, Figure \ref{lcs}). We do not see any continuum eclipses in these light curves ($\S$\ref{section:limits}, Figure \ref{ecllc}). Using the binary mass function and $\sin i \simeq 1$ ($\S$\ref{section:phoebe}), it is clear that V723 Mon has a companion of mass $2.6 ~M_\odot < M_{\rm comp} < 3.6~M_\odot$ for $0.5~ M_\odot < M_{\rm giant} < 1.5~ M_\odot$. We modeled the light curves with \verb"PHOEBE" using constraints on the period, radial velocities and stellar temperature to derive an inclination of $i=87.0^\circ{}^{+1.7^\circ}_{-1.4^\circ} $, a mass ratio of $q\simeq0.33\pm0.02$, a companion mass of $M_{\rm comp}=3.04\pm 0.06~M_\odot$, a stellar radius of $R_{\rm giant}=24.9\pm0.7~R_\odot$, and a giant mass of $M_{\rm giant}=1.00\pm0.07~ M_\odot$ consistent with the earlier estimates ($\S$\ref{section:phoebe}, Table \ref{tab:params}). 

We also identify a significant blue veiling component, both through line veiling and second light contributions to the $B$ and $V$-band light curves ($\S$ \ref{section:veiling}, Figure \ref{veiledsedfit}). The veiling component contributes ${\sim}63\%$ and ${\sim}24\%$ of the total flux in the $B$-band and $V$-band respectively. The SED of the veiling component decays rapidly towards both bluer and redder wavelengths, strongly inconsistent with a stellar SED. Given that we do not see eclipses in the light curves, we infer that the veiling component has to be diffuse. 

We find no evidence for a luminous stellar companion and can rule out both single and binary main sequence companions based on the SED and limits on eclipses ($M_{\rm single}<0.80~M_\odot$, $M_{\rm binary}<1.51~M_\odot$) from the light curves ($\S$\ref{section:limits}). The SED and the absence of continuum eclipses imply that the companion mass must be dominated by a compact object even if it is a binary ($\S$\ref{section:limits}, $\S$\ref{section:v723disc}, Figure \ref{fig:llimit}). 

Once the spectrum of the red giant is subtracted, we also find evidence of Balmer $\rm H\alpha$ and $\rm H\beta$ emission ($\S$\ref{section:balmer}, Figure \ref{spec}). The morphology of the Balmer emission lines is complicated and its origin is unclear. Even though we observe eclipses of the Balmer lines when the dark companion passes behind the giant, the velocity scales seem too low to be associated with an accretion disk. We also detect this system in the X-rays with $L_{\rm X}\simeq7.6\times10^{29}~\rm ergs~s^{-1}$ ($L/L_{\rm edd}{\sim}10^{-9}$) using $Swift$ XRT data ($\S$\ref{section:xrays}). 

The simplest explanation for the dark companion is a single compact object, most likely a black hole, in the ``mass gap'', making V723 Mon the host to the closest black hole yet discovered ($\S$\ref{section:v723disc}, Table \ref{tab:scenarios}). Prior to this discovery, A0620-00 (V616 Mon) was the closest confirmed black hole at an estimated distance of ${\sim}1.6$~kpc \citep{Gelino2001}. However, more exotic scenarios can also be plausible explanations, including a neutron star--neutron star binary and a white dwarf--neutron star binary. 

To better understand this unique system, further comprehensive multi-wavelength observations are necessary. In particular, UV observations from the $Hubble$ space telescope will constrain the nature of the veiling component, and X-ray light curves will be useful to understand the nature of the compact object in this system. Future data releases from \textit{Gaia} will also confirm the orbital inclination. 

We can very crudely estimate the expected number of similar systems based on the fraction of the thin disk mass from which V723~Mon was selected. We assume a simple exponential disk model with density
\begin{equation}
    \rho=\rho_0~e^{-R/R_d - |z|/h},
\end{equation} where $R_{\rm d}\approx3$~kpc is the disk scale length \citep{Amores2017} and the numerical values of the disk scale height $h$ and density normalization $\rho_0$ are not needed. The total disk mass is
$4\pi \rho_0 h R_d^2$. If
we assume that V723~Mon was selected from a cylinder of radius $R$ at the Galactocentric radius of the
Sun, $R_{\rm \sun} \simeq 8$~kpc \citep{Gravity2019,Stanek1998}, the survey examined a mass of approximately 
$4\pi h R^2 \exp(-R_{\rm sun}/R_d)$, so the fraction of the disk mass surveyed is approximately
\begin{equation}
    \left( { R \over R_d}\right)^2 \exp(-R_{\rm sun}/R_d) \simeq 0.008
\end{equation}
if we assume $R \simeq 1$~kpc, as this encompasses most of the systems in the SB9 catalog.  The fraction drops to $0.002$ if we use the distance $R \simeq 0.5$~kpc to V723~Mon.  This implies that the Galaxy
might contain some $100$-$1000$ similar systems.  Since ellipsoidal variability is only possible for a limited range of semimajor axes, it is not surprising that this estimate is larger than the number of mass transfer systems but smaller than estimates of the total number ($10^3$ to $10^4$) of non-interacting black hole binaries in the Galaxy based on population synthesis models (see, for e.g., \citealt{Breivik2017,Shao2019}). 

A number of large spectroscopy projects such as APOGEE \citep{Majewski2017} and LAMOST \citep{Cui2012} are in the process of physically characterizing (kinematics, temperature, abundances, etc.) millions of Galactic stars. These surveys frequently obtain their spectra in multiple visits, providing sparse radial velocity (RV) curves for huge numbers of stars. A particularly important synergy is the ability to combine photometric surveys with these spectroscopic surveys to search for non-interacting compact object binaries like V723 Mon. For example, for the vast majority of these relatively bright stars, the ASAS-SN survey (\citealt{Shappee2014,Kochanek2017, Jayasinghe2018,Jayasinghe2020}) will supply all-sky, well-sampled light curves spanning multiple years. If we make a conservative assumption that ASAS-SN can characterize the variability of most giants up to ${\sim}3$~kpc away (not accounting for extinction), there maybe  ${\sim}20$ red giants with non-interacting companions that have ASAS-SN light curves. However, there is a significant cost to confirming these systems. In particular, a well sampled set of RV measurements is required to accurately measure the mass function and to constrain the properties of any companion. Nonetheless, as the spectroscopic surveys expand from a few $10^5$ to a few $10^6$ stars during the next 5 years, this approach will become a major probe of compact object binaries.

\begin{table}
	\centering
	\caption{Swift UVM2 observations}
	\label{tab:swift}
\begin{tabular}{rrrrr}
		\hline
		 JD & Date & Phase & UVM2 (mag) & $\sigma$ (mag)\\
		\hline
2459144.439 & 2020-10-21 & 0.975 & 14.10 & 0.04\\
2459150.607 & 2020-10-28 & 0.078 & 14.06 & 0.04\\
2459155.040 & 2020-11-01 & 0.152 & 14.07 & 0.04\\
2459158.708 & 2020-11-05 & 0.213 & 14.17 & 0.04\\
2459162.689 & 2020-11-09 & 0.279 & 14.12 & 0.04\\
2459166.865 & 2020-11-13 & 0.345 & 14.32 & 0.06\\
2459172.832 & 2020-11-19 & 0.449 & 14.10 & 0.04\\
2459173.769 & 2020-11-20 & 0.464 & 14.16 & 0.04\\
2459175.497 & 2020-11-22 & 0.493 & 14.12 & 0.04\\
2459175.753 & 2020-11-22 & 0.497 & 14.13 & 0.04\\
2459176.624 & 2020-11-23 & 0.512 & 14.13 & 0.04\\
2459184.188 & 2020-11-30 & 0.634 & 14.06 & 0.04\\
2459203.850 & 2020-12-20 & 0.966 & 14.04 & 0.04\\
2459209.830 & 2020-12-26 & 0.066 & 14.09 & 0.04\\
2459237.510 & 2021-01-23 & 0.528 & 14.07 & 0.04\\
2459240.102 & 2021-01-25 & 0.571 & 14.07 & 0.04\\
2459241.949 & 2021-01-27 & 0.602 & 14.07 & 0.04\\
2459244.948 & 2021-01-30 & 0.652 & 14.11 & 0.04\\
2459246.145 & 2021-01-31 & 0.672 & 14.03 & 0.04\\
2459249.855 & 2021-02-04 & 0.734 & 14.07 & 0.04\\
2459252.309 & 2021-02-06 & 0.774 & 14.03 & 0.04\\
2459254.172 & 2021-02-08 & 0.806 & 14.04 & 0.04\\
2459255.822 & 2021-02-10 & 0.833 & 14.06 & 0.04\\
2459259.617 & 2021-02-14 & 0.896 & 14.05 & 0.04\\
2459262.011 & 2021-02-16 & 0.936 & 14.03 & 0.04\\
\hline
\end{tabular}
\end{table}

\begin{table}
	\centering
	\caption{STELLA RV observations}
	\label{tab:stellarv}
\begin{tabular}{rrrr}
		\hline
		 BJD & Phase & RV ($\rm km s^{-1}$) & $\sigma_{RV}$ ($\rm km s^{-1}$)\\
		\hline
2454065.57926 & 0.242 & $-$63.359 & 0.243 \\
2454066.65035 & 0.260  & $-$63.407 & 0.199 \\
2454073.62965 & 0.376 & $-$47.100   & 0.667 \\
2454073.66318 & 0.377 & $-$43.862 & 0.305 \\
2454092.53351 & 0.692 & 61.584  & 0.199 \\
2454096.49062 & 0.758 & 66.656  & 0.104 \\
2454101.53952 & 0.842 & 58.195  & 0.166 \\
2454106.54614 & 0.926 & 34.825  & 0.288 \\
2454110.49315 & 0.991 & 8.326   & 0.633 \\
2454116.54293 & 0.092 & $-$34.096 & 0.191 \\
2454119.51924 & 0.142 & $-$48.571 & 0.782 \\
2454122.50426 & 0.192 & $-$58.969 & 0.199 \\
2454125.53651 & 0.242 & $-$63.511 & 0.229 \\
2454134.55945 & 0.393 & $-$40.037 & 0.199 \\
2454146.52391 & 0.592 & 35.495  & 0.177 \\
2454147.48687 & 0.609 & 41.043  & 0.246 \\
2454152.52001 & 0.693 & 61.471  & 0.142 \\
2454162.44243 & 0.858 & 54.887  & 0.153 \\
2454209.37594 & 0.641 & 50.398  & 0.692 \\
2454359.72466 & 0.149 & $-$50.362 & 0.294 \\
2454364.72854 & 0.233 & $-$63.249 & 0.217 \\
2454369.71403 & 0.316 & $-$57.753 & 0.264 \\
2454377.71952 & 0.450  & $-$19.751 & 0.245 \\
2454380.69831 & 0.499 & 0.271   & 0.21  \\
2454383.73052 & 0.550  & 20.050   & 0.215 \\
2454388.73313 & 0.633 & 48.436  & 0.228 \\
2454393.70367 & 0.716 & 64.911  & 0.191 \\
2454405.66540  & 0.916 & 38.520   & 0.221 \\
2454420.72099 & 0.167 & $-$54.873 & 0.242 \\
2454439.66728 & 0.483 & $-$7.008  & 0.146 \\
2454455.56730  & 0.748 & 67.228  & 0.144 \\
2454460.61450  & 0.833 & 59.969  & 0.138 \\
2454462.62496 & 0.866 & 52.822  & 0.163 \\
2454465.56605 & 0.915 & 38.728  & 0.214 \\
2454467.57168 & 0.949 & 26.208  & 0.212 \\
2454474.52033 & 0.065 & $-$23.485 & 0.115 \\
2454480.53118 & 0.165 & $-$54.127 & 0.16  \\
2454491.51022 & 0.348 & $-$51.196 & 0.239 \\
2454497.51601 & 0.448 & $-$20.500   & 0.175 \\
2454507.49658 & 0.615 & 42.842  & 0.177 \\
2454532.44126 & 0.031 & $-$10.068 & 0.107 \\
2454562.37656 & 0.530  & 12.045  & 0.183 \\
2454711.73134 & 0.022 & $-$5.892  & 0.171 \\
2454712.73270  & 0.039 & $-$13.559 & 0.17  \\
2454713.72543 & 0.055 & $-$20.512 & 0.196 \\
2454714.72726 & 0.072 & $-$26.952 & 0.15  \\
2454722.70809 & 0.205 & $-$60.739 & 0.196 \\
2454725.73812 & 0.256 & $-$63.261 & 0.225 \\
2454730.74308 & 0.339 & $-$53.547 & 0.235 \\
2454746.71606 & 0.606 & 40.325  & 0.148 \\
2454753.62038 & 0.721 & 65.033  & 0.113 \\
2454762.70971 & 0.873 & 51.111  & 0.195 \\
2454777.67078 & 0.122 & $-$43.374 & 0.194 \\
2454779.70948 & 0.156 & $-$52.195 & 0.205 \\
2454781.55314 & 0.187 & $-$58.178 & 0.45  \\
2454804.64456 & 0.572 & 28.742  & 0.251 \\
2454811.50714 & 0.687 & 60.808  & 0.145 \\
2454871.42332 & 0.686 & 60.469  & 0.176 \\
2454872.43280  & 0.703 & 63.085  & 0.185 \\
2454928.39897 & 0.637 & 49.463  & 0.154 \\
2455170.50320  & 0.676 & 58.817  & 0.164 \\
2455170.54417 & 0.677 & 58.970   & 0.18  \\
2455171.68616 & 0.696 & 62.342  & 0.166 \\
\hline
\end{tabular}
\end{table}
\addtocounter{table}{-1}
\begin{table}
	\centering
	\caption{STELLA RV observations, continued}
\begin{tabular}{rrrr}
		\hline
		 BJD & Phase & RV ($\rm km s^{-1}$) & $\sigma_{RV}$ ($\rm km s^{-1}$)\\
		\hline
2455172.48258 & 0.709 & 64.109  & 0.091 \\
2455172.66102 & 0.712 & 64.566  & 0.118 \\
2455204.57655 & 0.244 & $-$63.338 & 0.167 \\
2455208.54129 & 0.311 & $-$58.753 & 0.153 \\
2455209.59017 & 0.328 & $-$55.705 & 0.195 \\
2455210.52913 & 0.344 & $-$52.276 & 0.203 \\
2455211.48735 & 0.360  & $-$48.608 & 0.205 \\
2455212.58559 & 0.378 & $-$43.778 & 0.195 \\
2455213.43953 & 0.392 & $-$39.561 & 0.214 \\
2455214.46909 & 0.409 & $-$34.428 & 0.193 \\
2455215.50881 & 0.427 & $-$28.606 & 0.167 \\
2455216.46226 & 0.443 & $-$22.608 & 0.142 \\
2455218.48622 & 0.476 & $-$9.364  & 0.136 \\
2455220.51537 & 0.510  & 4.421   & 0.138 \\
2455222.51638 & 0.544 & 17.986  & 0.141 \\
2455224.54100   & 0.577 & 30.681  & 0.172 \\
2455231.52035 & 0.694 & 62.259  & 0.143 \\
2455238.50895 & 0.811 & 63.532  & 0.172 \\
2455239.46454 & 0.826 & 61.403  & 0.221 \\
2455240.50720  & 0.844 & 58.147  & 0.139 \\
2455266.45443 & 0.277 & $-$62.580  & 0.148 \\
2455285.40569 & 0.593 & 36.069  & 0.148 \\
2455290.39565 & 0.676 & 58.852  & 0.114 \\
2455297.37772 & 0.793 & 65.281  & 0.144\\
\hline
\end{tabular}
\end{table}

\begin{table*}
	\centering
	\caption{Spectroscopic observations from HIRES, MODS and PEPSI.}
	\label{tab:spec}
\begin{tabular}{rrrrrr}
		\hline
		 BJD [TDB] & Date & Phase & RV ($\rm km s^{-1}$) & $\sigma_{RV}$ ($\rm km s^{-1}$) & Instrument\\
		\hline
2459143.09532 & 2020-10-20 & 0.952 & 24.39 & 0.10 & HIRES \\
2459153.99899 & 2020-10-31 & 0.134 & $-$47.28 & 0.10 & HIRES \\
2459162.02289 & 2020-11-08 & 0.268 & $-$63.27 & 0.10 & HIRES \\
2459188.01112 & 2020-12-04 & 0.702 & 62.67 & 0.10 & HIRES \\
2459208.01826 & 2020-12-24 & 0.036 & $-$12.23 & 0.10 & HIRES \\
2459208.89679 & 2020-12-25 & 0.050 & $-$17.51 & 0.10 & HIRES \\
2459209.98247 & 2020-12-26 & 0.068 & $-$24.92 & 0.10 & HIRES \\
2459171.94982 & 2020-11-18 & 0.434 & --- & --- & MODS \\
2459173.86921 & 2020-11-20 & 0.466 & --- & --- & MODS \\
2459174.89034 & 2020-11-21 & 0.483 & --- & --- & MODS \\
2459175.77657 & 2020-11-22 & 0.498 & --- & --- & MODS \\
2459183.77896 & 2020-11-30 & 0.631 & 48.74 & 0.25 & PEPSI \\
\hline
\end{tabular}
\end{table*}

\section*{Acknowledgements}

We thank the referee, Dr. Jerome Orosz, for his useful comments on this manuscript.

We thank Dr. Jennifer Johnson, Dr. Marc Pinsonneault, Dr. Jim Fuller and Dr. Kento Masuda for useful discussions on this manuscript. We thank Dr. Jay Strader for a careful reading of this manuscript. We thank Dr. Rick Pogge for his help with obtaining the LBT/MODS spectra.

The ASAS-SN team at OSU is supported by the Gordon and Betty Moore
Foundation through grant GBMF5490 to the Ohio State University, and NSF grant AST-1908570.

TJ, KZS and CSK are supported by NSF grants AST-1814440 and 
AST-1908570. TAT is supported in part by NASA grant 80NSSC20K0531. TAT acknowledges previous support from Scialog Scholar grant 24216 from the Research Corporation, from which this effort germinated.  J.T.H. is supported by NASA award 80NSSC21K0136.
Support for JLP is provided in part by the
Ministry of Economy, Development, and Tourism's Millennium Science 
Initiative through grant IC120009, awarded to The Millennium Institute 
of Astrophysics, MAS. D.H. acknowledges support from the Alfred P. Sloan Foundation, the National Aeronautics and Space Administration (80NSSC18K1585, 80NSSC19K0379), and the National Science Foundation (AST-1717000). CB acknowledges support from the National Science Foundation grant AST-1909022.

Parts of this research were supported by the Australian Research Council Centre of Excellence for All Sky Astrophysics in 3 Dimensions (ASTRO 3D), through project number CE170100013.

The LBT is an international collaboration among institutions in the
United States, Italy and Germany. LBT Corporation partners are: The
University of Arizona on behalf of the Arizona Board of Regents;
Istituto Nazionale di Astrofisica, Italy; LBT Beteiligungsgesellschaft,
Germany, representing the Max-Planck Society, The Leibniz Institute for
Astrophysics Potsdam, and Heidelberg University; The Ohio State
University, representing OSU, University of Notre Dame, University of
Minnesota and University of Virginia.

STELLA and PEPSI were made possible by funding through the State of
Brandenburg (MWFK) and the German Federal Ministry of Education and
Research (BMBF) through their Verbundforschung grants 05AL2BA1/3 and
05A08BAC.

This research is based on observations made with the \textit{Neil Gehrels Swift Observatory}, obtained from the MAST data archive at the Space Telescope Science Institute, which is operated by the Association of Universities for Research in Astronomy, Inc., under NASA contract NAS 5-26555. This paper includes data collected with the \textit{TESS} mission, obtained from the MAST data archive at the Space Telescope Science Institute (STScI). Funding for the TESS mission is provided by the NASA Explorer Program. STScI is operated by the Association of Universities for Research in Astronomy, Inc., under NASA contract NAS 5-26555.

Some of the data presented herein were obtained at the W. M. Keck Observatory, which is operated as a scientific partnership among the California Institute of Technology, the University of California and the National Aeronautics and Space Administration. The Observatory was made possible by the generous financial support of the W. M. Keck Foundation.

The authors wish to recognize and acknowledge the very significant cultural role and reverence that the summit of Maunakea has always had within the indigenous Hawaiian community.  We are most fortunate to have the opportunity to conduct observations from this mountain.

We thank the ASAS and KELT projects for making their light curve data publicly available.
This research has made use of the VizieR catalogue access tool, CDS, Strasbourg, France. 
This research also made use of Astropy, a community-developed core Python package for 
Astronomy \citep{astropy:2013,astropy:2018}.

\section*{Data Availability}

The data underlying this article will be shared on reasonable request to the corresponding author.



\bibliographystyle{mnras}
\bibliography{refbh} 

\begin{thebibliography}{}
\makeatletter
\relax
\def\mn@urlcharsother{\let\do\@makeother \do\$\do\&\do\#\do\^\do\_\do\%\do\~}
\def\mn@doi{\begingroup\mn@urlcharsother \@ifnextchar [ {\mn@doi@}
  {\mn@doi@[]}}
\def\mn@doi@[#1]#2{\def\@tempa{#1}\ifx\@tempa\@empty \href
  {http://dx.doi.org/#2} {doi:#2}\else \href {http://dx.doi.org/#2} {#1}\fi
  \endgroup}
\def\mn@eprint#1#2{\mn@eprint@#1:#2::\@nil}
\def\mn@eprint@arXiv#1{\href {http://arxiv.org/abs/#1} {{\tt arXiv:#1}}}
\def\mn@eprint@dblp#1{\href {http://dblp.uni-trier.de/rec/bibtex/#1.xml}
  {dblp:#1}}
\def\mn@eprint@#1:#2:#3:#4\@nil{\def\@tempa {#1}\def\@tempb {#2}\def\@tempc
  {#3}\ifx \@tempc \@empty \let \@tempc \@tempb \let \@tempb \@tempa \fi \ifx
  \@tempb \@empty \def\@tempb {arXiv}\fi \@ifundefined
  {mn@eprint@\@tempb}{\@tempb:\@tempc}{\expandafter \expandafter \csname
  mn@eprint@\@tempb\endcsname \expandafter{\@tempc}}}

\bibitem[\protect\citeauthoryear{{Abbott} et~al.,}{{Abbott}
  et~al.}{2016}]{Abbott2016}
{Abbott} B.~P.,  et~al., 2016, \mn@doi [\prl] {10.1103/PhysRevLett.116.061102},
  \href {https://ui.adsabs.harvard.edu/abs/2016PhRvL.116f1102A} {116, 061102}

\bibitem[\protect\citeauthoryear{{Abbott} et~al.,}{{Abbott}
  et~al.}{2017}]{Abbott2017NS}
{Abbott} B.~P.,  et~al., 2017, \mn@doi [\prl] {10.1103/PhysRevLett.119.161101},
  \href {https://ui.adsabs.harvard.edu/abs/2017PhRvL.119p1101A} {119, 161101}

\bibitem[\protect\citeauthoryear{{Abbott} et~al.,}{{Abbott}
  et~al.}{2020}]{Abbott2020}
{Abbott} R.,  et~al., 2020, \mn@doi [\apjl] {10.3847/2041-8213/ab960f}, \href
  {https://ui.adsabs.harvard.edu/abs/2020ApJ...896L..44A} {896, L44}

\bibitem[\protect\citeauthoryear{{Abdul-Masih} et~al.,}{{Abdul-Masih}
  et~al.}{2020}]{Abdul-Masih2020}
{Abdul-Masih} M.,  et~al., 2020, \mn@doi [\nat] {10.1038/s41586-020-2216-x},
  \href {https://ui.adsabs.harvard.edu/abs/2020Natur.580E..11A} {580, E11}

\bibitem[\protect\citeauthoryear{{Adams} \& {Kochanek}}{{Adams} \&
  {Kochanek}}{2015}]{Adams2015}
{Adams} S.~M.,  {Kochanek} C.~S.,  2015, \mn@doi [\mnras]
  {10.1093/mnras/stv1409}, \href
  {https://ui.adsabs.harvard.edu/abs/2015MNRAS.452.2195A} {452, 2195}

\bibitem[\protect\citeauthoryear{{Alard}}{{Alard}}{2000}]{alard2000}
{Alard} C.,  2000, \mn@doi [\aaps] {10.1051/aas:2000214}, \href
  {https://ui.adsabs.harvard.edu/abs/2000A&AS..144..363A} {144, 363}

\bibitem[\protect\citeauthoryear{{Am{\^o}res}, {Robin}  \&
  {Reyl{\'e}}}{{Am{\^o}res} et~al.}{2017}]{Amores2017}
{Am{\^o}res} E.~B.,  {Robin} A.~C.,   {Reyl{\'e}} C.,  2017, \mn@doi [\aap]
  {10.1051/0004-6361/201628461}, \href
  {https://ui.adsabs.harvard.edu/abs/2017A&A...602A..67A} {602, A67}

\bibitem[\protect\citeauthoryear{{Antoniadis} et~al.,}{{Antoniadis}
  et~al.}{2013}]{2013Sci...340..448A}
{Antoniadis} J.,  et~al., 2013, \mn@doi [Science] {10.1126/science.1233232},
  \href {https://ui.adsabs.harvard.edu/abs/2013Sci...340..448A} {340, 448}

\bibitem[\protect\citeauthoryear{{Arras}, {Burkart}, {Quataert}  \&
  {Weinberg}}{{Arras} et~al.}{2012}]{Arras2012}
{Arras} P.,  {Burkart} J.,  {Quataert} E.,   {Weinberg} N.~N.,  2012, \mn@doi
  [\mnras] {10.1111/j.1365-2966.2012.20756.x}, \href
  {https://ui.adsabs.harvard.edu/abs/2012MNRAS.422.1761A} {422, 1761}

\bibitem[\protect\citeauthoryear{{Asai}, {Dotani}, {Hoshi}, {Tanaka},
  {Robinson}  \& {Terada}}{{Asai} et~al.}{1998}]{Asai1998}
{Asai} K.,  {Dotani} T.,  {Hoshi} R.,  {Tanaka} Y.,  {Robinson} C.~R.,
  {Terada} K.,  1998, \mn@doi [\pasj] {10.1093/pasj/50.6.611}, \href
  {https://ui.adsabs.harvard.edu/abs/1998PASJ...50..611A} {50, 611}

\bibitem[\protect\citeauthoryear{{Astropy Collaboration} et~al.,}{{Astropy
  Collaboration} et~al.}{2013}]{astropy:2013}
{Astropy Collaboration} et~al., 2013, \mn@doi [\aap]
  {10.1051/0004-6361/201322068}, \href
  {http://adsabs.harvard.edu/abs/2013A%26A...558A..33A} {558, A33}

\bibitem[\protect\citeauthoryear{{Astropy Collaboration} et~al.,}{{Astropy
  Collaboration} et~al.}{2018}]{astropy:2018}
{Astropy Collaboration} et~al., 2018, \mn@doi [aj] {10.3847/1538-3881/aabc4f},
  \href {https://ui.adsabs.harvard.edu/abs/2018AJ....156..123A} {156, 123}

\bibitem[\protect\citeauthoryear{{Bailer-Jones}, {Rybizki}, {Fouesneau},
  {Mantelet}  \& {Andrae}}{{Bailer-Jones} et~al.}{2018}]{BailerJones2018}
{Bailer-Jones} C.~A.~L.,  {Rybizki} J.,  {Fouesneau} M.,  {Mantelet} G.,
  {Andrae} R.,  2018, \mn@doi [\aj] {10.3847/1538-3881/aacb21}, \href
  {https://ui.adsabs.harvard.edu/abs/2018AJ....156...58B} {156, 58}

\bibitem[\protect\citeauthoryear{{Beech}}{{Beech}}{1985}]{Beech1985}
{Beech} M.,  1985, \mn@doi [\apss] {10.1007/BF00660911}, \href
  {https://ui.adsabs.harvard.edu/abs/1985Ap&SS.117...69B} {117, 69}

\bibitem[\protect\citeauthoryear{{Blanco-Cuaresma}}{{Blanco-Cuaresma}}{2019}]{Blanco-Cuaresma2019}
{Blanco-Cuaresma} S.,  2019, \mn@doi [\mnras] {10.1093/mnras/stz549}, \href
  {https://ui.adsabs.harvard.edu/abs/2019MNRAS.486.2075B} {486, 2075}

\bibitem[\protect\citeauthoryear{{Blanco-Cuaresma}, {Soubiran}, {Heiter}  \&
  {Jofr{\'e}}}{{Blanco-Cuaresma} et~al.}{2014}]{Blanco-Cuaresma2014}
{Blanco-Cuaresma} S.,  {Soubiran} C.,  {Heiter} U.,   {Jofr{\'e}} P.,  2014,
  \mn@doi [\aap] {10.1051/0004-6361/201423945}, \href
  {https://ui.adsabs.harvard.edu/abs/2014A&A...569A.111B} {569, A111}

\bibitem[\protect\citeauthoryear{{Bodensteiner} et~al.,}{{Bodensteiner}
  et~al.}{2020}]{Bodensteiner2020}
{Bodensteiner} J.,  et~al., 2020, \mn@doi [\aap] {10.1051/0004-6361/202038682},
  \href {https://ui.adsabs.harvard.edu/abs/2020A&A...641A..43B} {641, A43}

\bibitem[\protect\citeauthoryear{{Boothroyd} \& {Sackmann}}{{Boothroyd} \&
  {Sackmann}}{1988}]{Boothroyd1988}
{Boothroyd} A.~I.,  {Sackmann} I.~J.,  1988, \mn@doi [\apj] {10.1086/166322},
  \href {https://ui.adsabs.harvard.edu/abs/1988ApJ...328..641B} {328, 641}

\bibitem[\protect\citeauthoryear{{Breeveld} et~al.,}{{Breeveld}
  et~al.}{2010}]{breeveld10}
{Breeveld} A.~A.,  et~al., 2010, \mn@doi [\mnras]
  {10.1111/j.1365-2966.2010.16832.x}, \href
  {https://ui.adsabs.harvard.edu/abs/2010MNRAS.406.1687B} {406, 1687}

\bibitem[\protect\citeauthoryear{{Breivik}, {Chatterjee}  \&
  {Larson}}{{Breivik} et~al.}{2017}]{Breivik2017}
{Breivik} K.,  {Chatterjee} S.,   {Larson} S.~L.,  2017, \mn@doi [\apjl]
  {10.3847/2041-8213/aa97d5}, \href
  {https://ui.adsabs.harvard.edu/abs/2017ApJ...850L..13B} {850, L13}

\bibitem[\protect\citeauthoryear{{Bressan}, {Marigo}, {Girardi}, {Salasnich},
  {Dal Cero}, {Rubele}  \& {Nanni}}{{Bressan} et~al.}{2012}]{Bressan2012}
{Bressan} A.,  {Marigo} P.,  {Girardi} L.,  {Salasnich} B.,  {Dal Cero} C.,
  {Rubele} S.,   {Nanni} A.,  2012, \mn@doi [\mnras]
  {10.1111/j.1365-2966.2012.21948.x}, \href
  {https://ui.adsabs.harvard.edu/abs/2012MNRAS.427..127B} {427, 127}

\bibitem[\protect\citeauthoryear{{Brown}, {Gilliland}, {Noyes}  \&
  {Ramsey}}{{Brown} et~al.}{1991}]{Brown1991}
{Brown} T.~M.,  {Gilliland} R.~L.,  {Noyes} R.~W.,   {Ramsey} L.~W.,  1991,
  \mn@doi [\apj] {10.1086/169725}, \href
  {https://ui.adsabs.harvard.edu/abs/1991ApJ...368..599B} {368, 599}

\bibitem[\protect\citeauthoryear{{Burrows} et~al.,}{{Burrows}
  et~al.}{2005}]{Burrows2005}
{Burrows} D.~N.,  et~al., 2005, \mn@doi [\ssr] {10.1007/s11214-005-5097-2},
  \href {https://ui.adsabs.harvard.edu/abs/2005SSRv..120..165B} {120, 165}

\bibitem[\protect\citeauthoryear{{Cardelli}, {Clayton}  \& {Mathis}}{{Cardelli}
  et~al.}{1989}]{Cardelli1989}
{Cardelli} J.~A.,  {Clayton} G.~C.,   {Mathis} J.~S.,  1989, \mn@doi [\apj]
  {10.1086/167900}, \href
  {https://ui.adsabs.harvard.edu/abs/1989ApJ...345..245C} {345, 245}

\bibitem[\protect\citeauthoryear{{Casares}, {Charles}, {Naylor}  \&
  {Pavlenko}}{{Casares} et~al.}{1993}]{Casares1993}
{Casares} J.,  {Charles} P.~A.,  {Naylor} T.,   {Pavlenko} E.~P.,  1993,
  \mn@doi [\mnras] {10.1093/mnras/265.4.834}, \href
  {https://ui.adsabs.harvard.edu/abs/1993MNRAS.265..834C} {265, 834}

\bibitem[\protect\citeauthoryear{{Castelli} \& {Kurucz}}{{Castelli} \&
  {Kurucz}}{2003}]{Castelli2003}
{Castelli} F.,  {Kurucz} R.~L.,  2003, in {Piskunov} N.,  {Weiss} W.~W.,
  {Gray} D.~F.,  eds,  Vol. 210, Modelling of Stellar Atmospheres. p.~A20
  (\mn@eprint {arXiv} {astro-ph/0405087})

\bibitem[\protect\citeauthoryear{{Champion} et~al.,}{{Champion}
  et~al.}{2008}]{Champion2008}
{Champion} D.~J.,  et~al., 2008, \mn@doi [Science] {10.1126/science.1157580},
  \href {https://ui.adsabs.harvard.edu/abs/2008Sci...320.1309C} {320, 1309}

\bibitem[\protect\citeauthoryear{{Choi}, {Dotter}, {Conroy}, {Cantiello},
  {Paxton}  \& {Johnson}}{{Choi} et~al.}{2016}]{Choi2016}
{Choi} J.,  {Dotter} A.,  {Conroy} C.,  {Cantiello} M.,  {Paxton} B.,
  {Johnson} B.~D.,  2016, \mn@doi [\apj] {10.3847/0004-637X/823/2/102}, \href
  {https://ui.adsabs.harvard.edu/abs/2016ApJ...823..102C} {823, 102}

\bibitem[\protect\citeauthoryear{{Claret} \& {Bloemen}}{{Claret} \&
  {Bloemen}}{2011}]{Claret2011}
{Claret} A.,  {Bloemen} S.,  2011, \mn@doi [\aap]
  {10.1051/0004-6361/201116451}, \href
  {https://ui.adsabs.harvard.edu/abs/2011A&A...529A..75C} {529, A75}

\bibitem[\protect\citeauthoryear{{Conroy} et~al.,}{{Conroy}
  et~al.}{2020}]{Conroy2020}
{Conroy} K.~E.,  et~al., 2020, \mn@doi [\apjs] {10.3847/1538-4365/abb4e2},
  \href {https://ui.adsabs.harvard.edu/abs/2020ApJS..250...34C} {250, 34}

\bibitem[\protect\citeauthoryear{{Cromartie} et~al.,}{{Cromartie}
  et~al.}{2020}]{Cromartie2020}
{Cromartie} H.~T.,  et~al., 2020, \mn@doi [Nature Astronomy]
  {10.1038/s41550-019-0880-2}, \href
  {https://ui.adsabs.harvard.edu/abs/2020NatAs...4...72C} {4, 72}

\bibitem[\protect\citeauthoryear{{Cui} et~al.,}{{Cui} et~al.}{2012}]{Cui2012}
{Cui} X.-Q.,  et~al., 2012, \mn@doi [Research in Astronomy and Astrophysics]
  {10.1088/1674-4527/12/9/003}, \href
  {https://ui.adsabs.harvard.edu/abs/2012RAA....12.1197C} {12, 1197}

\bibitem[\protect\citeauthoryear{{Demircan}}{{Demircan}}{1987}]{Demircan1987}
{Demircan} O.,  1987, \mn@doi [\apss] {10.1007/BF00641634}, \href
  {https://ui.adsabs.harvard.edu/abs/1987Ap&SS.137..195D} {137, 195}

\bibitem[\protect\citeauthoryear{{Din{\c{c}}er}, {Bailyn}, {Miller-Jones},
  {Buxton}  \& {MacDonald}}{{Din{\c{c}}er} et~al.}{2018}]{Dincer2018}
{Din{\c{c}}er} T.,  {Bailyn} C.~D.,  {Miller-Jones} J. C.~A.,  {Buxton} M.,
  {MacDonald} R. K.~D.,  2018, \mn@doi [\apj] {10.3847/1538-4357/aa9a46}, \href
  {https://ui.adsabs.harvard.edu/abs/2018ApJ...852....4D} {852, 4}

\bibitem[\protect\citeauthoryear{{Dotter}}{{Dotter}}{2016}]{Dotter2016}
{Dotter} A.,  2016, \mn@doi [\apjs] {10.3847/0067-0049/222/1/8}, \href
  {https://ui.adsabs.harvard.edu/abs/2016ApJS..222....8D} {222, 8}

\bibitem[\protect\citeauthoryear{{ESA}}{{ESA}}{1997}]{ESA1997}
{ESA} .,  1997, VizieR Online Data Catalog, \href
  {https://ui.adsabs.harvard.edu/abs/1997yCat.1239....0E} {p. I/239}

\bibitem[\protect\citeauthoryear{{Eastman}, {Siverd}  \& {Gaudi}}{{Eastman}
  et~al.}{2010}]{Eastman2010}
{Eastman} J.,  {Siverd} R.,   {Gaudi} B.~S.,  2010, \mn@doi [\pasp]
  {10.1086/655938}, \href
  {https://ui.adsabs.harvard.edu/abs/2010PASP..122..935E} {122, 935}

\bibitem[\protect\citeauthoryear{{Eaton}}{{Eaton}}{2008}]{Eaton2008}
{Eaton} J.~A.,  2008, \mn@doi [\apj] {10.1086/588270}, \href
  {https://ui.adsabs.harvard.edu/abs/2008ApJ...681..562E} {681, 562}

\bibitem[\protect\citeauthoryear{{El-Badry} \& {Quataert}}{{El-Badry} \&
  {Quataert}}{2020a}]{El-Badry2020b}
{El-Badry} K.,  {Quataert} E.,  2020a, arXiv e-prints, \href
  {https://ui.adsabs.harvard.edu/abs/2020arXiv200611974E} {p. arXiv:2006.11974}

\bibitem[\protect\citeauthoryear{{El-Badry} \& {Quataert}}{{El-Badry} \&
  {Quataert}}{2020b}]{El-Badry2020a}
{El-Badry} K.,  {Quataert} E.,  2020b, \mn@doi [\mnras]
  {10.1093/mnrasl/slaa004}, \href
  {https://ui.adsabs.harvard.edu/abs/2020MNRAS.493L..22E} {493, L22}

\bibitem[\protect\citeauthoryear{{Elitzur} \& {Ivezi{\'c}}}{{Elitzur} \&
  {Ivezi{\'c}}}{2001}]{Elitzur2001}
{Elitzur} M.,  {Ivezi{\'c}} {\v{Z}}.,  2001, \mn@doi [\mnras]
  {10.1046/j.1365-8711.2001.04706.x}, \href
  {https://ui.adsabs.harvard.edu/abs/2001MNRAS.327..403E} {327, 403}

\bibitem[\protect\citeauthoryear{{Esin}, {McClintock}  \& {Narayan}}{{Esin}
  et~al.}{1997}]{Esin1997}
{Esin} A.~A.,  {McClintock} J.~E.,   {Narayan} R.,  1997, \mn@doi [\apj]
  {10.1086/304829}, \href
  {https://ui.adsabs.harvard.edu/abs/1997ApJ...489..865E} {489, 865}

\bibitem[\protect\citeauthoryear{{Farr}, {Sravan}, {Cantrell}, {Kreidberg},
  {Bailyn}, {Mandel}  \& {Kalogera}}{{Farr} et~al.}{2011}]{Farr2011}
{Farr} W.~M.,  {Sravan} N.,  {Cantrell} A.,  {Kreidberg} L.,  {Bailyn} C.~D.,
  {Mandel} I.,   {Kalogera} V.,  2011, \mn@doi [\apj]
  {10.1088/0004-637X/741/2/103}, \href
  {https://ui.adsabs.harvard.edu/abs/2011ApJ...741..103F} {741, 103}

\bibitem[\protect\citeauthoryear{{Foreman-Mackey}, {Hogg}, {Lang}  \&
  {Goodman}}{{Foreman-Mackey} et~al.}{2013}]{Foreman-Mackey2013}
{Foreman-Mackey} D.,  {Hogg} D.~W.,  {Lang} D.,   {Goodman} J.,  2013, \mn@doi
  [\pasp] {10.1086/670067}, \href
  {https://ui.adsabs.harvard.edu/abs/2013PASP..125..306F} {125, 306}

\bibitem[\protect\citeauthoryear{{Gabriel} et~al.,}{{Gabriel}
  et~al.}{2004}]{gabriel04}
{Gabriel} C.,  et~al., 2004, {The XMM-Newton SAS - Distributed Development and
  Maintenance of a Large Science Analysis System: A Critical Analysis}.
p.~759

\bibitem[\protect\citeauthoryear{{Gagn{\'e}} et~al.,}{{Gagn{\'e}}
  et~al.}{2018}]{Gagne2018}
{Gagn{\'e}} J.,  et~al., 2018, \mn@doi [\apj] {10.3847/1538-4357/aaae09}, \href
  {https://ui.adsabs.harvard.edu/abs/2018ApJ...856...23G} {856, 23}

\bibitem[\protect\citeauthoryear{{Gahm}, {Walter}, {Stempels}, {Petrov}  \&
  {Herczeg}}{{Gahm} et~al.}{2008}]{Gahm2008}
{Gahm} G.~F.,  {Walter} F.~M.,  {Stempels} H.~C.,  {Petrov} P.~P.,   {Herczeg}
  G.~J.,  2008, \mn@doi [\aap] {10.1051/0004-6361:200809488}, \href
  {https://ui.adsabs.harvard.edu/abs/2008A&A...482L..35G} {482, L35}

\bibitem[\protect\citeauthoryear{{Gaia Collaboration} et~al.,}{{Gaia
  Collaboration} et~al.}{2018}]{GaiaMAIN}
{Gaia Collaboration} et~al., 2018, \mn@doi [\aap]
  {10.1051/0004-6361/201833051}, \href
  {https://ui.adsabs.harvard.edu/abs/2018A&A...616A...1G} {616, A1}

\bibitem[\protect\citeauthoryear{{Gaia Collaboration}, {Brown}, {Vallenari},
  {Prusti}, {de Bruijne}, {Babusiaux}  \& {Biermann}}{{Gaia Collaboration}
  et~al.}{2020}]{GaiaEDR3MAIN}
{Gaia Collaboration} {Brown} A.~G.~A.,  {Vallenari} A.,  {Prusti} T.,  {de
  Bruijne} J.~H.~J.,  {Babusiaux} C.,   {Biermann} M.,  2020, arXiv e-prints,
  \href {https://ui.adsabs.harvard.edu/abs/2020arXiv201201533G} {p.
  arXiv:2012.01533}

\bibitem[\protect\citeauthoryear{{Gelino}, {Harrison}  \& {Orosz}}{{Gelino}
  et~al.}{2001}]{Gelino2001}
{Gelino} D.~M.,  {Harrison} T.~E.,   {Orosz} J.~A.,  2001, \mn@doi [\aj]
  {10.1086/323714}, \href
  {https://ui.adsabs.harvard.edu/abs/2001AJ....122.2668G} {122, 2668}

\bibitem[\protect\citeauthoryear{{Giesers} et~al.,}{{Giesers}
  et~al.}{2018}]{Giesers2018}
{Giesers} B.,  et~al., 2018, \mn@doi [\mnras] {10.1093/mnrasl/slx203}, \href
  {https://ui.adsabs.harvard.edu/abs/2018MNRAS.475L..15G} {475, L15}

\bibitem[\protect\citeauthoryear{{Giesers} et~al.,}{{Giesers}
  et~al.}{2019}]{Giesers2019}
{Giesers} B.,  et~al., 2019, \mn@doi [\aap] {10.1051/0004-6361/201936203},
  \href {https://ui.adsabs.harvard.edu/abs/2019A&A...632A...3G} {632, A3}

\bibitem[\protect\citeauthoryear{{Gomel}, {Faigler}  \& {Mazeh}}{{Gomel}
  et~al.}{2020}]{Gomel2020}
{Gomel} R.,  {Faigler} S.,   {Mazeh} T.,  2020, arXiv e-prints, \href
  {https://ui.adsabs.harvard.edu/abs/2020arXiv200811209G} {p. arXiv:2008.11209}

\bibitem[\protect\citeauthoryear{{Gondoin}}{{Gondoin}}{2007}]{Gondoin2007}
{Gondoin} P.,  2007, \mn@doi [\aap] {10.1051/0004-6361:20066751}, \href
  {https://ui.adsabs.harvard.edu/abs/2007A&A...464.1101G} {464, 1101}

\bibitem[\protect\citeauthoryear{{Gontcharov} \& {Mosenkov}}{{Gontcharov} \&
  {Mosenkov}}{2017}]{Gontcharov2017}
{Gontcharov} G.~A.,  {Mosenkov} A.~V.,  2017, \mn@doi [\mnras]
  {10.1093/mnras/stx2219}, \href
  {https://ui.adsabs.harvard.edu/abs/2017MNRAS.472.3805G} {472, 3805}

\bibitem[\protect\citeauthoryear{{Gonz{\'a}lez Hern{\'a}ndez} \&
  {Casares}}{{Gonz{\'a}lez Hern{\'a}ndez} \& {Casares}}{2010}]{GH2010}
{Gonz{\'a}lez Hern{\'a}ndez} J.~I.,  {Casares} J.,  2010, \mn@doi [\aap]
  {10.1051/0004-6361/201014088}, \href
  {https://ui.adsabs.harvard.edu/abs/2010A&A...516A..58G} {516, A58}

\bibitem[\protect\citeauthoryear{{Gr{\"a}fener}, {Owocki}  \&
  {Vink}}{{Gr{\"a}fener} et~al.}{2012}]{Grafner2012}
{Gr{\"a}fener} G.,  {Owocki} S.~P.,   {Vink} J.~S.,  2012, \mn@doi [\aap]
  {10.1051/0004-6361/201117497}, \href
  {https://ui.adsabs.harvard.edu/abs/2012A&A...538A..40G} {538, A40}

\bibitem[\protect\citeauthoryear{{Gravity Collaboration} et~al.,}{{Gravity
  Collaboration} et~al.}{2019}]{Gravity2019}
{Gravity Collaboration} et~al., 2019, \mn@doi [\aap]
  {10.1051/0004-6361/201935656}, \href
  {https://ui.adsabs.harvard.edu/abs/2019A&A...625L..10G} {625, L10}

\bibitem[\protect\citeauthoryear{{Gray} \& {Corbally}}{{Gray} \&
  {Corbally}}{1994}]{Gray1994}
{Gray} R.~O.,  {Corbally} C.~J.,  1994, \mn@doi [\aj] {10.1086/116893}, \href
  {https://ui.adsabs.harvard.edu/abs/1994AJ....107..742G} {107, 742}

\bibitem[\protect\citeauthoryear{{Green}, {Schlafly}, {Zucker}, {Speagle}  \&
  {Finkbeiner}}{{Green} et~al.}{2019}]{Green2019}
{Green} G.~M.,  {Schlafly} E.,  {Zucker} C.,  {Speagle} J.~S.,   {Finkbeiner}
  D.,  2019, \mn@doi [\apj] {10.3847/1538-4357/ab5362}, \href
  {https://ui.adsabs.harvard.edu/abs/2019ApJ...887...93G} {887, 93}

\bibitem[\protect\citeauthoryear{{Griffin}}{{Griffin}}{2010}]{Griffin2010}
{Griffin} R.~F.,  2010, The Observatory, \href
  {https://ui.adsabs.harvard.edu/abs/2010Obs...130...60G} {130, 60}

\bibitem[\protect\citeauthoryear{{Griffin}}{{Griffin}}{2014}]{Griffin2014}
{Griffin} R.~F.,  2014, The Observatory, \href
  {https://ui.adsabs.harvard.edu/abs/2014Obs...134..109G} {134, 109}

\bibitem[\protect\citeauthoryear{{Gustafsson}, {Edvardsson}, {Eriksson},
  {J{\o}rgensen}, {Nordlund}  \& {Plez}}{{Gustafsson}
  et~al.}{2008}]{Gustafsson2008}
{Gustafsson} B.,  {Edvardsson} B.,  {Eriksson} K.,  {J{\o}rgensen} U.~G.,
  {Nordlund} {\r{A}}.,   {Plez} B.,  2008, \mn@doi [\aap]
  {10.1051/0004-6361:200809724}, \href
  {https://ui.adsabs.harvard.edu/abs/2008A&A...486..951G} {486, 951}

\bibitem[\protect\citeauthoryear{{Hall}}{{Hall}}{1996}]{Hall1996}
{Hall} J.~C.,  1996, \mn@doi [\pasp] {10.1086/133724}, \href
  {https://ui.adsabs.harvard.edu/abs/1996PASP..108..313H} {108, 313}

\bibitem[\protect\citeauthoryear{{Hambsch}}{{Hambsch}}{2012}]{Hambsch2012}
{Hambsch} F.~J.,  2012, Journal of the American Association of Variable Star
  Observers (JAAVSO), \href
  {https://ui.adsabs.harvard.edu/abs/2012JAVSO..40.1003H} {40, 1003}

\bibitem[\protect\citeauthoryear{{Henden}, {Levine}, {Terrell}, {Welch},
  {Munari}  \& {Kloppenborg}}{{Henden} et~al.}{2018}]{Henden2018}
{Henden} A.~A.,  {Levine} S.,  {Terrell} D.,  {Welch} D.~L.,  {Munari} U.,
  {Kloppenborg} B.~K.,  2018, in American Astronomical Society Meeting
  Abstracts \#232. p. 223.06

\bibitem[\protect\citeauthoryear{{Hill}, {Fisher}  \& {Holmgren}}{{Hill}
  et~al.}{1989}]{Hill1989}
{Hill} G.,  {Fisher} W.~A.,   {Holmgren} D.,  1989, \aap, \href
  {https://ui.adsabs.harvard.edu/abs/1989A&A...218..152H} {218, 152}

\bibitem[\protect\citeauthoryear{{Horvat}, {Conroy}, {Pablo}, {Hambleton},
  {Kochoska}, {Giammarco}  \& {Pr{\v{s}}a}}{{Horvat} et~al.}{2018}]{Horvat2018}
{Horvat} M.,  {Conroy} K.~E.,  {Pablo} H.,  {Hambleton} K.~M.,  {Kochoska} A.,
  {Giammarco} J.,   {Pr{\v{s}}a} A.,  2018, \mn@doi [\apjs]
  {10.3847/1538-4365/aacd0f}, \href
  {https://ui.adsabs.harvard.edu/abs/2018ApJS..237...26H} {237, 26}

\bibitem[\protect\citeauthoryear{{Houk} \& {Swift}}{{Houk} \&
  {Swift}}{2000}]{Houk2000}
{Houk} N.,  {Swift} C.,  2000, VizieR Online Data Catalog, \href
  {https://ui.adsabs.harvard.edu/abs/2000yCat.3214....0H} {p. III/214}

\bibitem[\protect\citeauthoryear{{Howard} et~al.,}{{Howard}
  et~al.}{2010}]{Howard2010}
{Howard} A.~W.,  et~al., 2010, \mn@doi [\apj] {10.1088/0004-637X/721/2/1467},
  \href {https://ui.adsabs.harvard.edu/abs/2010ApJ...721.1467H} {721, 1467}

\bibitem[\protect\citeauthoryear{{Hynes}, {Robinson}  \& {Bitner}}{{Hynes}
  et~al.}{2005}]{Hynes2005}
{Hynes} R.~I.,  {Robinson} E.~L.,   {Bitner} M.,  2005, \mn@doi [\apj]
  {10.1086/431966}, \href
  {https://ui.adsabs.harvard.edu/abs/2005ApJ...630..405H} {630, 405}

\bibitem[\protect\citeauthoryear{{Irrgang}, {Geier}, {Kreuzer}, {Pelisoli}  \&
  {Heber}}{{Irrgang} et~al.}{2020}]{Irrgang2020}
{Irrgang} A.,  {Geier} S.,  {Kreuzer} S.,  {Pelisoli} I.,   {Heber} U.,  2020,
  \mn@doi [\aap] {10.1051/0004-6361/201937343}, \href
  {https://ui.adsabs.harvard.edu/abs/2020A&A...633L...5I} {633, L5}

\bibitem[\protect\citeauthoryear{{Ivezic} \& {Elitzur}}{{Ivezic} \&
  {Elitzur}}{1997}]{Ivezic1997}
{Ivezic} Z.,  {Elitzur} M.,  1997, \mn@doi [\mnras] {10.1093/mnras/287.4.799},
  \href {https://ui.adsabs.harvard.edu/abs/1997MNRAS.287..799I} {287, 799}

\bibitem[\protect\citeauthoryear{{Jansen} et~al.,}{{Jansen}
  et~al.}{2001}]{Jansen2001}
{Jansen} F.,  et~al., 2001, \mn@doi [\aap] {10.1051/0004-6361:20000036}, \href
  {https://ui.adsabs.harvard.edu/abs/2001A&A...365L...1J} {365, L1}

\bibitem[\protect\citeauthoryear{{Jayasinghe} et~al.,}{{Jayasinghe}
  et~al.}{2018}]{Jayasinghe2018}
{Jayasinghe} T.,  et~al., 2018, \mn@doi [\mnras] {10.1093/mnras/sty838}, \href
  {https://ui.adsabs.harvard.edu/abs/2018MNRAS.477.3145J} {477, 3145}

\bibitem[\protect\citeauthoryear{{Jayasinghe} et~al.,}{{Jayasinghe}
  et~al.}{2020}]{Jayasinghe2020}
{Jayasinghe} T.,  et~al., 2020, arXiv e-prints, \href
  {https://ui.adsabs.harvard.edu/abs/2020arXiv200610057J} {p. arXiv:2006.10057}

\bibitem[\protect\citeauthoryear{{Kanodia} \& {Wright}}{{Kanodia} \&
  {Wright}}{2018}]{Kanodia2018}
{Kanodia} S.,  {Wright} J.,  2018, \mn@doi [Research Notes of the American
  Astronomical Society] {10.3847/2515-5172/aaa4b7}, \href
  {https://ui.adsabs.harvard.edu/abs/2018RNAAS...2....4K} {2, 4}

\bibitem[\protect\citeauthoryear{{Kazarovets}, {Samus}, {Durlevich}, {Frolov},
  {Antipin}, {Kireeva}  \& {Pastukhova}}{{Kazarovets}
  et~al.}{1999}]{Kazarovets1999}
{Kazarovets} E.~V.,  {Samus} N.~N.,  {Durlevich} O.~V.,  {Frolov} M.~S.,
  {Antipin} S.~V.,  {Kireeva} N.~N.,   {Pastukhova} E.~N.,  1999, Information
  Bulletin on Variable Stars, \href
  {https://ui.adsabs.harvard.edu/abs/1999IBVS.4659....1K} {4659, 1}

\bibitem[\protect\citeauthoryear{{Kjeldsen} \& {Bedding}}{{Kjeldsen} \&
  {Bedding}}{1995}]{Kjeldsen1995}
{Kjeldsen} H.,  {Bedding} T.~R.,  1995, \aap, \href
  {https://ui.adsabs.harvard.edu/abs/1995A&A...293...87K} {293, 87}

\bibitem[\protect\citeauthoryear{{Kochanek} et~al.,}{{Kochanek}
  et~al.}{2017}]{Kochanek2017}
{Kochanek} C.~S.,  et~al., 2017, \mn@doi [\pasp] {10.1088/1538-3873/aa80d9},
  \href {https://ui.adsabs.harvard.edu/abs/2017PASP..129j4502K} {129, 104502}

\bibitem[\protect\citeauthoryear{{Kong}, {McClintock}, {Garcia}, {Murray}  \&
  {Barret}}{{Kong} et~al.}{2002}]{Kong2002}
{Kong} A. K.~H.,  {McClintock} J.~E.,  {Garcia} M.~R.,  {Murray} S.~S.,
  {Barret} D.,  2002, \mn@doi [\apj] {10.1086/339501}, \href
  {https://ui.adsabs.harvard.edu/abs/2002ApJ...570..277K} {570, 277}

\bibitem[\protect\citeauthoryear{{Kurucz}}{{Kurucz}}{1993}]{Kurucz1993}
{Kurucz} R.~L.,  1993, {SYNTHE spectrum synthesis programs and line data}

\bibitem[\protect\citeauthoryear{Lagarias, Reeds, Wright  \& Wright}{Lagarias
  et~al.}{1998}]{Lagarias1998}
Lagarias J.,  Reeds J.,  Wright M.,   Wright P.,  1998, \mn@doi [SIAM Journal
  on Optimization] {10.1137/S1052623496303470}, 9, 112

\bibitem[\protect\citeauthoryear{{Lattimer} \& {Prakash}}{{Lattimer} \&
  {Prakash}}{2001}]{Lattimer2001}
{Lattimer} J.~M.,  {Prakash} M.,  2001, \mn@doi [\apj] {10.1086/319702}, \href
  {https://ui.adsabs.harvard.edu/abs/2001ApJ...550..426L} {550, 426}

\bibitem[\protect\citeauthoryear{{Lenz} \& {Breger}}{{Lenz} \&
  {Breger}}{2005}]{Lenz2005}
{Lenz} P.,  {Breger} M.,  2005, \mn@doi [Communications in Asteroseismology]
  {10.1553/cia146s53}, \href
  {https://ui.adsabs.harvard.edu/abs/2005CoAst.146...53L} {146, 53}

\bibitem[\protect\citeauthoryear{{Liszt}}{{Liszt}}{2014}]{Liszt2014}
{Liszt} H.,  2014, \mn@doi [\apj] {10.1088/0004-637X/780/1/10}, \href
  {https://ui.adsabs.harvard.edu/abs/2014ApJ...780...10L} {780, 10}

\bibitem[\protect\citeauthoryear{{Liu}, {van Paradijs}  \& {van den
  Heuvel}}{{Liu} et~al.}{2006}]{Liu2006}
{Liu} Q.~Z.,  {van Paradijs} J.,   {van den Heuvel} E.~P.~J.,  2006, \mn@doi
  [\aap] {10.1051/0004-6361:20064987}, \href
  {https://ui.adsabs.harvard.edu/abs/2006A&A...455.1165L} {455, 1165}

\bibitem[\protect\citeauthoryear{{Liu} et~al.,}{{Liu} et~al.}{2019}]{Liu2019}
{Liu} J.,  et~al., 2019, \mn@doi [\nat] {10.1038/s41586-019-1766-2}, \href
  {https://ui.adsabs.harvard.edu/abs/2019Natur.575..618L} {575, 618}

\bibitem[\protect\citeauthoryear{{L{\'o}pez-Santiago}, {Montes},
  {G{\'a}lvez-Ortiz}, {Crespo-Chac{\'o}n}, {Mart{\'\i}nez-Arn{\'a}iz},
  {Fern{\'a}ndez-Figueroa}, {de Castro}  \& {Cornide}}{{L{\'o}pez-Santiago}
  et~al.}{2010}]{LS2010}
{L{\'o}pez-Santiago} J.,  {Montes} D.,  {G{\'a}lvez-Ortiz} M.~C.,
  {Crespo-Chac{\'o}n} I.,  {Mart{\'\i}nez-Arn{\'a}iz} R.~M.,
  {Fern{\'a}ndez-Figueroa} M.~J.,  {de Castro} E.,   {Cornide} M.,  2010,
  \mn@doi [\aap] {10.1051/0004-6361/200913437}, \href
  {https://ui.adsabs.harvard.edu/abs/2010A&A...514A..97L} {514, A97}

\bibitem[\protect\citeauthoryear{{Majewski} et~al.,}{{Majewski}
  et~al.}{2017}]{Majewski2017}
{Majewski} S.~R.,  et~al., 2017, \mn@doi [\aj] {10.3847/1538-3881/aa784d},
  \href {https://ui.adsabs.harvard.edu/abs/2017AJ....154...94M} {154, 94}

\bibitem[\protect\citeauthoryear{{Mardling} \& {Aarseth}}{{Mardling} \&
  {Aarseth}}{2001}]{Mardling2001}
{Mardling} R.~A.,  {Aarseth} S.~J.,  2001, \mn@doi [\mnras]
  {10.1046/j.1365-8711.2001.03974.x}, \href
  {https://ui.adsabs.harvard.edu/abs/2001MNRAS.321..398M} {321, 398}

\bibitem[\protect\citeauthoryear{{McDonald} \& {Zijlstra}}{{McDonald} \&
  {Zijlstra}}{2015}]{McDonald2015}
{McDonald} I.,  {Zijlstra} A.~A.,  2015, \mn@doi [\mnras]
  {10.1093/mnras/stv007}, \href
  {https://ui.adsabs.harvard.edu/abs/2015MNRAS.448..502M} {448, 502}

\bibitem[\protect\citeauthoryear{{Menou}, {Esin}, {Narayan}, {Garcia}, {Lasota}
   \& {McClintock}}{{Menou} et~al.}{1999}]{Menou1999}
{Menou} K.,  {Esin} A.~A.,  {Narayan} R.,  {Garcia} M.~R.,  {Lasota} J.-P.,
  {McClintock} J.~E.,  1999, \mn@doi [\apj] {10.1086/307443}, \href
  {https://ui.adsabs.harvard.edu/abs/1999ApJ...520..276M} {520, 276}

\bibitem[\protect\citeauthoryear{{M{\'e}sz{\'a}ros} et~al.,}{{M{\'e}sz{\'a}ros}
  et~al.}{2012}]{Meszaros2012}
{M{\'e}sz{\'a}ros} S.,  et~al., 2012, \mn@doi [\aj]
  {10.1088/0004-6256/144/4/120}, \href
  {https://ui.adsabs.harvard.edu/abs/2012AJ....144..120M} {144, 120}

\bibitem[\protect\citeauthoryear{{Morris}}{{Morris}}{1985}]{Morris1985}
{Morris} S.~L.,  1985, \mn@doi [\apj] {10.1086/163359}, \href
  {https://ui.adsabs.harvard.edu/abs/1985ApJ...295..143M} {295, 143}

\bibitem[\protect\citeauthoryear{{Morton}}{{Morton}}{2015}]{Morton2015}
{Morton} T.~D.,  2015, {isochrones: Stellar model grid package} (\mn@eprint
  {ascl} {1503.010})

\bibitem[\protect\citeauthoryear{{Narayan} \& {Yi}}{{Narayan} \&
  {Yi}}{1995}]{Narayan1995}
{Narayan} R.,  {Yi} I.,  1995, \mn@doi [\apj] {10.1086/176343}, \href
  {https://ui.adsabs.harvard.edu/abs/1995ApJ...452..710N} {452, 710}

\bibitem[\protect\citeauthoryear{{Neilsen}, {Steeghs}  \& {Vrtilek}}{{Neilsen}
  et~al.}{2008}]{Neilsen2008}
{Neilsen} J.,  {Steeghs} D.,   {Vrtilek} S.~D.,  2008, \mn@doi [\mnras]
  {10.1111/j.1365-2966.2007.12599.x}, \href
  {https://ui.adsabs.harvard.edu/abs/2008MNRAS.384..849N} {384, 849}

\bibitem[\protect\citeauthoryear{{Noyes}, {Hartmann}, {Baliunas}, {Duncan}  \&
  {Vaughan}}{{Noyes} et~al.}{1984}]{Noyes1984}
{Noyes} R.~W.,  {Hartmann} L.~W.,  {Baliunas} S.~L.,  {Duncan} D.~K.,
  {Vaughan} A.~H.,  1984, \mn@doi [\apj] {10.1086/161945}, \href
  {https://ui.adsabs.harvard.edu/abs/1984ApJ...279..763N} {279, 763}

\bibitem[\protect\citeauthoryear{{Onken} et~al.,}{{Onken}
  et~al.}{2019}]{Onken2019}
{Onken} C.~A.,  et~al., 2019, \mn@doi [\pasa] {10.1017/pasa.2019.27}, \href
  {https://ui.adsabs.harvard.edu/abs/2019PASA...36...33O} {36, e033}

\bibitem[\protect\citeauthoryear{{Orosz} \& {Hauschildt}}{{Orosz} \&
  {Hauschildt}}{2000}]{Orosz2000}
{Orosz} J.~A.,  {Hauschildt} P.~H.,  2000, \aap, \href
  {https://ui.adsabs.harvard.edu/abs/2000A&A...364..265O} {364, 265}

\bibitem[\protect\citeauthoryear{{{\"O}zel}, {Psaltis}, {Narayan}  \&
  {McClintock}}{{{\"O}zel} et~al.}{2010}]{Ozel2010}
{{\"O}zel} F.,  {Psaltis} D.,  {Narayan} R.,   {McClintock} J.~E.,  2010,
  \mn@doi [\apj] {10.1088/0004-637X/725/2/1918}, \href
  {https://ui.adsabs.harvard.edu/abs/2010ApJ...725.1918O} {725, 1918}

\bibitem[\protect\citeauthoryear{{Papaj}, {Krelowski}  \& {Wegner}}{{Papaj}
  et~al.}{1993}]{Papaj1993}
{Papaj} J.,  {Krelowski} J.,   {Wegner} W.,  1993, \aap, \href
  {https://ui.adsabs.harvard.edu/abs/1993A&A...273..575P} {273, 575}

\bibitem[\protect\citeauthoryear{{Penoyre} \& {Stone}}{{Penoyre} \&
  {Stone}}{2019}]{Penoyre2019}
{Penoyre} Z.,  {Stone} N.~C.,  2019, \mn@doi [\aj] {10.3847/1538-3881/aaf965},
  \href {https://ui.adsabs.harvard.edu/abs/2019AJ....157...60P} {157, 60}

\bibitem[\protect\citeauthoryear{{Pepper} et~al.,}{{Pepper}
  et~al.}{2007}]{Pepper2007}
{Pepper} J.,  et~al., 2007, \mn@doi [\pasp] {10.1086/521836}, \href
  {https://ui.adsabs.harvard.edu/abs/2007PASP..119..923P} {119, 923}

\bibitem[\protect\citeauthoryear{{Petigura}}{{Petigura}}{2015}]{Petigura2015}
{Petigura} E.~A.,  2015, PhD thesis, University of California, Berkeley

\bibitem[\protect\citeauthoryear{{Pinsonneault} et~al.,}{{Pinsonneault}
  et~al.}{2018}]{Pinsonneault2018}
{Pinsonneault} M.~H.,  et~al., 2018, \mn@doi [\apjs]
  {10.3847/1538-4365/aaebfd}, \href
  {https://ui.adsabs.harvard.edu/abs/2018ApJS..239...32P} {239, 32}

\bibitem[\protect\citeauthoryear{{Pogge} et~al.,}{{Pogge}
  et~al.}{2010}]{PoggeMODS}
{Pogge} R.~W.,  et~al., 2010, in Ground-based and Airborne Instrumentation for
  Astronomy III. p. 77350A, \mn@doi{10.1117/12.857215}

\bibitem[\protect\citeauthoryear{{Pojmanski}}{{Pojmanski}}{1997}]{Pojmanski1997}
{Pojmanski} G.,  1997, \actaa, \href
  {https://ui.adsabs.harvard.edu/abs/1997AcA....47..467P} {47, 467}

\bibitem[\protect\citeauthoryear{{Pojmanski}}{{Pojmanski}}{2002}]{Pojmanski2002}
{Pojmanski} G.,  2002, \actaa, \href
  {https://ui.adsabs.harvard.edu/abs/2002AcA....52..397P} {52, 397}

\bibitem[\protect\citeauthoryear{{Poole} et~al.,}{{Poole}
  et~al.}{2008}]{poole08}
{Poole} T.~S.,  et~al., 2008, \mn@doi [\mnras]
  {10.1111/j.1365-2966.2007.12563.x}, \href
  {https://ui.adsabs.harvard.edu/abs/2008MNRAS.383..627P} {383, 627}

\bibitem[\protect\citeauthoryear{{Pourbaix} et~al.,}{{Pourbaix}
  et~al.}{2004}]{Pourbaix2004}
{Pourbaix} D.,  et~al., 2004, \mn@doi [\aap] {10.1051/0004-6361:20041213},
  \href {https://ui.adsabs.harvard.edu/abs/2004A&A...424..727P} {424, 727}

\bibitem[\protect\citeauthoryear{{Poznanski}, {Prochaska}  \&
  {Bloom}}{{Poznanski} et~al.}{2012}]{Poznanski2012}
{Poznanski} D.,  {Prochaska} J.~X.,   {Bloom} J.~S.,  2012, \mn@doi [\mnras]
  {10.1111/j.1365-2966.2012.21796.x}, \href
  {https://ui.adsabs.harvard.edu/abs/2012MNRAS.426.1465P} {426, 1465}

\bibitem[\protect\citeauthoryear{{Price-Whelan} \& {Goodman}}{{Price-Whelan} \&
  {Goodman}}{2018}]{Price-Whelan2018}
{Price-Whelan} A.~M.,  {Goodman} J.,  2018, \mn@doi [\apj]
  {10.3847/1538-4357/aae264}, \href
  {https://ui.adsabs.harvard.edu/abs/2018ApJ...867....5P} {867, 5}

\bibitem[\protect\citeauthoryear{{Price-Whelan}, {Hogg}, {Foreman-Mackey}  \&
  {Rix}}{{Price-Whelan} et~al.}{2017}]{Price-Whelan2017}
{Price-Whelan} A.~M.,  {Hogg} D.~W.,  {Foreman-Mackey} D.,   {Rix} H.-W.,
  2017, \mn@doi [\apj] {10.3847/1538-4357/aa5e50}, \href
  {https://ui.adsabs.harvard.edu/abs/2017ApJ...837...20P} {837, 20}

\bibitem[\protect\citeauthoryear{{Pr{\v{s}}a} et~al.,}{{Pr{\v{s}}a}
  et~al.}{2016}]{Phoebe2016}
{Pr{\v{s}}a} A.,  et~al., 2016, \mn@doi [\apjs] {10.3847/1538-4365/227/2/29},
  \href {https://ui.adsabs.harvard.edu/abs/2016ApJS..227...29P} {227, 29}

\bibitem[\protect\citeauthoryear{{Quataert} \& {Narayan}}{{Quataert} \&
  {Narayan}}{1999}]{Quataert1999}
{Quataert} E.,  {Narayan} R.,  1999, \mn@doi [\apj] {10.1086/307439}, \href
  {https://ui.adsabs.harvard.edu/abs/1999ApJ...520..298Q} {520, 298}

\bibitem[\protect\citeauthoryear{{Ram{\'\i}rez}, {Allende Prieto}  \&
  {Lambert}}{{Ram{\'\i}rez} et~al.}{2013}]{Ramirez2013}
{Ram{\'\i}rez} I.,  {Allende Prieto} C.,   {Lambert} D.~L.,  2013, \mn@doi
  [\apj] {10.1088/0004-637X/764/1/78}, \href
  {https://ui.adsabs.harvard.edu/abs/2013ApJ...764...78R} {764, 78}

\bibitem[\protect\citeauthoryear{{Reimers}}{{Reimers}}{1975}]{Reimers1975}
{Reimers} D.,  1975, Memoires of the Societe Royale des Sciences de Liege,
  \href {https://ui.adsabs.harvard.edu/abs/1975MSRSL...8..369R} {8, 369}

\bibitem[\protect\citeauthoryear{{Ricker} et~al.,}{{Ricker}
  et~al.}{2015}]{Ricker2015}
{Ricker} G.~R.,  et~al., 2015, \mn@doi [Journal of Astronomical Telescopes,
  Instruments, and Systems] {10.1117/1.JATIS.1.1.014003}, \href
  {https://ui.adsabs.harvard.edu/abs/2015JATIS...1a4003R} {1, 014003}

\bibitem[\protect\citeauthoryear{{Rivinius}, {Baade}, {Hadrava}, {Heida}  \&
  {Klement}}{{Rivinius} et~al.}{2020}]{Rivinius2020}
{Rivinius} T.,  {Baade} D.,  {Hadrava} P.,  {Heida} M.,   {Klement} R.,  2020,
  \mn@doi [\aap] {10.1051/0004-6361/202038020}, \href
  {https://ui.adsabs.harvard.edu/abs/2020A&A...637L...3R} {637, L3}

\bibitem[\protect\citeauthoryear{{Robson}, {Cornish}  \& {Liu}}{{Robson}
  et~al.}{2019}]{Robson2019}
{Robson} T.,  {Cornish} N.~J.,   {Liu} C.,  2019, \mn@doi [Classical and
  Quantum Gravity] {10.1088/1361-6382/ab1101}, \href
  {https://ui.adsabs.harvard.edu/abs/2019CQGra..36j5011R} {36, 105011}

\bibitem[\protect\citeauthoryear{{Roming} et~al.,}{{Roming}
  et~al.}{2005}]{roming05}
{Roming} P. W.~A.,  et~al., 2005, \mn@doi [\ssr] {10.1007/s11214-005-5095-4},
  \href {https://ui.adsabs.harvard.edu/abs/2005SSRv..120...95R} {120, 95}

\bibitem[\protect\citeauthoryear{{Sch{\"o}nrich}, {Binney}  \&
  {Dehnen}}{{Sch{\"o}nrich} et~al.}{2010}]{Schonrich2010}
{Sch{\"o}nrich} R.,  {Binney} J.,   {Dehnen} W.,  2010, \mn@doi [\mnras]
  {10.1111/j.1365-2966.2010.16253.x}, \href
  {https://ui.adsabs.harvard.edu/abs/2010MNRAS.403.1829S} {403, 1829}

\bibitem[\protect\citeauthoryear{{Shahbaz}, {Russell}, {Zurita}, {Casares},
  {Corral-Santana}, {Dhillon}  \& {Marsh}}{{Shahbaz}
  et~al.}{2013}]{Shahbaz2013}
{Shahbaz} T.,  {Russell} D.~M.,  {Zurita} C.,  {Casares} J.,  {Corral-Santana}
  J.~M.,  {Dhillon} V.~S.,   {Marsh} T.~R.,  2013, \mn@doi [\mnras]
  {10.1093/mnras/stt1212}, \href
  {https://ui.adsabs.harvard.edu/abs/2013MNRAS.434.2696S} {434, 2696}

\bibitem[\protect\citeauthoryear{{Shao} \& {Li}}{{Shao} \&
  {Li}}{2019}]{Shao2019}
{Shao} Y.,  {Li} X.-D.,  2019, \mn@doi [\apj] {10.3847/1538-4357/ab4816}, \href
  {https://ui.adsabs.harvard.edu/abs/2019ApJ...885..151S} {885, 151}

\bibitem[\protect\citeauthoryear{{Shappee} et~al.,}{{Shappee}
  et~al.}{2014}]{Shappee2014}
{Shappee} B.~J.,  et~al., 2014, \mn@doi [\apj] {10.1088/0004-637X/788/1/48},
  \href {https://ui.adsabs.harvard.edu/abs/2014ApJ...788...48S} {788, 48}

\bibitem[\protect\citeauthoryear{{Shenar} et~al.,}{{Shenar}
  et~al.}{2020}]{Shenar2020}
{Shenar} T.,  et~al., 2020, \mn@doi [\aap] {10.1051/0004-6361/202038275}, \href
  {https://ui.adsabs.harvard.edu/abs/2020A&A...639L...6S} {639, L6}

\bibitem[\protect\citeauthoryear{{Siverd} et~al.,}{{Siverd}
  et~al.}{2012}]{Siverd2012}
{Siverd} R.~J.,  et~al., 2012, \mn@doi [\apj] {10.1088/0004-637X/761/2/123},
  \href {https://ui.adsabs.harvard.edu/abs/2012ApJ...761..123S} {761, 123}

\bibitem[\protect\citeauthoryear{{Skrutskie} et~al.,}{{Skrutskie}
  et~al.}{2006}]{Skrutskie2006}
{Skrutskie} M.~F.,  et~al., 2006, \mn@doi [\aj] {10.1086/498708}, \href
  {https://ui.adsabs.harvard.edu/abs/2006AJ....131.1163S} {131, 1163}

\bibitem[\protect\citeauthoryear{{Sneden}}{{Sneden}}{1973}]{Sneden1973}
{Sneden} C.,  1973, \mn@doi [\apj] {10.1086/152374}, \href
  {https://ui.adsabs.harvard.edu/abs/1973ApJ...184..839S} {184, 839}

\bibitem[\protect\citeauthoryear{{Soderblom}, {Stauffer}, {Hudon}  \&
  {Jones}}{{Soderblom} et~al.}{1993}]{Soderblom1993}
{Soderblom} D.~R.,  {Stauffer} J.~R.,  {Hudon} J.~D.,   {Jones} B.~F.,  1993,
  \mn@doi [\apjs] {10.1086/191767}, \href
  {https://ui.adsabs.harvard.edu/abs/1993ApJS...85..315S} {85, 315}

\bibitem[\protect\citeauthoryear{{Stanek} \& {Garnavich}}{{Stanek} \&
  {Garnavich}}{1998}]{Stanek1998}
{Stanek} K.~Z.,  {Garnavich} P.~M.,  1998, \mn@doi [\apjl] {10.1086/311539},
  \href {https://ui.adsabs.harvard.edu/abs/1998ApJ...503L.131S} {503, L131}

\bibitem[\protect\citeauthoryear{{Stassun} et~al.,}{{Stassun}
  et~al.}{2019}]{2019TIC}
{Stassun} K.~G.,  et~al., 2019, \mn@doi [\aj] {10.3847/1538-3881/ab3467}, \href
  {https://ui.adsabs.harvard.edu/abs/2019AJ....158..138S} {158, 138}

\bibitem[\protect\citeauthoryear{{Strader} et~al.,}{{Strader}
  et~al.}{2015}]{Strader2015}
{Strader} J.,  et~al., 2015, \mn@doi [\apjl] {10.1088/2041-8205/804/1/L12},
  \href {https://ui.adsabs.harvard.edu/abs/2015ApJ...804L..12S} {804, L12}

\bibitem[\protect\citeauthoryear{{Straizys} \& {Kuriliene}}{{Straizys} \&
  {Kuriliene}}{1981}]{Straizys1981}
{Straizys} V.,  {Kuriliene} G.,  1981, \mn@doi [\apss] {10.1007/BF00652936},
  \href {https://ui.adsabs.harvard.edu/abs/1981Ap&SS..80..353S} {80, 353}

\bibitem[\protect\citeauthoryear{{Strassmeier}, {Weber}, {Granzer}  \&
  {J{\"a}rvinen}}{{Strassmeier} et~al.}{2012}]{Strassmeier2012}
{Strassmeier} K.~G.,  {Weber} M.,  {Granzer} T.,   {J{\"a}rvinen} S.,  2012,
  \mn@doi [Astronomische Nachrichten] {10.1002/asna.201211719}, \href
  {https://ui.adsabs.harvard.edu/abs/2012AN....333..663S} {333, 663}

\bibitem[\protect\citeauthoryear{{Strassmeier} et~al.,}{{Strassmeier}
  et~al.}{2015}]{Strassmeier2015}
{Strassmeier} K.~G.,  et~al., 2015, \mn@doi [Astronomische Nachrichten]
  {10.1002/asna.201512172}, \href
  {https://ui.adsabs.harvard.edu/abs/2015AN....336..324S} {336, 324}

\bibitem[\protect\citeauthoryear{{Strassmeier}, {Ilyin}  \&
  {Steffen}}{{Strassmeier} et~al.}{2018}]{Strassmeier2018}
{Strassmeier} K.~G.,  {Ilyin} I.,   {Steffen} M.,  2018, \mn@doi [\aap]
  {10.1051/0004-6361/201731631}, \href
  {https://ui.adsabs.harvard.edu/abs/2018A&A...612A..44S} {612, A44}

\bibitem[\protect\citeauthoryear{{Suwa}, {Yoshida}, {Shibata}, {Umeda}  \&
  {Takahashi}}{{Suwa} et~al.}{2018}]{Suwa2018}
{Suwa} Y.,  {Yoshida} T.,  {Shibata} M.,  {Umeda} H.,   {Takahashi} K.,  2018,
  \mn@doi [\mnras] {10.1093/mnras/sty2460}, \href
  {https://ui.adsabs.harvard.edu/abs/2018MNRAS.481.3305S} {481, 3305}

\bibitem[\protect\citeauthoryear{{Swihart} et~al.,}{{Swihart}
  et~al.}{2018}]{Swihart2018}
{Swihart} S.~J.,  et~al., 2018, \mn@doi [\apj] {10.3847/1538-4357/aadcab},
  \href {https://ui.adsabs.harvard.edu/abs/2018ApJ...866...83S} {866, 83}

\bibitem[\protect\citeauthoryear{{Ter Braak}}{{Ter Braak}}{2006}]{TerBraak2006}
{Ter Braak} C. J.~F.,  2006, \mn@doi [Statistics and Computing]
  {10.1007/s11222-006-8769-1}, \href
  {https://ui.adsabs.harvard.edu/abs/2006S&C....16..239T} {16, 239}

\bibitem[\protect\citeauthoryear{{Thompson} et~al.,}{{Thompson}
  et~al.}{2019}]{Thompson2019}
{Thompson} T.~A.,  et~al., 2019, \mn@doi [Science] {10.1126/science.aau4005},
  \href {https://ui.adsabs.harvard.edu/abs/2019Sci...366..637T} {366, 637}

\bibitem[\protect\citeauthoryear{{Tremblay}, {Cummings}, {Kalirai},
  {G{\"a}nsicke}, {Gentile-Fusillo}  \& {Raddi}}{{Tremblay}
  et~al.}{2016}]{Tremblay2016}
{Tremblay} P.~E.,  {Cummings} J.,  {Kalirai} J.~S.,  {G{\"a}nsicke} B.~T.,
  {Gentile-Fusillo} N.,   {Raddi} R.,  2016, \mn@doi [\mnras]
  {10.1093/mnras/stw1447}, \href
  {https://ui.adsabs.harvard.edu/abs/2016MNRAS.461.2100T} {461, 2100}

\bibitem[\protect\citeauthoryear{{Tsantaki}, {Andreasen}, {Teixeira}, {Sousa},
  {Santos}, {Delgado-Mena}  \& {Bruzual}}{{Tsantaki}
  et~al.}{2018}]{Tsantaki2018}
{Tsantaki} M.,  {Andreasen} D.~T.,  {Teixeira} G.~D.~C.,  {Sousa} S.~G.,
  {Santos} N.~C.,  {Delgado-Mena} E.,   {Bruzual} G.,  2018, \mn@doi [\mnras]
  {10.1093/mnras/stx2564}, \href
  {https://ui.adsabs.harvard.edu/abs/2018MNRAS.473.5066T} {473, 5066}

\bibitem[\protect\citeauthoryear{Tsantaki, Andreasen  \& Teixeira}{Tsantaki
  et~al.}{2020}]{Tsantaki2020}
Tsantaki M.,  Andreasen D.,   Teixeira G.,  2020, \mn@doi [Journal of Open
  Source Software] {10.21105/joss.02048}, 5, 2048

\bibitem[\protect\citeauthoryear{{Vallely}, {Kochanek}, {Stanek}, {Fausnaugh}
  \& {Shappee}}{{Vallely} et~al.}{2020}]{Vallely2020}
{Vallely} P.~J.,  {Kochanek} C.~S.,  {Stanek} K.~Z.,  {Fausnaugh} M.,
  {Shappee} B.~J.,  2020, arXiv e-prints, \href
  {https://ui.adsabs.harvard.edu/abs/2020arXiv201006596V} {p. arXiv:2010.06596}

\bibitem[\protect\citeauthoryear{{Verbunt} \& {Phinney}}{{Verbunt} \&
  {Phinney}}{1995}]{Verbunt1995}
{Verbunt} F.,  {Phinney} E.~S.,  1995, \aap, \href
  {https://ui.adsabs.harvard.edu/abs/1995A&A...296..709V} {296, 709}

\bibitem[\protect\citeauthoryear{{Vogt} et~al.,}{{Vogt}
  et~al.}{1994}]{Vogt1994}
{Vogt} S.~S.,  et~al., 1994, in {Crawford} D.~L.,  {Craine} E.~R.,  eds,
  Society of Photo-Optical Instrumentation Engineers (SPIE) Conference Series
  Vol. 2198, Instrumentation in Astronomy VIII. p.~362,
  \mn@doi{10.1117/12.176725}

\bibitem[\protect\citeauthoryear{{Wagner}, {Kreidl}, {Howell}  \&
  {Starrfield}}{{Wagner} et~al.}{1992}]{Wagner1992}
{Wagner} R.~M.,  {Kreidl} T.~J.,  {Howell} S.~B.,   {Starrfield} S.~G.,  1992,
  \mn@doi [\apjl] {10.1086/186680}, \href
  {https://ui.adsabs.harvard.edu/abs/1992ApJ...401L..97W} {401, L97}

\bibitem[\protect\citeauthoryear{{Watson}, {Henden}  \& {Price}}{{Watson}
  et~al.}{2006}]{Watson2006}
{Watson} C.~L.,  {Henden} A.~A.,   {Price} A.,  2006, Society for Astronomical
  Sciences Annual Symposium, \href
  {https://ui.adsabs.harvard.edu/abs/2006SASS...25...47W} {25, 47}

\bibitem[\protect\citeauthoryear{{Weber}, {Granzer}, {Strassmeier}  \&
  {Woche}}{{Weber} et~al.}{2008}]{Weber2008}
{Weber} M.,  {Granzer} T.,  {Strassmeier} K.~G.,   {Woche} M.,  2008, in
  {Bridger} A.,  {Radziwill} N.~M.,  eds,  Society of Photo-Optical
  Instrumentation Engineers (SPIE) Conference Series Vol. 7019, Advanced
  Software and Control for Astronomy II. p. 70190L, \mn@doi{10.1117/12.790687}

\bibitem[\protect\citeauthoryear{{Wilson} \& {Sofia}}{{Wilson} \&
  {Sofia}}{1976}]{Wilson1976}
{Wilson} R.~E.,  {Sofia} S.,  1976, \mn@doi [\apj] {10.1086/154062}, \href
  {https://ui.adsabs.harvard.edu/abs/1976ApJ...203..182W} {203, 182}

\bibitem[\protect\citeauthoryear{{Wright} et~al.,}{{Wright}
  et~al.}{2010}]{Wright2010AJ}
{Wright} E.~L.,  et~al., 2010, \mn@doi [\aj] {10.1088/0004-6256/140/6/1868},
  \href {https://ui.adsabs.harvard.edu/abs/2010AJ....140.1868W} {140, 1868}

\bibitem[\protect\citeauthoryear{{Wu} et~al.,}{{Wu}
  et~al.}{2015}]{Jianfeng2015}
{Wu} J.,  et~al., 2015, \mn@doi [\apj] {10.1088/0004-637X/806/1/92}, \href
  {https://ui.adsabs.harvard.edu/abs/2015ApJ...806...92W} {806, 92}

\bibitem[\protect\citeauthoryear{{Zurita}, {Casares}  \& {Shahbaz}}{{Zurita}
  et~al.}{2003}]{Zurita2003}
{Zurita} C.,  {Casares} J.,   {Shahbaz} T.,  2003, \mn@doi [\apj]
  {10.1086/344534}, \href
  {https://ui.adsabs.harvard.edu/abs/2003ApJ...582..369Z} {582, 369}

\bibitem[\protect\citeauthoryear{{van Belle} et~al.,}{{van Belle}
  et~al.}{1999}]{vanBelle1999}
{van Belle} G.~T.,  et~al., 1999, \mn@doi [\aj] {10.1086/300677}, \href
  {https://ui.adsabs.harvard.edu/abs/1999AJ....117..521V} {117, 521}

\makeatother
\end{thebibliography}



\clearpage
\appendix
\section{The origin of the radial velocity residuals}
\label{app:tertiary}

S12 interpreted V723~Mon as a triple system where the companion is a binary
with period $P_{\rm in} = P_{\rm orb}/3 \simeq 20$~days.  G14 argued against this hypothesis,
but there are significant RV residuals at $P_{\rm orb}/2$ or $P_{\rm orb}/3$ depending
on whether the model for the orbit of the giant is circular or elliptical (see \S\ref{section:orbit}
and Fig~\ref{rvs}).  We know from the limits on stellar companions in \S\ref{section:limits}
that a star cannot dominate the mass of the companion, but 
there are also two independent arguments against any such binary, stellar or not.
We do, however, have a hypothesis for the origin of these RV residuals.

The first argument against a $\sim 20$~day inner binary is the dynamical 
stability argument raised by G14.  For our nominal parameters, the
outer orbit has a total mass of $\simeq 4M_\odot$ (in round numbers) and a 
semi-major axis of $100~ R_\odot$ while the inner orbit has a mass of 
$3~ M_\odot$. The $45~ R_\odot$ semi-major axis of a $20$~day
binary is very close to the Roche limit around the companion of $49~ R_\odot$.
A rough estimate of the largest semi-major axis that could be dynamically
stable given the outer orbit and the mass ratios is $32~ R_\odot$ based
on \cite{Mardling2001}.  In short, G14 was correct to hypothesize that
such an orbit should be dynamically unstable.

We experimented with numerically integrating planar 3-body orbits starting
from nominally circular orbits as initial conditions for $\sim 100$ orbits of the outer binary
and  varying mass ratios of the inner binary. High mass ratio inner
binaries were generally very unstable, presumably because 
the lighter star in the inner binary is trying to maintain an orbital
radius close to the full semi-major axis of $45~ R_\odot$ in this limit.  
Equal mass inner binaries tended to be more stable, presumably because
each star now only has an orbit of $\sim 23~ R_\odot$ about the center of
mass of the inner binary.  Nonetheless, many trials resulted in the 
destruction of the system well before $100$ orbits were completed. These
results were not unique to picking a truly resonant $P_{\rm orb}/3$ period for 
the inner binary.  
We suspect, but do not investigate here, that essentially all such inner
binaries are unstable on long time scales.  The inner binary would have
to have a period shorter than $\sim 12$~days to satisfy the dynamical stability
criterion, although such systems may still be secularly unstable.

The other problem with longer period inner binaries is that they perturb
the outer orbit and produce time varying tidal forces on the
giant even if stable.  In particular, the wider inner binaries
would generally drive the outer orbit to be significantly elliptical in
our numerical experiments even if they were stable over the 100 orbits.
Making the companion a binary also means that the tidal forces on the
giant are time variable.  If the companion is a single star, the
amplitude of the elliptical variability depends on $(R_{\rm giant}/a)^3$
while if it is an equal mass binary it depends on 
\begin{equation}
       { x \over 2 } \left( { R_{\rm giant} \over d_1 }\right)^2
   + { 1-x \over 2 } \left( { R_{\rm giant} \over d_2 }\right)^3
\end{equation}
where $d_1$ and $d_2$ are the distances to the two stars which comprise
fractions $x$ and $1-x$ of the mass of the inner binary.  Compared
to a single companion, there is a fractional fluctuating tidal
amplitude for circular orbits of order
\begin{equation}
     3\left(1-2x\right) { a_{in} \over a} \cos \omega_t t +
     { 15 \over 4 }\left(1-3x+3x^2\right) {a_{in}^2 \over a^2} \cos 2\omega_t t
\end{equation}
where the frequency $\omega_t = \omega_{\rm in}-\omega$ is the frequency
difference between the inner and outer binaries. In particular, for an equal 
mass ($x=1/2$) inner binary, the peak-to-peak fractional change in the tidal forcing is
\begin{equation}
     { 15 \over 8 } \left( {a_{\rm in} \over a } \right)^2,
\end{equation}
so a $P_{\rm in}=20$~day period inner binary would produce $\sim 38\%$ peak-to-peak
fluctuations in the tidal force with a period of $15$~days.  No such residuals are
seen in the residuals from the ellipsoidal model of the light curves at this or any
similar period (see Appendix~\ref{app:shortimes}). To have fractional fluctuations in the tidal forcing  smaller than $f = 0.1 f_{10}$ requires an equal mass inner binary to be more compact
than $a_{\rm in} < 23 f_{10}^{1/2} R_\odot$ and to have a period $P_{\rm in}<7.4f_{10}^{3/2}$~days.

These arguments appear to strongly rule out the inner binary proposed by S12.
We instead suspect that the RV residuals and much of the evidence for a binary
companion are driven by the consequences of making RV observations of a high amplitude
ELL variable. This effect has previously been discussed \citep{Hill1989, Eaton2008} in the context of contact/semi-detached eclipsing binaries and ellipsoidal variables in close orbits \citep{Wilson1976}. The very similar issues for observations of stars with
dynamical tides have also been discussed (e.g. \citealt{Arras2012}, \citealt{Penoyre2019}) 
recently.

A tidally locked star is simply rotating at the orbital frequency,
so the rotation velocity at the surface scales with the cylindrical radius from the
rotation axis.   Thus, the rotational velocity is larger on the long axis of the
star than on the short axis.  When viewed along a principal axis, there is the
usual cancellation of the contributions from the parts of the star rotating towards
and away from the observer.  However, when viewed at an intermediate direction, the
contribution of one sign of the rotation comes from the slower moving short axis
while the contribution from the other sign comes from the faster moving long
axis.  This makes a contribution to the observed radial velocities with a period
$P_{\rm orb}/2$ because it is a $m=2$ perturbation just like the tidal distortions. If
the orbit is genuinely circular, a fit to the RV curve will instead yield a
small eccentricity to try to remove the contribution from this effect (see \citealt{Eaton2008}).  This
should mean that the dominant velocity residual will now have a period $P_{\rm orb}/3$,
and we think this is the likely origin of the $P_{\rm orb}/3$ signal found by \citet{Strassmeier2012}.  

\section{Short timescale variability}
\label{app:shortimes}

We searched for additional variability on three broad time scales.  First, we looked for extra variability on the time scales of weeks or longer in the residuals of the ASAS and KELT light curves after subtracting the best model for the ellipsoidal variability. Next, we looked for variability on the time scale of days in the residuals of the TESS light curve. Finally, we searched for very short time scale variability during several periods of very high cadence ROAD observations.

As discussed in Appendix~\ref{app:tertiary}, a sufficiently wide binary secondary will produce strong fluctuations in the strength of the tides on the giant at periods with frequencies that are either the difference in frequency between the inner and outer binary $w_t$ or twice that frequency, with the perturbations at $w_t$ requiring an un-equal mass binary companion.  In particular, for the $20$~day period inner binary proposed by S12, we would expect strong fluctuating tides on periods of 30 and
15 days, although the 30 day period could be partly absorbed into the ELL fit. We looked for periodic signals in the residuals of the fits to the ASAS and KELT light curves using  \verb"Period04" \citep{Lenz2005}, and found none that were significant (a signal-to-noise ratio $>5$).  Given the level of the residuals and the expected amplitude of the perturbations for a $20$~day inner binary, this appears to strongly rule out such a companion.

\begin{figure*}
	\includegraphics[width=0.90\textwidth]{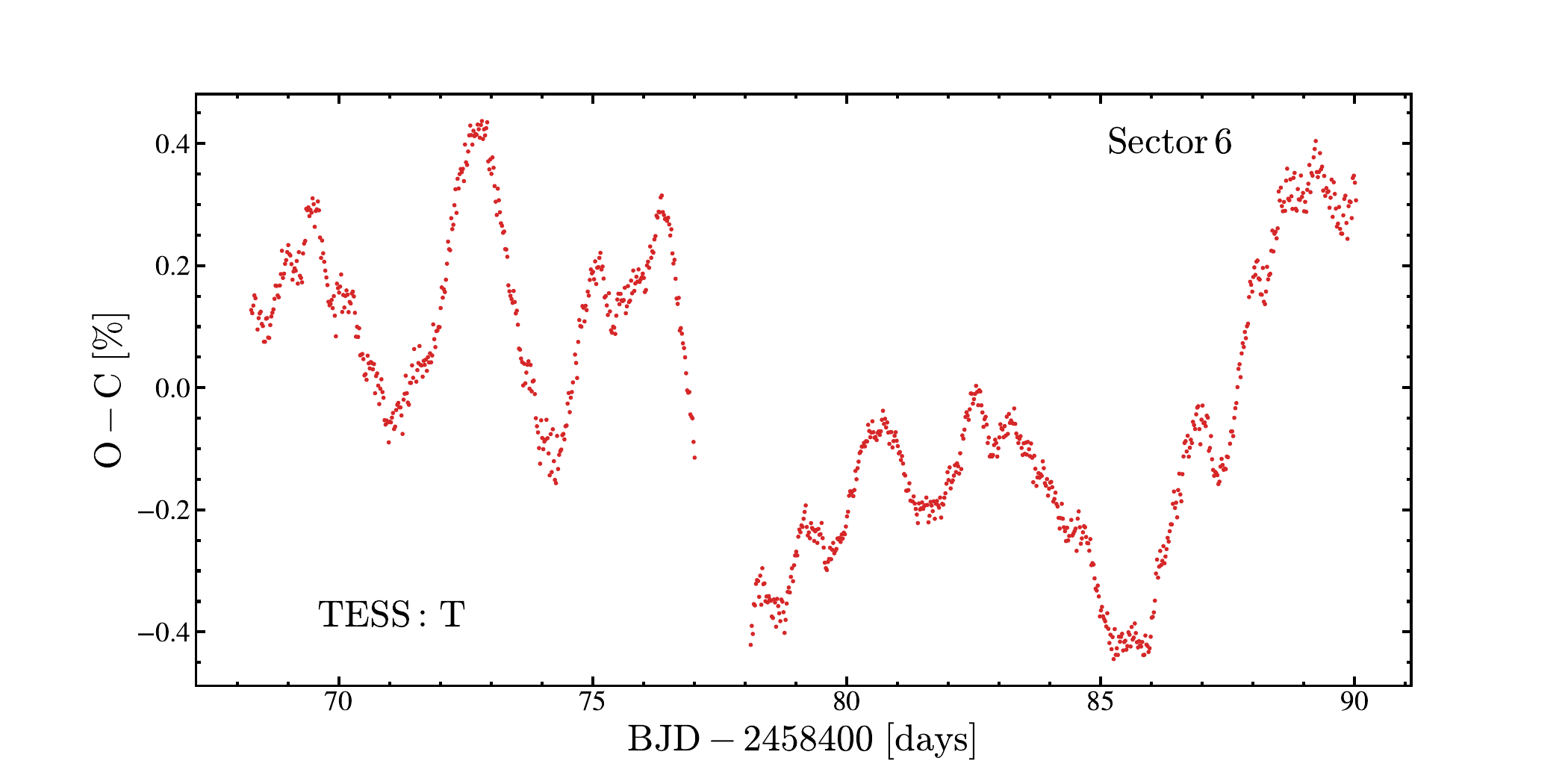}
    \caption{The light curve residuals for the \textit{TESS} sector 6 data after the PHOEBE model is subtracted.}
    \label{tessresid}
\end{figure*}

We see clear evidence of structure in the \textit{TESS} residuals on short time scales (Figure \ref{tessresid}) with an amplitude of $0.4\%$.  At least in this short TESS light curve, we find no evidence for a significant periodic signal on a time scale of days.  These time scales are interesting because the roughly correspond to the expected frequency of maximum power for asteroseismic oscillations. This frequency is defined by
\begin{equation}
    \nu_{\rm max}=\nu_{\rm max,\odot}\bigg(\frac{g}{g_\odot}\bigg)\bigg(\frac{T_{\rm eff}}{T_{\rm eff,\odot}}\bigg)^{-1/2},
\end{equation} where $\nu_{\rm max,\odot}=3100~\mu$Hz, $T_{\rm eff,\odot}=5777$~K and $g_\odot=2.7\times10^4~\rm cm^2/s$ \citep{Brown1991,Kjeldsen1995}. Given the spectroscopic parameters in $\S$\ref{section:giant}, we estimate that $\nu_{\rm max}\approx6.8~\mu$Hz, which corresponds to a period of $P_{\rm osc}\approx1.7$~days. With no clear peaks in the periodogram of the existing TESS data, we cannot presently use asteroseismology as an additional probe of the mass of the giant, but this may change with further analysis and additional TESS data.

Finally, many compact-object systems (e.g., \citealt{Wagner1992,Zurita2003,Shahbaz2013}) show rapid variability on time scales of minutes to hours.  We used the ROAD telescope \citep{Hambsch2012} to continuously observe the system for about 2 hours on multiple nights, obtaining in total more than 3,000 10-sec $B$-band exposures. We have reduced these data using the ASAS-SN version of the \citet{alard2000} image subtraction software. We do not find any evidence for short ($<$1 hour) timescale $B$-band variability, with the rms scatter during individual nights below $0.01\;$mag. Given that the ``second light'' contribution to the total $B$-band flux of V723 Mon is about 60\%, this translates to the short-timescale variability of this component being less that $\sim2$\%. However, on longer timescales of hours to days $B$-band variability at the $\sim2$\% level is still possible (on top of the observed ellipsoidal modulations discussed in $\S$\ref{section:phoebe}).

\bsp	
\label{lastpage}
\end{document}